\documentclass[12pt]{JHEP}

\usepackage{epsfig}
\usepackage{latexsym}
\usepackage{cite}
\newcommand{\newsection}{    
\setcounter{equation}{0}
\section}
\renewcommand{\appendix}[1]{
    \addtocounter{section}{1}
    \setcounter{equation}{0}
    \renewcommand{\thesection}{\Alph{section}}
    \section*{Appendix \thesection\protect\indent #1}
    \addcontentsline{toc}{section}{Appendix \thesection\ \ \ #1}
}
\newcommand\encadremath[1]{\vbox{\hrule\hbox{\vrule\kern8pt
\vbox{\kern8pt \hbox{$\displaystyle #1$}\kern8pt}
\kern8pt\vrule}\hrule}}
\def\enca#1{\vbox{\hrule\hbox{
\vrule\kern8pt\vbox{\kern8pt \hbox{$\displaystyle #1$}
\kern8pt} \kern8pt\vrule}\hrule}}

\newcommand\figureframex[3]{
\begin{figure}[bth]
\hrule\hbox{\vrule\kern8pt
\vbox{\kern8pt \vbox{
\begin{center}
{\mbox{\epsfxsize=#1.truecm\epsfbox{#2}}}
\end{center}
\caption{#3}
}\kern8pt}
\kern8pt\vrule}\hrule
\end{figure}
}
\newcommand\figureframey[3]{
\begin{figure}[bth]
\hrule\hbox{\vrule\kern8pt
\vbox{\kern8pt \vbox{
\begin{center}
{\mbox{\epsfysize=#1.truecm\epsfbox{#2}}}
\end{center}
\caption{#3}
}\kern8pt}
\kern8pt\vrule}\hrule
\end{figure}
}

\newcommand{\eq}[1]{eq.~(\ref{#1})}

\newcommand{\beq}{\begin{equation}}
\newcommand{\eeq}{\end{equation}}
\newcommand{\bea}{\begin{eqnarray}}
\newcommand{\eea}{\end{eqnarray}}

\newcommand\eop{\vspace*{\fill}\pagebreak}

\newcommand{\vs}{\vspace{0.7cm}}
\renewcommand{\and}{{\qquad {\rm and} \qquad}}

\newcommand{\virg}{{\qquad , \qquad}}

\newcommand{\tr}{{\,\rm tr}\:}

\newcommand{\Res}{\mathop{\,\rm Res\,}}

\newcommand{\td}[1]{{\tilde{#1}}}

\newcommand{\om}{\omega}

\renewcommand{\d}{{{\partial}}}

\newcommand{\Pint}{{\int\kern -1.em -\kern-.25em}}

\renewcommand{\a}{o}

\newcommand{\ovl}{\overline}
\newcommand{\bfx}{{\mathbf x}}

\newcommand{\bfp}{{\mathbf p}}

\preprint{SPhT-T05/045, ccsd-00004752, math-ph/0504058}

\title{Topological expansion of the 2-matrix model correlation functions:
diagrammatic rules for a residue formula}

\author{B.\ Eynard, N. \ Orantin \\
Service de Physique Th\'eorique de Saclay, CEA/DSM/SPhT,\\
Unit\'e de Recherche associ\'ee au CNRS (URA D2306), CEA Saclay,\\
F-91191 Gif-sur-Yvette Cedex, France.\\
 E-mail: eynard@spht.saclay.cea.fr, orantin@spht.saclay.cea.fr}

\abstract{We solve the loop equations of the hermitian 2-matrix
model to all orders in the topological $1/N^2$ expansion,
i.e. we obtain all non-mixed correlation functions, in terms of residues on an algebraic curve.
We give two representations of those residues as Feynman-like graphs,
one of them involving only cubic vertices. }


\begin{document}






%





\vspace{26pt}
\pagestyle{plain}
\setcounter{page}{1}


\newsection{Introduction}

The purpose of this article is to generalize the  method invented
in \cite{eynloop1mat}, for the 2-matrix model. The method of
\cite{eynloop1mat} is a diagrammatic technique for computing
correlation functions of the 1-matrix model in terms of residues
on some algebraic curve.

\smallskip

Random matrix models play an important role in physics and
mathematics \cite{Mehta}, and have a wealth of applications which
are too long to list here. In this article, we consider ``formal''
random matrix integrals, which are known to be generating
functions for counting some classes of discrete surfaces
\cite{ZJDFG, thooft, BIPZ, courseynard, eynhabilit}.

The partition function, free energy and correlation functions are all generating functions enumerating some kinds of graphs (respectively closed graphs, connected closed graphs, open graphs),
which graphs can be seen as discrete surfaces.

In the formal model, the size $N$ of matrices, is just a complex parameter, it needs not be an integer, and all observables (free energy, correlation functions)
always have a $1/N$ expansion, because for each power of the expansion parameters, there is only a finite number of graphs with a given power of $N$.
The power of $N$ in a graph is its Euler characteristic, and thus the $1/N$ expansion is known as the ``topological expansion'' discovered by 't Hooft \cite{thooft}.
In the formal model, $N$ is thus an expansion parameter, and working order by order in $N$ enumerates only discrete surfaces of a given topology \cite{BIPZ}.
An efficient method for dealing with this formal model is to consider the Schwinger-Dyson equations, called
loop equations in this context \cite{ZJDFG, staudacher}.

To large $N$ limit (i.e. planar topologies), the solution of loop equations is known to be related to Toda hierarchy \cite{Virasoro,KMMM, ZJZ, PZJ}.
For this reason, the large $N$ expansion of matrix models plays an important role in integrable systems, and in many areas of physics \cite{Kos}.
It was understood by \cite{DV} that the low energy effective action of some string theory models is also described by matrix models.

In the beginning, formal matrix models were considered omlyin their 1-cut phase, because a potential which is a small deformation of a quadratic one,
must have only one well, i.e. the variables perturbatively explore only one well.
However, a $N\times N$ matrix has $N$ eigenvalues, and even though each of them can explore perturbatively only one well, they do not need explore all the same well.
That gives ``multicut'' solutions of matrix models, where the number of eigenvalues near each extremum of the potential is fixed (fixed filling fractions).
Multicut solutions play an important role in string theory, as they describe multi-particle states \cite{DV,DW}.
Multicut solutions correspond to enumerating surfaces with contact terms, which can be called ``foam of surfaces'' as described in \cite{BDE, eynhabilit}.

\medskip

The link between formal matrix models (which always have a $1/N$ expansion) and convergent matrix integrals (which have a $1/N$ expansion only in the 1-cut case under certain assumptions),
has been better understood after the work of \cite{BDE}.
We emphasize again, that the results developed in this article concern the formal matrix model with fixed filling fractions, and should not be applied to
convergent matrix model directly.

\medskip

Recently, it has progressively become clear that large $N$
expansion of random matrix models has a strong link with algebraic
geometry \cite{KazMar}. The free energy and correlation functions have been
computed in terms of properties of an algebraic curve. The large
$N$ limit of the 1-point correlation function (called the
resolvent) is solution of an algebraic equation, which thus
defines an algebraic curve. There have been many works which
computed free energy and correlation functions in terms of that
algebraic curve. The leading order resolvent and free energy were
computed in the 1-cut case (algebraic curve of genus zero) in the
pioneering work of \cite{BIPZ}, then some recursive method
for computing correlation functions and free energy to all orders
in $1/N$ were invented by \cite{ACM,ACKM}. Those methods were first
limited to 1-matrix case and 1-cut.

Then for 1-matrix several works have dealt with multicut: Akeman
and Ambj{\o}rn found the first subleading term for the multicut
resolvent and the 2-cut free energy \cite{Ak96,AkAm}, Chekhov \cite{Chekh}
and one of the authors together with Kokotov and Korotkin \cite{EKK}
found simultaneously the first subleading term for the multi-cut free energy
. Then a (non-recursive) diagrammatic method was
invented in \cite{eynloop1mat} to find all correlation functions
to all orders, in the multicut case.

\medskip

The 1-matrix model, corresponds to hyper elliptical curves only. In order to have more general algebraic curves, one needs at least a 2-matrix model.
For the 2-matrix models, the loop equations have been known since \cite{staudacher},
and have been written in a concise form in \cite{eynchain, eynchaint, eynmultimat}.
They have been used to find the subleading term of the free energy, first in the genus zero case in \cite{eynm2m}, then in the genus 1 case in \cite{eynm2mg1},
and with arbitrary genus in \cite{EKK}.
The purpose of this article is to generalize the diagrammatic method of \cite{eynloop1mat} for the computation of non-mixed correlation functions
in the 2-matrix case. We solve the loop equations and present their solutions (the non-mixed correlation function's expansion) under two different diagrammatic forms.
We first build a cubic diagrammatic representation before presenting an effective non cubic theory.

\bigskip

{\bf Outline of the article:}
\begin{itemize}
\item In section 2, we introduce the model and our notations.

\item Section 3 is dedicated to the derivation of loop equations. We derive the fundamental "master loop equation"
before deriving loop equations whose solutions are non-mixed correlation functions

\item In section 4, we show how a compact Riemann surface arises from the leading order of the master loop equation
and present notations and tools of algebraic geometry needed for the computation of correlation functions.

\item In section 5, we present a diagrammatic solution of the loop equations as cubic Feynman-like graphs.

\item Section 6 is dedicated to the presentation of another representation of the non-mixed correlation functions
as graphs of a non cubic effective theory.

\item In section 7, we study the example of the gaussian case corresponding to the 1-matrix model limit.

\end{itemize}

\newsection{Definitions and notations}

\subsection{Definition of the formal 2-matrix model with fixed filling fractions}
\label{secdef}

In this article, we are interested in the study of the formal-two-matrix-model
and the computation of a whole family of observables.
The partition function $Z$ is the formal matrix integral:
\beq\label{defZ}
Z:=\int_{H_n\times H_n} dM_1 dM_2\, e^{-N Tr(V_1(M_1) + V_2(M_2) - M_1 M_2 )}
\eeq
where $M_1$ and $M_2$ are two $N \times N$ hermitian matrices,
$dM_1$ and  $dM_2$ the products of Lebesgue measures of the real components of $M_1$ and $M_2$ respectively,
and $V_1$ and $V_2$ two polynomial potentials of degree $d_1+1$ and
$d_2+1$ respectively :
\beq\label{defVpot}
V_1(x) = \sum_{k=1}^{d_1+1} {g_k\over k} x^k
\virg
V_2(y) = \sum_{k=1}^{d_2+1} {\td{g}_k\over k} y^k
\eeq

Formal integral means it
is computed as the formal power series expansion order by order in
the $g_k$'s (see \cite{ZJDFG,thooft,BIPZ}) of a matrix
integral, where the non-quadratic terms in the potentials $V_1$
and $V_2$ are treated as perturbations near quadratic potentials.
Such a perturbative expansion can be performed only near local
extrema of $V_1(x)+V_2(y)-xy$, i.e. near points such that: \beq
V'_1(\xi_i)=\eta_i \virg V'_2(\eta_i)=\xi_i \eeq which has $d_1
d_2$ solutions. Therefore, if $\ovl{M}_1$ and $\ovl{M}_2$ are
diagonal matrices, whose diagonal entries are some $\xi_i$'s
(resp. $\eta_i$'s), $(\ovl{M}_1,\ovl{M}_2)$ is a local extremum of
$\tr (V_1(M_1)+V_2(M_2)-M_1 M_2)$ around which we can perform a
perturbative expansion.

The choice of such an extremum, around which the perturbative
series is computed, is equivalent to the choice of the number of
eigenvalues near each pair $(\xi_i,\eta_i)$, $i=1,\dots, d_1 d_2$,
i.e. the data of $d_1 d_2$ integers $n_i$ such that:
\beq
\sum_{i=1}^{d_1 d_2} n_i=N
\eeq
This means, that we can choose
some contours ${\cal C}_i$, $i=1,\dots, d_1 d_2$, such that the following equality holds
order by order in the perturbative expansion:
\beq\label{fixfrac}
\left<{1\over 2i\pi}\oint_{{\cal C}_i} \tr {dx\over x-M_1}\right> =-n_i \eeq

The numbers ${n_i \over N}$ are called filling fractions.
Thus, in the formal model, filling fractions are fixed parameters.

\bigskip
{\bf Fat graphs and discrete random surfaces }

\smallskip

Once filling fractions are chosen, we perform the perturbative expansion.
Each term of that formal expansion is an
expectation value of a gaussian integral, and using Wick's
theorem, each term can be represented by a Feynman graph. Because
the integration variables are matrices, the graphs are ``fat
graphs'', which have a 2-dimensional structure. The Hermitean
matrix models thus enumerate oriented surfaces (other matrix
ensembles can enumerate non-oriented surfaces). This Formal
expansion equivalent to an enumerating function of Feynman graphs
is a standard tool in physics \cite{ZJDFG,thooft}. Random matrices have thus played a
role in all theories where one needs to sum over surfaces, i.e.
string theory and quantum gravity (i.e. statistical physics on a
random lattice).

Following this interpretation, the loop equations \cite{staudacher} can be
understood as relationships linking surfaces of different genus
and different number of boundaries.

\subsection{Notations}

\subsubsection{Notation for sets of variables}

We will consider functions of many variables $x_1,x_2,x_3,\dots, x_k$, or of a subset of those variables.
In that purpose we introduce the following notations:

Let $K$ be a $k-$upple of integers:
\beq
K=(i_1,i_2,\dots, i_k)
\eeq
We denote $k=|K|$ the length (or cardinal) of $K$.
For any $j\leq |K|$, we denote $K_j$ the set of all $j-$upples (i.e. subsets of length $j$) contained in $K$:
\beq
K_j:=\{J \subset K \,\,\, , \,\, |J|=j \}
\eeq

We define the following $k-$upple of complex numbers:
\beq
\bfx_K:=(x_{i_1},x_{i_2},\dots, x_{i_k})
\eeq

\subsubsection{Correlation functions}

For a given $k$, we define the correlation function:
\beq
\overline{w}_{k}(x_1,\dots,x_k) := N^{k-2}\left< \prod_{i=1}^k \tr{1\over x_{i}-M_1}\right>_c
\virg
\eeq

i.e., with the previous notations:
\beq\label{defwbarkl}
\overline{w}_{|K|}(\bfx_K) := N^{|K|-2}\left< \prod_{r=1}^{|K|} \tr{1\over x_{i_r}-M_1}\right>_c
\virg
\eeq

where the formal average $\langle . \rangle$ is computed with the measure in \eq{defZ},
 and the subscript $c$ means connected part (cumulant).

Those correlation functions can be expanded as formal series in ${1 \over N^2}$ in the large N limit:
\beq\label{defwklh}
\overline{w}_{k}(\bfx_K) = \sum_{h=0}^\infty {1\over N^{2h}}\,\overline{w}_{k}^{(h)}(\bfx_K)
\eeq

The purpose of this article is to compute $\overline{w}_{k}^{(h)}(\bfx_K)$  as residues on an algebraic curve
and represent it with Feynman-like graphs of a cubic field theory on the curve.

\medskip

We also define the following auxiliary functions:
\beq\label{defubarkl}
\overline{u}_{k}(x,y;\bfx_K) := N^{|K|-1} \left< \tr {1\over x-M_1} {V'_2(y)-V'_2(M_2)\over y-M_2}\,\,
\prod_{r=1}^{|K|} \tr{1\over x_{i_r}-M_1}\right>_c
\eeq

\beq\label{defpkl}
p_{k}(x,y;\bfx_K) := N^{|K|-1} \left< \tr {V'_1(x)-V'_1(M_1)\over x-M_1} {V'_2(y)-V'_2(M_2)\over y-M_2}\,\,
\prod_{r=1}^{|K|} \tr{1\over x_{i_r}-M_1}\right>_c
\eeq

\beq\label{defakl}
a_{k}(x;\bfx_K) := N^{|K|-1} \left< \tr {1\over x-M_1} V'_2(M_2)\,\,
\prod_{r=1}^{|K|} \tr{1\over x_{i_r}-M_1}\right>_c
\eeq
Notice that $\overline{u}_{k,}(x,y;\bfx_K)$ is a polynomial in $y$ of degree $d_2-1$, and
$p_{k}(x,y;\bfx_K)$ is a polynomial in $x$ of degree $d_1-1$ and in $y$ of degree $d_2-1$.

It is convenient to renormalize those functions, and define:
\beq\label{defukl}
u_{k}(x,y;\bfx_K):=\overline{u}_{k}(x,y;\bfx_K)  -\delta_{k,0}(V'_2(y)-x)
\eeq
and
\beq\label{defwkl}
w_{k}(\bfx_K):=\overline{w}_{k}(\bfx_K)  + {\delta_{k,2}  \over (x_1-x_2)^2}
\eeq

Let us remark that all those functions have the same kind of topological expansion as $\overline{w}_{k}(\bfx_K)$ and one
defines $p_{k}^{(h)}(x,y;\bfx_K)$ and $u_{k}^{(h)}(x,y;\bfx_K)$ as well like in \eq{defwklh}.

\vs

We define the function:
\beq\label{defY}
Y(x):=V'_1(x)-w_{1}(x)
\eeq
which we see below, describes the algebraic curve.

The ${1 \over N^2}$ expansion of such correlation functions is
known to enumerate discrete surfaces of a given topology,
whose polygons carry a spin + or - (Ising model on a random surface \cite{Kazakov,Kos}), see \cite{eynhabilit} for the multicut case i.e. foam of Ising surfaces.

The $\overline{w}_{k}^{(h)}$ are generating functions enumerating genus $h$ discrete surfaces with $k$ boundaries of spin $+$.

As an example, $\overline{w}_{2}^{(3)}$ enumerates surfaces of genus 3 with 2 boundaries:
\beq
\overline{w}_{2}^{(3)}= \begin{array}{r}
{\epsfxsize 7cm\epsffile{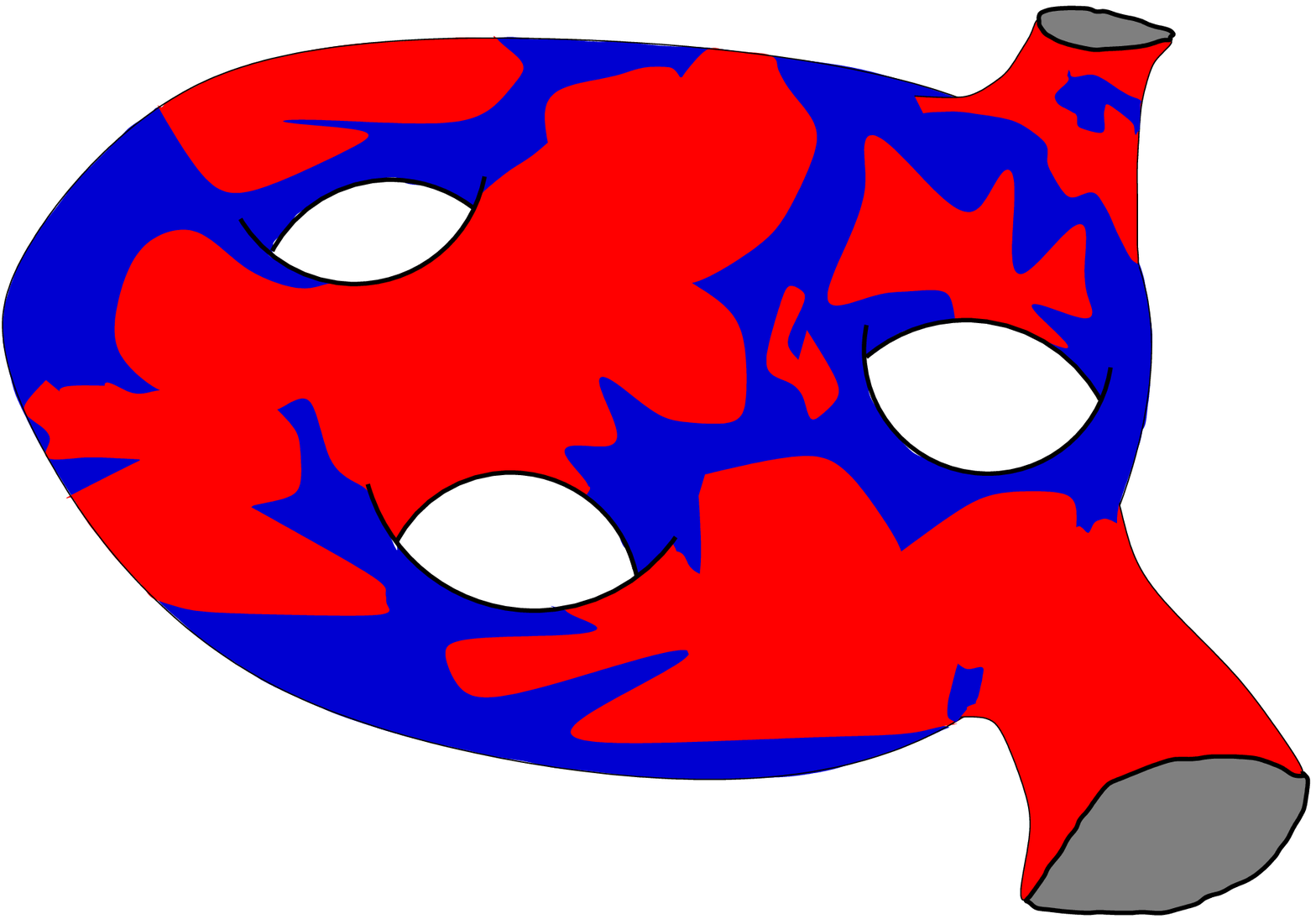}}
\end{array}
\eeq

Notice that the question of boundaries with non uniform spin, i.e. with changes of boundary conditions has been solved to leading order only in \cite{EOtrmixte}.

\newsection{Loop equations}

There exist several methods for computing the free energy and correlation functions, the one we consider here is the ``loop equation'' method, which is nothing
but Schwinger-Dyson, or Ward identities \cite{ZJDFG, staudacher}.
They implement the Virasoro or W-algebra constraints on the partition function \cite{KazMar, MMM}, i.e. the fact that the matrix integral is left unchanged under a change of variable.
The loop equations are valid in the formal model, order by order in the expansion parameters.

For the 2-matrix model, loop equations have been known since \cite{staudacher}, and written
in a more systematic way in \cite{eynchain,eynchaint,eynmultimat,KazMar}.

\subsection{The master loop equation}

It is well known that in the large $N$ limit, loop equations imply an algebraic equation for the functions $w_{1}$,
i.e. for the function $Y(x)$, called the master loop equation.
Let us briefly recall how to derive it (see \cite{eynmultimat}):

$\bullet$ the change of variables $M_2 \rightarrow M_2 + \epsilon \frac{1}{x-M_1}$ implies:
\beq\label{chvara}
0=a_{0}(x) - x \ovl{w}_{1}(x) + 1
\eeq

$\bullet$ the change of variables $M_1 \rightarrow M_1 + \epsilon  \frac{1}{x-M_1}\frac{V_2'(y)-V_2'(M_2)}{y-M_2} $ implies:
\bea\label{chvaru}
\ovl{w}_{1}(x) \ovl{u}_{0}(x,y) +{1\over N^2}u_{1}(x,y;x) &=& V'_1(x)\ovl{u}_{0}(x,y) - p_{0}(x,y) - y \ovl{u}_{0}(x,y)\cr
&& + V'_2(y) w_{1}(x) - a_{0}(x) \cr
\eea

i.e., putting everything together:
\beq
(y-Y(x))u_{0}(x,y)+{1\over N^2}u_{1}(x,y;x) = (V'_2(y)-x) (V'_1(x)-y) - p_{0}(x,y)   +1
\eeq

We define:
\beq\label{defExy}
E(x,y) = ( V_2'(y) -x ) ( V_1'(x)-y) - p_{0}(x,y) + 1
\eeq
The {\em master loop equation} is thus:
\beq\label{masterloopallgenus}\encadremath{
(y-Y(x))u_{0}(x,y)+{1\over N^2}u_{1}(x,y;x) = E(x,y)}
\eeq

where $E(x,y)$ is a polynomial of degree $d_1+1$ in $x$ and $d_2+1$ in y.

\subsection{Loop equations for correlation functions}

We now derive the loop equations
which allow to compute recursively the k-point non-mixed correlation
functions.

$\bullet$ The change of variables $\delta M_2 ={1\over x-M_1}
\,\,\prod_{i=1}^k \tr {1\over x_i-M_1}$ implies (see
\cite{eynmultimat}): \beq a_{k}(x;\bfx_K) = x\,
\ovl{w}_{k+1}(x,\bfx_K) - N^2 \ovl{w}_{k}(\bfx_K)\eeq

$\bullet$ The change of variables $\delta M_1 ={1\over x-M_1}
{V'_2(y)-V'_2(M_2)\over y-M_2}\,\,\prod_{i=1}^k \tr {1\over
x_i-M_1} $ implies (see \cite{eynmultimat}):

\bea
&& w_{1}(x) \,\ovl{u}_{k}(x,y;\bfx_K) +\sum_{j=0}^{k-1} \sum_{J\in
K_j}  \ovl{u}_{j}(x,y;\bfx_J)\, \ovl{w}_{k-j+1}(x,\bfx_{K-J}) \cr
&& + {1 \over N^2} \ovl{u}_{k+1}(x,y;x,\bfx_{K}) \cr
&& +
\sum_{j=1}^k {\d\over \d x_j}\,
{\ovl{u}_{k-1}(x,y;\bfx_{K-\{j\}})-\ovl{u}_{k-1}(x_j,y;\bfx_{K-\{j\}})
\over x-x_j} \cr &=& V'_1(x)\ovl{u}_{k,0}(x,y;\bfx_K) -
p_{k}(x,y;\bfx_K) \cr && - y \ovl{u}_{k}(x,y;\bfx_K) + V'_2(y)
\ovl{w}_{k+1}(x,\bfx_K) - a_{k}(x;\bfx_K) \cr \eea

i.e. for $k\geq 1$:
 \beq \label{loop1}
 \encadremath{\begin{array}{rcl}
 (y-Y(x)) \,u_{k}(x,y;\bfx_K)
&=&-\sum_{j=0}^{k-1} \sum_{J\in K_j} u_{j}(x,y;\bfx_J)\,
w_{k-j+1}(x,\bfx_{K-J}) \cr && - {1 \over N^2}
u_{k+1}(x,y;x,\bfx_{K}) \cr
 && + \sum_{j=1}^k {\d\over \d
x_j}\, {u_{k-1}(x_j,y;\bfx_{K-\{j\}})  \over x-x_j}   -
p_{k}(x,y;\bfx_K) \cr
\end{array}}
\eeq

The purpose of this article is to solve \eq{loop1} and compute $\overline{w}_k^{(h)}$ for
all $k$ and $h$.

\newsection{Leading order and algebraic geometry}

\subsection{Leading order of the master loop equation}

To large $N$ leading order, the master loop equation
\eq{masterloopallgenus} reads:

\beq\encadremath{\label{masterloopu} (y-Y(x))u_{0}(x,y) = E(x,y)
}\eeq

Since $u_{0}(x,y)$ is a polynomial in $y$, it has no singularity
for y finite and the LHS vanishes for $y=Y(x)$, i.e.:

\beq\label{masterloopeq} E(x,Y(x))=0 \eeq This defines an
algebraic curve $E(x,y)=0$.

Notice that to leading order we have: \beq\label{uoo} u_{0}(x,y) =
{E(x,y)\over y-Y(x)} \eeq and \beq\label{uooY} u_{0}(x,Y(x)) =
E_y(x,Y(x)) \eeq

\subsection{Introduction to some algebraic geometry}

We use notations similar to \cite{Fay} or \cite{Farkas}. Some useful hints for understanding this
section can be found in {\emph{Appendix A}}.

Let us parameterize the curve $E(x,y)=0$ with a running point p of
a compact Riemann surface ${\cal E}$. It means that we define two
meromorphic functions $x(p)$ and $y(p)$ on ${\cal E}$ such that:
\beq E(x,y)=0 \Leftrightarrow \exists p \in {\cal E} \,\,\,\,\,
x=x(p) \,\, , \,\, y=y(p) \eeq

The functions $x$ and $y$ are not bijective. Indeed, since
$E(x,y)$ is a polynomial of degree $d_2+1$ in $y$, it has $d_2+1$
solutions, i.e. for a given $x$, there exist $d_2+1$ points $p$ on
$ {\cal E} $ such that $x(p)=x$. Thus, the Riemann surface is made
of $d_2+1$ $x$-sheets, respectively $d_1+1$ $y$-sheets. Hence,
from now on, we use these notations:
\beq x(p) = x \Leftrightarrow
p = p^{j}(x) \,\,\,\,\, \hbox{for} \,\,\,\, j=0,\dots,d_2 \eeq

\beq y(p) = y \Leftrightarrow p = \td{p}^{j}(x) \,\,\,\,\,
\hbox{for} \,\,\,\, j=0,\dots,d_1 \eeq

We will most often omit the
exponent 0 corresponding to the physical sheet: $p=p^{0}$.

For instance, one can write $E(x,y)$ as:
\bea E(x(p),y(q)) &=& -
g_{d_1+1} \times \prod_{i=0}^{d_1} (x(p)-x(\td{q}^{i}(y))) \cr &=&
- \td{g}_{d_2+1} \times \prod_{i=0}^{d_2} (y(q)-y(p^{i}(x))) \eea

\vs

Considering that the $w_{k}^{(h)}$'s, $u_{k}^{(h)}$'s and
$p_{k}^{(h)}$'s are multivalued functions in their arguments $x$,
we now work with differentials monovalued on the Riemann surface.
Let us write the differentials: \beq\label{defWk}
W_{k+1}(p,\bfp_K) := w_{k+1}(x(p),\bfx(p_K)) dx(p) \prod_{i=1}^k
dx(p_i) \eeq \beq\label{defUk} U_{k}(p,y;\bfp_K) :=
u_{k}(x(p),y;\bfx(p_K)) dx(p) \prod_{i=1}^k dx(p_i) \eeq
\beq\label{defPk} P_{k}(x,y;\bfp_K) := p_{k}(x,y;\bfx(p_K))
\prod_{i=1}^k dx(p_i) \eeq

{\bf Note:} In the following, the arguments of a function will be
called $x(p)$ or $y(r)$ if the function is defined on the basis,
and $p$ or $r$ if the function is defined on the Riemann surface -
and so multivalued on the basis-. \vs

Let us now review the notations we use in this article to denote some basic objects. For definitions and details, we refer the reader to {\emph{Appendix A}} and \cite{Fay} or \cite{Farkas}.

\bigskip

$\bullet${\bf Canonical cycles:} ${\cal A}_i$,
${\cal B}_i$ for $i=1,\dots, g$ where $g$ is the genus of the compact Riemann surface ${\cal{E}}$ ($0\leq g \leq d_1d_2 -1$),
such that:
\beq
{\cal A}_i \cap {\cal B}_i = \delta_{i,j}
\eeq

\bigskip
$\bullet${\bf Branch points in $x$:} They are the zeroes of $dx$ on the surface. We denote them by $a_i$, $i=1, \dots , d_2+1+2g$.

\bigskip
$\bullet${\bf Bergmann kernel:} It is the unique bilinear differential with only one double pole
at $p=q$ satisfying:
\beq\label{defB}
B(p,q)\mathop\sim_{p\to q} {dx(p)dx(q)\over (x(p)-x(q))^2}+{\rm
finite} \quad {\rm and} \quad \forall i \,\,\,\oint_{{p\in{\cal
A}_i}} B(p,q) = 0 \eeq

\bigskip
$\bullet${\bf Abelian differential of third kind:} It is the differential defined by
$dS_{q,r}(p)  = \int_{q'=r}^{q} B(p,q')$. Notice that it has the following properties:
\beq\label{defdS} \Res_{p
\to q} dS_{q,r}(p) = 1 = -\Res_{p \to r} dS_{q,r}(p) \quad {\rm
and} \quad \forall i \,\,\,\oint_{{{\cal A}_i}} dS_{q,r}(p) = 0
\eeq

\subsection{Fixed filling fractions}
To large $N$ leading order, the loop equation \eq{masterloopeq} is
an algebraic equation: \beq E(x,Y(x))=0 \eeq 
The coefficients of $E$ are determined using filling fractions.
Since $w_{1}(x) = V'_1(x)-Y(x)$, \eq{fixfrac} gives (up to a redefinition of ${\cal A}_i$):

\beq\label{deffillingfractions} {1\over 2i\pi}\oint_{{\cal A}_i} y
dx =-{1\over 2i\pi}\oint_{{\cal A}_i} x dy =\epsilon_i \eeq 

Let us recall that (see section \ref{secdef}) the
$\epsilon_i$'s are called filling fractions, and they are given
parameters (moduli) of the model. They don't depend on the
potential or on any other parameter.

In particular, since all correlation functions
$w_{k}(x_1,\dots,x_k)$ are obtained by derivation of $w_{1}$ with
respect to the potential $V_1$ (\cite{ACKM}), we have for $k\geq 2$:
\beq\label{vanishingAcycles} {1\over 2i\pi}\oint_{{\cal A}_i}
w_{k}(x_1,\dots,x_k) dx_1 =0 \eeq

Equation \eq{deffillingfractions} together with the large $x$ and $y$
behaviors \eq{largex} and \eq{largey}, are sufficient to
determine completely all the coefficients of the polynomial
$E(x,y)$, and thus the leading large $N$ resolvent $w_{1}(x)$.

\medskip
In what follows, we assume that the leading resolvent, i.e. the
function $Y(x)$ is known, and we refer the reader to
the existing literature on that topic, for instance
\cite{MarcoF,eynmultimat,KazMar,Kri}.

\newsection{Diagrammatic solution as cubic
graphs}

In this section we present a first way of describing the solution
of the loop equation \eq{loop1} by trivalent diagrams whose $h$ loop level
corresponds to the $h$-th term $W_k^{(h)}$ of the topological
expansion.

\subsection{Solution in the planar limit}

Before considering the full ${1 \over N^2}$ expansion, let us
focus on the structure of the leading terms corresponding to planar fat graphs.
Thus the $1/N^2$
terms in the loop equations are omited.

From now on and particularly in this paragraph, we drop the genus
zero exponent ${(0)}$ when it is clear that we deal with the
planar limit, i.e. $w_k^{(0)}(\bfx_K)\to w_k(\bfx_K)$.

\smallskip
Up to now, the loop equations were written in terms of multivalued functions. It is more
appropriate to write them in terms of meromorphic differentials on the Riemann surface.
Thus, one writes \eq{loop1} in the planar limit as follows:

\beq\label{eqUk}
\begin{array}{rcl}
(y(r)-y(p)) U_{k}(p,y(r);\bfp_K)  &=& - \sum_{j=0}^{k-1} \sum_{J\in K_j} {U_{j}(p,y(r);\bfp_J)\, W_{k-j+1}(p,\bfp_{K-J}) \over dx(p)} \cr
&& + \sum_{j=1}^k  d_{p_j}\left( {{U}_{k-1}(p_j,y(r);\bfp_{K-\{j\}})  \over x(p)-x(p_j)}\,{dx(p)\over dx(p_j)}\right) \cr
&&  - P_{k}(x(p),y(r);\bfp_K) dx(p)
\end{array}
\eeq

Starting from \eq{eqUk}, we determine $W_k$ and
$U_k$ for any $k$ by recursion on $k$.

Let us assume that one knows $W_{j}(\bfp_J )$ for $j\leq k$ and
$U_{j}(p,\bfp_J )$ for $j \leq k-1$. The first step consists in
the determination of $W_{k+1}(p,\bfp_K)$ as a function of the
lower order correlation functions. The second step leads to the
computation of $U_{k}(p,\bfp_K)$. Once this is done, one knows
the correlation functions one order upper. The initial terms $W_2$ and $U_1$
can be found in the literature \cite{MarcoF,eynmultimat,KazMar}
and are rederived in {\emph{Appendix B}}.

\subsubsection{Determination of $W_{k+1}$ for $k\geq 2$}

If one chooses $r=p$ in \eq{eqUk}, one gets (using \eq{uoo} and \eq{uooY}):
\beq\label{eqWk}
\begin{array}{lll}
 E_y(x(p),y(p)) W_{k+1}(p,\bfp_{K})
&=&  - P_{k}(x(p),y(p);\bfp_K) \, dx(p) \cr
 &&  -\sum_{j=1}^{k-1} \sum_{J\in K_j}  {U_{j}(p,y(p);\bfp_J)\, W_{k-j+1}(p,\bfp_{K-J}) \over dx(p)} \cr
&& + \sum_{j=1}^k d_{p_j}\left({{U}_{k-1}(p_j,y(p);\bfp_{K-\{j\}})\over x(p)-x(p_j)}\, {dx(p)\over dx(p_j)} \right) \cr
\end{array}
\eeq

Notice that the two equations \eq{eqUk} and \eq{eqWk} imply by recursion, that $W_k$ and $U_k$ are
indeed meromorphic differentials on the curve, in all their variables.

\medskip

We define:
\beq\label{defRik}
\forall (i,j)\qquad
R_{k}^i(p^{j},p_K) := {U_{k}(p^{j},y(p^{i});p_K) \over
E_y(x(p^{j}),y(p^{i})) dx(p^{j})}
\eeq

Note that we have already obtained (see \eq{uoo}) that: \beq
R_{0}^i(p^{l}) = \delta_{i,l}  \eeq

Using \eq{defdS}, the Cauchy formula gives: \beq {W}_{k+1}(p,\bfp_K) = -\Res_{p'\to
p} {W}_{k+1}(p',\bfp_{K}) dS_{p',\a}(p) \eeq
where $\a \in {\cal E}$ is an arbitrary point on the Riemann surface.

The integrand has poles in $p'$ only at $p'=p$ and the branch points
$p'=a_s$ (this can be proven recursively by differentiating wrt the potential ${\partial \over \partial V_1}$).
Using Riemann bilinear identity
\eq{RiemannbilinearId}, we can then move the integration contour
and get: \beq {W}_{k+1}(p,\bfp_{K}) =  \sum_s \Res_{p'\to a_s}
{W}_{k+1}(p',\bfp_{K})  dS_{p',\a}(p) \eeq

We now introduce the loop equation \eq{eqWk} inside this
expression and remark that only one term has poles when $p'\to
a_s$. Thus ${W}_{k+1}(p,\bfp_{K})$ can be written:
\bea\label{recW} {W}_{k+1}(p,\bfp_{K}) &=& -\sum_s \Res_{p'\to
a_s} \sum_{j=1}^{k-1} \sum_{J\in K_j}
{{U}_{j}(p',y(p');\bfp_J)\over  E_y(x(p'),y(p')) }
{{W}_{k-j+1}(p',\bfp_{K-J}) \over dx(p')} dS_{p',\a}(p) \cr &=&
-\sum_s \Res_{p'\to a_s} \sum_{j=1}^{k-1}\sum_{J\in K_j}
{R}_{j}^0(p',\bfp_J)  {W}_{k-j+1}(p',\bfp_{K-J}) dS_{p',\a}(p) \cr
\eea

Notice that $U_{k}(p,y;\bfp_K)$ is a polynomial in y whose degree is
equal to $d_2-1$. Considering its $d_2$ values for $y=y(p^{i})$
with $i\in [1,d_2]$, the interpolation formula reads:

\beq \forall y \,\,\,\, {(y-y(p)) U_{k}(p,y;\bfp_K) \over
E(x(p),y)} = - \sum_{i=1}^{d_2} {U_{k}(p,y(p^{i});\bfp_K)
(y(p)-y(p^{i})) \over (y-y(p^{i})) E_y(x(p),y(p^{i}))} \eeq

for $y=y(p)$, this gives: \beq R_{k}^0(p,\bfp_K) = -
\sum_{i=1}^{d_2} R_{k}^i(p,\bfp_K) \eeq

So, in \eq{recW}, one obtains the recursive formula for
$W_{k}(\bfp_K)$: \beq\encadremath{\label{recWk}
W_{k+1}(p,\bfp_{K})=   \sum_{i=1}^{d_2} \sum_{j=1}^{k-1}
\sum_{J\in K_j} \sum_s \Res_{p'\to a_s} {R}_{j}^i(p';\bfp_J)
{W}_{k-j+1}(p',\bfp_{K-J}) dS_{p',\a}(p) }\eeq

The sum over $j$ represents the summation over all partitions of
$K$ into two subsets $J$ and $K-J$.

\subsubsection{Determination of $R_{k}^i$}
In this section, we find a
recursion formula for $R_{k}^i$.

For this purpose, one needs to know an intermediate expression
defining the different $U_{k}$'s as well as a relation linking the
value of \beq \sum_{j=0}^{k-1} U_{j}(p^{i},y(p);\bfp_J)
W_{k-j+1}(p^{i},\bfp_{K-J}) \eeq for different $i$'s.

Let us rewrite here \eq{eqUk}: \bea\label{eqUkbis} (y(r)-y(q))
U_{k}(q,y(r);\bfp_K)  &=& - \sum_{j=0}^{k-1} \sum_{J\in K_j}
\frac{1}{dx(q)}\, U_{j}(q,y(r);\bfp_J)\, W_{k-j+1}(q,\bfp_{K-J})
\cr && + \sum_{j=1}^k  d_{p_j}\left(
{{U}_{k-1}(p_j,y(r);\bfp_{K-\{j\}})  \over
x(q)-x(p_j)}\,{dx(q)\over dx(p_j)}\right) \cr &&  -
P_{k}(x(q),y(r);\bfp_K) dx(q) \eea

In what follows, we use the
properties of rational functions defined on the basis and not on the
Riemann surface (for some more details, see the case $k=1$ in {\emph{Appendix B}}).

For $r=q=p^{i}$, \eq{eqUkbis} reads:

\bea\label{eqRkint} 0  &=& - \sum_{j=0}^{k-1} \sum_{J\in K_j}
\frac{1}{dx(p^{i})}\, U_{j}(p^{i},y(p^{i});\bfp_J)\,
W_{k-j+1}(p^{i},\bfp_{K-J}) \cr && + \sum_{j=1}^k  d_{p_j}\left(
{{U}_{k-1}(p_j,y(p^{i});\bfp_{K-\{j\}})  \over
x(p^{i})-x(p_j)}\,{dx(p^{i})\over dx(p_j)}\right) \cr &&  -
P_{k}(x(p^{i}),y(p^{i});\bfp_K) dx(p^{i}) \cr &=& -
\sum_{j=0}^{k-1} \sum_{J\in K_j}  \frac{1}{dx(p)}\,
U_{j}(p^{i},y(p^{i});\bfp_J)\, W_{k-j+1}(p^{i},\bfp_{K-J})
\cr && + \sum_{j=1}^k  d_{p_j}\left(
{{U}_{k-1}(p_j,y(p^{i});\bfp_{K-\{j\}})  \over
x(p)-x(p_j)}\,{dx(p)\over dx(p_j)}\right) \cr &&  -
P_{k}(x(p),y(p^{i});\bfp_K) dx(p) \eea

where we have used that
$x(p)=x(p^{i})$.

Now, write \eq{eqUkbis} with $r=p^{i}$ and $q=p$:

\bea &&
(y(p^{i})-y(p)) U_{k}(p,y(p^{i});\bfp_K) \cr &=& -
\sum_{j=0}^{k-1} \sum_{J\in K_j}  \frac{1}{dx(p)}\,
U_{j}(p,y(p^{i});\bfp_J)\, W_{k-j+1}(p,\bfp_{K-J}) \cr && +
\sum_{j=1}^k  d_{p_j}\left(
{{U}_{k-1}(p_j,y(p^{i});\bfp_{K-\{j\}})  \over
x(p)-x(p_j)}\,{dx(p)\over dx(p_j)}\right) \cr &&  -
P_{k}(x(p),y(p^{i});\bfp_K) dx(p) \eea

and inserting \eq{eqRkint} we get:

\bea\label{Uk} && (y(p^{i})-y(p))
U_{k}(p,y(p^{i});\bfp_K) \cr &=& - \sum_{j=0}^{k-1} \sum_{J\in
K_j}  \frac{1}{dx(p)}\, U_{j}(p,y(p^{i});\bfp_J)\,
W_{k-j+1}(p,\bfp_{K-J}) \cr && + \sum_{j=0}^{k-1} \sum_{J\in K_j}
\frac{1}{dx(p)}\, U_{j}(p^{i},y(p^{i});\bfp_J)\,
W_{k-j+1}(p^{i},\bfp_{K-J}) \eea

This formula is in principle sufficient to compute the $U_{k}$'s
recursively, and then, one can compute the $R_{k}^i$'s. However,
what we need in order to get diagrammatic rules, is a closed recursion
relation for the $R^i_{k}$'s themselves. In order to achieve this
aim, we show that:

\medskip

{\emph{Lemma:}} for any $k\geq 1$, one has:

\bea\label{UWW}
U_{k}(p,y;\bfp_K) &=& {E(x(p),y) dx(p)\over
y-y(p)}\sum_{r=1}^{d_2} \sum_{K_1\cup\dots\cup K_r=K}
\sum_{j_1\neq j_2\neq \dots \neq j_r=1}^{d_2}\cr && \qquad
\prod_{t=1}^r {W_{|K_t|+1}(p^{j_t},\bfp_{K_t})\over
(y-y(p^{(j_t)}))\,dx(p)}\cr \eea

where the sum over
$K_1\cup\dots\cup K_r=K$ is a sum over all partitions of $K$ into
$r$ subsets.

{\emph{Proof:}} It can be proven easily by recursive action of
${\partial/\partial V_1}$, as in \cite{ACKM}, however, in order to
have a self-contained method, we want to derive it here only from
the loop equations \eq{loop1}.

The proof works by recursion on $k$. It is proven in {\emph{Appendix B}} for $k=1$.
Let us assume that, it holds for any $l \leq
k-1$.

Notice, that since both sides of \eq{UWW} are polynomials of $y$,
of degree $d_2-1$, it is sufficient to prove that the equality
holds for $d_2$ values of $y$, namely, it is sufficient to prove
it for $y=y(p^{i})$, $i=1,\dots,d_2$. Therefore, one has to
prove that:

\bea {U_{k}(p,y(p^{i});\bfp_K) \over dx(p)} &=&
{E_y(x(p^{i}),y(p^{i})) \over y(p^{i})-y(p)}
\sum_{r=1}^{d_2} \sum_{K_1\cup\dots\cup K_r=K} \sum_{j_1\neq
j_2\neq \dots \neq j_{r-1}\neq 0,i} \cr &&
{W_{|K_r|+1}(p^{i},\bfp_{K_r})\over \,dx(p)}\,\prod_{t=1}^{r-1}
{W_{|K_t|+1}(p^{j_t},\bfp_{K_t})\over (y-y(p^{j_t}))\,dx(p)}
\eea where only the sums in which one of the $j_t$'s is equal to
$i$ contribute.

The recursion hypothesis for $j\leq k-1$, and any $J\in K_j$
gives:

\bea {U_{j}(p^{i},y(p^{i});\bfp_J)\over dx(p)} &=&
E_y(x(p^{i}),y(p^{i})) \sum_{r=1}^{d_2} \sum_{J_1\cup\dots\cup
J_r=J} \sum_{j_1\neq j_2\neq \dots \neq j_r\neq i} \cr && \qquad
\prod_{t=1}^r {W_{|J_t|+1}(p^{j_t},\bfp_{J_t})\over
(y(p^{i})-y(p^{j_t}))\,dx(p)} \cr \eea

In order to compute
$U_{j}(p,y(p^{i});\bfp_J)$, one has to keep only terms in the
sum such that there exists a $t$ such that $j_t=i$, i.e.

\bea
{U_{j}(p,y(p^{i});\bfp_J)\over dx(p)} &=&
E_y(x(p^{i}),y(p^{i}))\, \sum_{r=1}^{d_2}
\sum_{J_1\cup\dots\cup J_r=J} \sum_{j_1\neq j_2\neq \dots \neq
j_{r-1}\neq 0,i} \cr && \qquad
{W_{|J_{r}|+1}(p^{i},\bfp_{J_{r}})\over
(y(p^{i})-y(p))\,dx(p)}\,\prod_{t=1}^{r-1}
{W_{|J_t|+1}(p^{j_t},\bfp_{J_t})\over
(y(p^{i})-y(p^{j_t}))\,dx(p)} \cr \eea

Insert that into \eq{Uk}:

\bea && {(y(p^{i})-y(p))
U_{k}(p,y(p^{i});\bfp_K)} \cr &=& - E_y(x(p^{i}),y(p^{i}))
\sum_{j=0}^{k-1} \sum_{J\in K_j} \sum_{r=1}^{d_2}
\sum_{J_1\cup\dots\cup J_r=J} \sum_{j_1\neq j_2\neq \dots \neq
j_{r-1}\neq 0,i} \cr && \qquad W_{k-j+1}(p,\bfp_{K-J})
{W_{|J_{r}|+1}(p^{i},\bfp_{J_{r}})\over
(y(p^{i})-y(p))\,dx(p)}\,\prod_{t=1}^{r-1}
{W_{|J_t|+1}(p^{j_t},\bfp_{J_t})\over
(y(p^{i})-y(p^{j_t}))\,dx(p)} \cr && +
E_y(x(p^{i}),y(p^{i}))\,\sum_{j=0}^{k-1} \sum_{J\in K_j}
\sum_{r=1}^{d_2} \sum_{J_1\cup\dots\cup J_r=J} \sum_{j_1\neq
j_2\neq \dots \neq j_r\neq i} \cr && \qquad
W_{k-j+1}(p^{i},\bfp_{K-J}) \prod_{t=1}^r
{W_{|J_t|+1}(p^{j_t},\bfp_{J_t})\over
(y(p^{i})-y(p^{j_t}))\,dx(p)} \cr \eea

The difference between
these two summation, keeps only $j_t\neq 0,i$, thus:

\bea &&
U_{k}(p,y(p^{i});\bfp_K) \cr &=&
E_y(x(p^{i}),y(p^{i}))\,dx(p)\,\sum_{j=0}^{k-1} \sum_{J\in
K_j} \sum_{r=1}^{d_2} \sum_{J_1\cup\dots\cup J_r=J} \sum_{j_1\neq
j_2\neq \dots \neq j_r\neq i,0} \cr && \qquad
{W_{k-j+1}(p^{i},\bfp_{K-J})\over (y(p^{i})-y(p))\,dx(p)}
\prod_{t=1}^r {W_{|J_t|+1}(p^{j_t},\bfp_{J_t})\over
(y(p^{i})-y(p^{j_t}))\,dx(p)} \cr \eea

 i.e. we have proven the
lemma for $k$, for $y=y(p^{i})$, and since both sides are
polynomials in $y$ of degree $d_2-1$, the equality holds for all
$y$.

\begin{flushright}
$\bullet$
\end{flushright}

\bigskip

{\emph{Theorem:}} For all $k\geq 1$, one has:
\beq\label{equality}
\begin{array}{lll}
&& \sum_{i=1}^{d_2} \sum_{j=0}^{k-1} \sum_{J\in K_j}
U_{j}(p^{i},y(p);\bfp_J) W_{k-j+1}(p^{i},\bfp_{K-J}) \cr &=&
\sum_{j=1}^{k-1} \sum_{J\in K_j} U_{j}(p,y(p);\bfp_J)
W_{k-j+1}(p,\bfp_{K-J})
\end{array}
\eeq

{\emph{Proof of the theorem:}} Let us simply perform some basic
rearrangements: \bea &&\sum_{i=1}^{d_2} \sum_{j=0}^{k-1}
\sum_{J\in K_j} U_{j}(p^{i},y(p);\bfp_J)
W_{k-j+1}(p^{i},\bfp_{K-J}) \cr &=& \sum_{K_1\bigcup L = K}
\sum_{j_1=1}^{d_2} W_{|K_1|+1}(p^{j_1},\bfp_{K_1})
U_{|L|+1}(p^{j_1},y(p);\bfp_{L})\cr & = & {E_y(x(p),y(p))} dx(p)
\sum_{K_1\bigcup L = K} \sum_{j_1=1}^{d_2} \sum_{r=1}^{d_2}
\sum_{K_2\cup\dots\cup K_{r+1}=L}  \sum_{j_2\neq j_3\neq \dots
\neq j_{r} \in [1,d_2]-\{j_1\} } \cr &&
W_{|K_1|+1}(p^{j_1},\bfp_{K_1})
{W_{|K_{r+1}|+1}(p,\bfp_{K_{r+1}})\over (y(p)-y(p^{j_1}))}
 \prod_{a=2}^r {W_{|K_a|+1}(p^{j_a},\bfp_{K_a}) \over (y(p)-y(p^{j_a})) dx(p)}\cr
& = & {E_y(x(p),y(p))} dx(p) \sum_{r=1}^{d_2}
\sum_{K_1\cup\dots\cup K_{r+1}=K} \sum_{j_1\neq j_2\neq \dots \neq
j_{r}=1}^{d_2}\cr & & \prod_{a=1}^r
{W_{|K_a|+1}(p^{j_a},\bfp_{K_a})
W_{|K_{r+1}|+1}(p,\bfp_{K_{r+1}}) \over (y(p)-y(p^{j_a}))
dx(p)}\cr &=& \sum_{K_{r+1}\bigcup J = K}
W_{|K_{r+1}|+1}(p,\bfp_{K_{r+1}}) U_{|J|}(p,y(p);\bfp_J)\cr \eea
\begin{flushright}
$\bullet$
\end{flushright}
\vs

This identity simplifies \eq{Uk} which becomes now: \bea &&
(y(p^{i})-y(p)) R_k^i(p,\bfp_K) dx(p) = \cr &&
W_{k+1}(p^{i},\bfp_{K}) + \sum_{j=1}^{k-1} \sum_{J\in K_j}
\sum_{l\neq 0,i} {U_{j}(p^{l},y(p^{i});\bfp_J)
W_{k-j+1}(p^{l},\bfp_{K-J}) \over  E_y(x(p),y(p^{i})) dx(p)}\cr
\eea

One can now write down the final recursion formula for
$R_k^i(p,\bfp_K)$ in these terms:

\beq\label{recUk}\encadremath{
\begin{array}{rcl}
R_k^i(p,\bfp_K) &=& {W_{k+1}(p^{i},\bfp_{K}) \over
(y(p^{i})-y(p)) dx(p)} \cr && + \sum_{j=1}^{k-1} \sum_{J\in K_j}
\sum_{l\neq 0,i} {R_j^{i}(p^{l},\bfp_J)
W_{k-j+1}(p^{l},\bfp_{K-J}) \over (y(p^{i})-y(p)) dx(p)}\cr
\end{array}
}\eeq

\bigskip
The relations  \eq{recW} and \eq{recUk} allow to compute recursively $W_k$ for any $k$.
This solution can be represented by binary trees as it is presented in section (\ref{cub}).

\subsection{Solution for any genus}

In the previous paragraph, one has kept only the leading terms when
performing the changes of variables to obtain the Schwinger-Dyson
equations. Let us now write the ${1 \over N^2}$ corrective term
for the same changes of variables so that we write a system of
equations giving the whole ${1 \over N^2}$ expansion. One obtains
the following loop equations :

\beq\label{eqUklg0}
\begin{array}{lll}
&& (y(r)-y(p)) {U}_{k}(p,y(r);\bfp_K) \cr &=&  -
P_{k}(x(p),y(r);\bfp_K) dx(p) - \sum_{j=0}^{k-1} {1 \over dx(p)}
{U}_{j}(p,y(r);\bfp_J) {W}_{k-j+1}(p,\bfp_{K-J}) \cr && - {1 \over
N^2} {U_{k+1}(p,y(r);p,\bfp_k) \over dx(p)} + \sum_j d_{p_j}
\left({{U}_{k-1}(p_j,y(r);\bfp_{K-\{j\}}) \over x(p)-x(p_j)}
{dx(p) \over dx(p_j)} \right) \cr
\end{array}
\eeq

For the following, one should remind the expression of the
function $Y(x(p))$: \beq Y(x):=V'_1(x)-\overline{w}_{1}(x) \eeq

Then, for $h\geq 1$: \beq Y^{(h)}(x(p)) = -{W_{1}^{(h)}(p)\over
dx(p)} \eeq

Consider now the ${1 \over N^2}$ expansion of this equation order
by order. The genus h term (corresponding to the ${1 \over N^{2
h}}$ term) gives: \beq\label{eqUklgh}
\begin{array}{lll}
&& (y(r)-y(p)) {U}_{k}^{(h)}(p,y(r);\bfp_K) - \sum_{m=1}^h
Y^{(m)}(x(p)) {U}_{k}^{(h-m)}(p,y(r);\bfp_K) \cr &=& -
P_{k}^{(h)}(x(p),y(r);\bfp_K) dx(p) \cr && - \sum_{m=0}^h
\sum_{j=0}^{k-1} {1 \over dx(p)} {U}_{j}^{(m)}(p,y(r);\bfp_J)
{W}_{k-j+1}^{(h-m)}(p,\bfp_{K-J}) \cr && -
{U_{k+1}^{(h-1)}(p,y(r);p,\bfp_k) \over dx(p)} + \sum_j  d_{p_j}
\left( {{U}_{k-1}^{(h)}(p_j,y(r);\bfp_{K-\{j\}}) \over
x(p)-x(p_j)} {dx(p) \over dx(p_j)} \right) \cr
\end{array}
\eeq

When $y(r)=y(p)$:

\beq\label{eqWklg0}
\begin{array}{lll}
&& \sum_{m=1}^h Y^{(m)}(x(p)) {U}_{k}^{(h-m)}(p,y(p);\bfp_K) \cr
&=& P_{k}^{(h)}(x(p),y(p);\bfp_K) dx(p) + \sum_{m=0}^h
\sum_{j=0}^{k-1}  {1 \over dx(p)} {U}_{j}^{(m)}(p,y(p);\bfp_J)
{W}_{k-j+1}^{(h-m)}(p,\bfp_{K-J}) \cr && +
{U_{k+1}^{(h-1)}(p,y(p);p,\bfp_k) \over dx(p)} - \sum_j d_{p_j}
\left(  {{U}_{k-1}^{(h)}(p_j,y(p);\bfp_{K-\{j\}}) \over
x(p)-x(p_j)} {dx(p) \over dx(p_j)} \right) \cr
\end{array}
\eeq

These two equations are the generalization of \eq{eqUk} and
\eq{eqWk} for any genus in the topological expansion. With all
these tools, we are now able to compute all the terms of the
${1\over N^2}$ expansion of non mixed traces.

In this section, we proceed in two steps to compute the
correlation function $W_{k}^{(h)}$ for any k and any h, and
represent it as a Feynman graph with h loops. The first step
consists in the determination of a recursive relation for
$W_{k}^{(h)}$, whereas the second one gives $R_{k}^{i,(h)}:={U_{k}^{(h)}(p^{j},y(p^{i});p_K) \over
E_y(x(p^{j}),y(p^{i})) dx(p^{j})}$
considered the lower order terms known.

For the following, let h and k be two given positive integers. Let
us consider $W_{j}^{(m)}$ known for any j if $m<h$ and any $j\leq
k$ if $m=h$. One also assume that $R_{j}^{i,(m)}$ is known for any
i and any j if $m<h$ and any $j< k$ if $m=h$. Starting from these
assumptions, one computes $W_{k+1}^{(h)}$ and $R_{k}^{i,(h)}$,
what will allow to know any term recursively.

\subsubsection{A recursive formula for $W_{k+1}^{(h)}$}
Let us remind \eq{eqWklg0} in a more suitable way to emphasize
that it allows us to compute $W_{k+1}^{(h)}(p,p_K)$ with our
assumption:

\beq\label{6110}
\begin{array}{lll}
&& W_{k+1}^{(h)}(p,\bfp_K) U_{0}(p,y(p)) = \cr && -
\sum_{m=0}^{h-1} W_{1}^{(h-m)}(p) {U}_{k}^{(m)}(p,y(p);\bfp_K) \cr
&& - P_{k}^{(h)}(p,y(p);\bfp_K) dx(p)^2 \cr && - \sum_{m=0}^h
\sum_{j=0, m+j \neq 0}^{k-1}  {U}_{j}^{(m)}(p,y(p);\bfp_J)
{W}_{k-j+1}^{(h-m)}(p,\bfp_{K-J}) \cr && -
U_{k+1}^{(h-1)}(p,y(p);p,\bfp_k) + \sum_j \sum_j d_{p_j} \left(
{{U}_{k-1}^{(h)}(p_j,y(p);\bfp_{K-\{j\}}) \over x(p)-x(p_j)}
{dx(p) \over dx(p_j)} \right) dx(p) \cr
\end{array}
\eeq

Remark that the RHS contains only known terms except
$P_{k}^{(h)}(p,y(p);\bfp_K)$. Fortunately, it plays no role in
Cauchy formula.

Indeed, we write the Cauchy formula, move the integration contour
and vanish integrals around the cycles thanks to the Riemann bilinear identity \eq{RiemannbilinearId}. This gives: \bea
{W}_{k+1}^{(h)}(p,\bfp_{K}) &=& - \Res_{p'\to p}
{W}_{k+1}^{(h)}(p',\bfp_{K})  dS_{p',\a}(p) \cr & = & \sum_s
\Res_{p'\to a_s} {W}_{k+1}^{(h)}(p',\bfp_{K})  dS_{p',\a}(p) \eea

We now introduce \eq{6110} inside this formula and keep only terms
which have poles at the branch points: \beq
\begin{array}{l}
W_{k+1}^{(h)}(p,\bfp_K) = \cr
 - \sum_{m=0}^{h-1} \sum_s \Res_{p'\to a_s} W_{1}^{(h-m)}(p') {R}_{k}^{(m)}(p';\bfp_K) dS_{p',o}(p)\cr
 - \sum_{m=0}^h \sum_{j=0, m+j \neq 0}^{k-1}  \sum_s \Res_{p'\to a_s} {R}_{j}^{(m)}(p';\bfp_J) {W}_{k-j+1}^{(h-m)}(p',\bfp_{K-J}) dS_{p',o}(p)\cr
 - \sum_s \Res_{p'\to a_s} R_{k+1}^{(h-1)}(p';p',\bfp_k) dS_{p',o}(p)
\end{array}
\eeq

For convenience, let us note: \beq W_{1}^{(0)}(p)\equiv 0 \eeq

Then, the recursive definition of $W_{k+1}^{(h)}(p,p_K)$ reads:
\beq\label{soluce1}\encadremath{
\begin{array}{l}
W_{k+1}^{(h)}(p,\bfp_K) = \cr
 \sum_{i=1}^{d_2}  \sum_{m=0}^h \sum_{j=0, m+j \neq 0}^{k}  \sum_s \Res_{p'\to a_s} {R}_{j}^{i,(m)}(p';\bfp_J) {W}_{k-j+1}^{(h-m)}(p',\bfp_{K-J}) dS_{p',o}(p)\cr
 + \sum_{i=1}^{d_2}  \sum_s \Res_{p'\to a_s} R_{k+1}^{i,(h-1)}(p';p',\bfp_K) dS_{p',o}(p)\cr
\end{array}
}\eeq

\subsubsection{A recursive formula for $R_{k}^{i,(h)}$}
The second step consists in the derivation of an equivalent
formula for $R_{k}^{i,(h)}$. We proceed in the same way as for the
genus 0 case: we use the rational properties of some of the
correlation functions to write the recursive formula, with the aid
of a relation similar to \eq{equality}.

Let $G_k^{(h)}(x(q),y(r))$ be :

\bea
G_k^{(h)}(x(q),y(r)) &=& (y(r)-y(q)) {U}_{k}^{(h)}(q,y(r);\bfp_K) + {U_{k+1}^{(h-1)}(q,y(r);q,p_k) \over dx(q)} \cr
&& + \sum_{m=1}^h
\sum_{j=0}^{k} {1 \over dx(q)} {U}_{j}^{(m)}(q,y(r);\bfp_J)
{W}_{k-j+1}^{(h-m)}(q,\bfp_{K-J}) \cr
&& + \sum_{j=0}^{k-1} {1 \over dx(q)} {U}_{j}(q,y(r);\bfp_J)
{W}_{k-j+1}^{(h)}(q,\bfp_{K-J}) \cr
\eea

The loop equation \eq{eqUklgh} shows that $G_k^{(h)}(x(q),y(r))$ is a rational function in $x(q)$
and a polynomial in $y(r)$.

Thus, one has:
\beq
G_k^{(h)}(x(p^i),y(p^i)) = G_k^{(h)}(x(p),y(p^i))
\eeq

which can be written:

\beq\label{solUkh}
\begin{array}{rcl}
 (y(p^{i})-y(p)) {U}_{k}^{(h)}(p,y(p^{i});\bfp_K) &=&\sum_{m=0}^{h} \sum_{j=0}^{k}  {{W}_{j+1}^{(m)}(p^{i},\bfp_J) {U}_{k-j}^{(h-m)}(p^{i},y(p^{i});\bfp_{K-J}) \over dx(p)}\cr
&& + {U_{k+1}^{(h-1)}(p^{i},y(p^{i});p^{i},\bfp_k) \over
dx}\cr && - \sum_{m=0}^{h} \sum_{j=0}^{k}
{{W}_{j+1}^{(m)}(p,\bfp_{J})
{U}_{k-j}^{(h-m)}(p,y(p^{i});\bfp_{K-J}) \over dx(p)}\cr && -
{U_{k+1}^{(h-1)}(p,y(p^{i});p,\bfp_k) \over dx}\cr
\end{array}
\eeq

We now establish a relation similar to \eq{equality} in order to
present our recursive formula in such a way that it can be
graphically interpreted.

In order to achieve this aim, one has to determine an explicit
intermediate formula for $U_{k}^{(h)}(p,y;p_K)$. Let us assume
that (for the proof, see {\emph{Appendix C}}): \beq\label{eqUW}
\begin{array}{l}
U_{k}^{(h)}(p,y(p^{i});\bfp_K)= \cr {E_y(x,y(p^{i})) \over
y(p^{i})-y(p)} \sum_{r=1}^{min(d_2,k+h)} \sum_{ K_1 \bigcup
\dots \bigcup K_r = K}  \sum_{h_{\alpha} = 0}^h
\sum_{k_\alpha=|K_\alpha|}^{k+h} \sum_{j_{\alpha,\beta} \neq
j_{\alpha',\beta'} \in [1,d_2]-\{i\}} {1 \over \Omega} \cr
{W_{k_1+1}^{(h_1)}(p^{i}, \bfp_{K_1} ,p^{j_{1,1}}, \dots
,p^{j_{1,k_1-|K_1|}}) \left(\prod_{\alpha=2}^{r}
W_{k_{\alpha}+1}^{(h_{\alpha})}(p^{j_{\alpha,0}},\bfp_{K_{\alpha}}
,p^{j_{\alpha,1}}, \dots
,p^{j_{\alpha,k_\alpha-|K_\alpha|}})\right) \over
dx(p)^{r-k-1+\sum k_\alpha} \prod_{\alpha,\beta}
y(p^{i})-y(p^{j_{\alpha,\beta}})}\cr
\end{array}
\eeq

where $\Omega = \prod_{\alpha} (k_\alpha-|K_\alpha|)!$ is a symmetry factor and one has the following constraints:
\begin{itemize}
\item $\sum_{\a} (h_{\alpha}+k_{\alpha}) = h+k$; \item $0 \leq
|K_{\alpha}| \leq k_{\alpha}$

\end{itemize}

One should note that the only external parameter entering these
constraints is $k+h$.

It is now possible to derive an equality equivalent to
\eq{equality}. One shows -- in {\emph{Appendix D}} -- that:
\bea
\label{equality2} &&\sum_{m=0}^{h} \sum_{j=0; mj\neq kh}^{k}
{W}_{j+1}^{(m)}(p,\bfp_J) {U}_{k-j}^{(h-m)}(p,y(p);\bfp_{K-J}) +
{U_{k+1}^{(h-1)}(p,y(p);p,\bfp_k) } \cr &=& \sum_{i=1}^{d_2}
\sum_{m=0}^{h} \sum_{j=0; mj\neq kh}^{k}
{W}_{j+1}^{(m)}(p^{i},\bfp_J)
{U}_{k-j}^{(h-m)}(p^{i},y(p);\bfp_{K-J}) \cr && +
\sum_{i=1}^{d_2} {U_{k+1}^{(h-1)}(p^{i},y(p);p^{i},\bfp_k) }
\cr \eea

This equality allows us to write: \beq
\begin{array}{l}
(y(p^{i})-y(p)) {U}_{k}^{(h)}(p,y(p^{i});\bfp_K) = \cr
\sum_{m=0}^{h} \sum_{j=0; mj\neq kh}^{k}  \sum_{l\neq 0,i}
{{W}_{j+1}^{(m)}(p^{l},\bfp_J)
{U}_{k-j}^{(h-m)}(p^{l},y(p^{i});\bfp_{K-J}) \over dx(p)}\cr
 + \sum_{l\neq 0,i} {U_{k+1}^{(h-1)}(p^{l},y(p^{i});p^{l},\bfp_k) \over dx(p)} + W_{k+1}^{(h)}(p^{i},\bfp_K) E_y(x,y(p^{i}))\cr
\end{array}
\eeq

That is to say: \beq\label{soluce2}\encadremath{
\begin{array}{rcl}
R_{k}^{i,(h)}(p,\bfp_K) &=& \sum_{m=0}^{h} \sum_{j=0; mj\neq
kh}^{k}  \sum_{l\neq 0,i} {{W}_{j+1}^{(m)}(p^{l},\bfp_J)
{R}_{k-j}^{i,(h-m)}(p^{l};\bfp_{K-J}) \over (y(p^{i})-y(p))
dx(p) }\cr && + \sum_{l\neq 0,i}
{R_{k+1}^{i,(h-1)}(p^{l};p^{l},\bfp_k) \over (y(p^{i})-y(p))
dx(p)} + {W_{k+1}^{(h)}(p^{i},\bfp_K) \over (y(p^{i})-y(p))
dx(p) }\cr
\end{array}
}\eeq

\subsection{Diagrammatic solution: a cubic theory}
\label{cub}

This section is the principal part of the article. We define a
correspondence between the correlation functions and a system of
Feynman-like graphs. To every $k$-point function of genus $h$, we
associate a graph with $k$ external legs and $h$ loops and
\eq{soluce1} and \eq{soluce2} become two relations describing these
graphs as functions of graphs with less legs or loops thanks to some rules we introduce in this part.

First of all, let us represent diagrammatically \eq{prop1} and
\eq{prop2} as the propagators of the theory:

\beq W_{2}(p,q) = \begin{array}{l} {\epsfxsize
3.5cm\epsffile{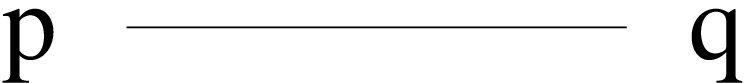}}
\end{array}
\eeq

and

\beq R_{1}^i(p,p_1) = \begin{array}{l} {\epsfxsize
3.5cm\epsffile{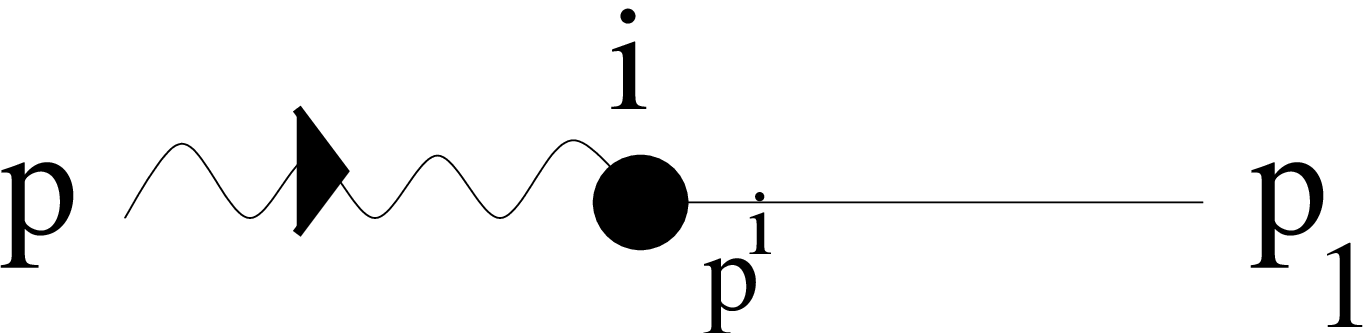}}
\end{array}
\eeq

These two diagrams represent the basis of the whole
representation: they allow to draw the $k>2$ correlation
functions.

Note that the second propagator can also be seen has a vertex of valence 2,
and this is the way it will be presented in the diagrammatic rules.

\smallskip

Let us now introduce the whole diagrammatic representation:

Let $R_{k}^{i,(h)}$, and $W_{k+1}^{(h)}$ respectively, be
represented as white and black disks with h holes and $k$ external legs
(remember that $W_{k+1}^{(h)}$
is the generating function of discrete surfaces with $k+1$ boundaries and $h$ holes):

\beq W_{k+1}^{(h)}(p,p_K)  := \begin{array}{l} {\epsfxsize
4cm\epsffile{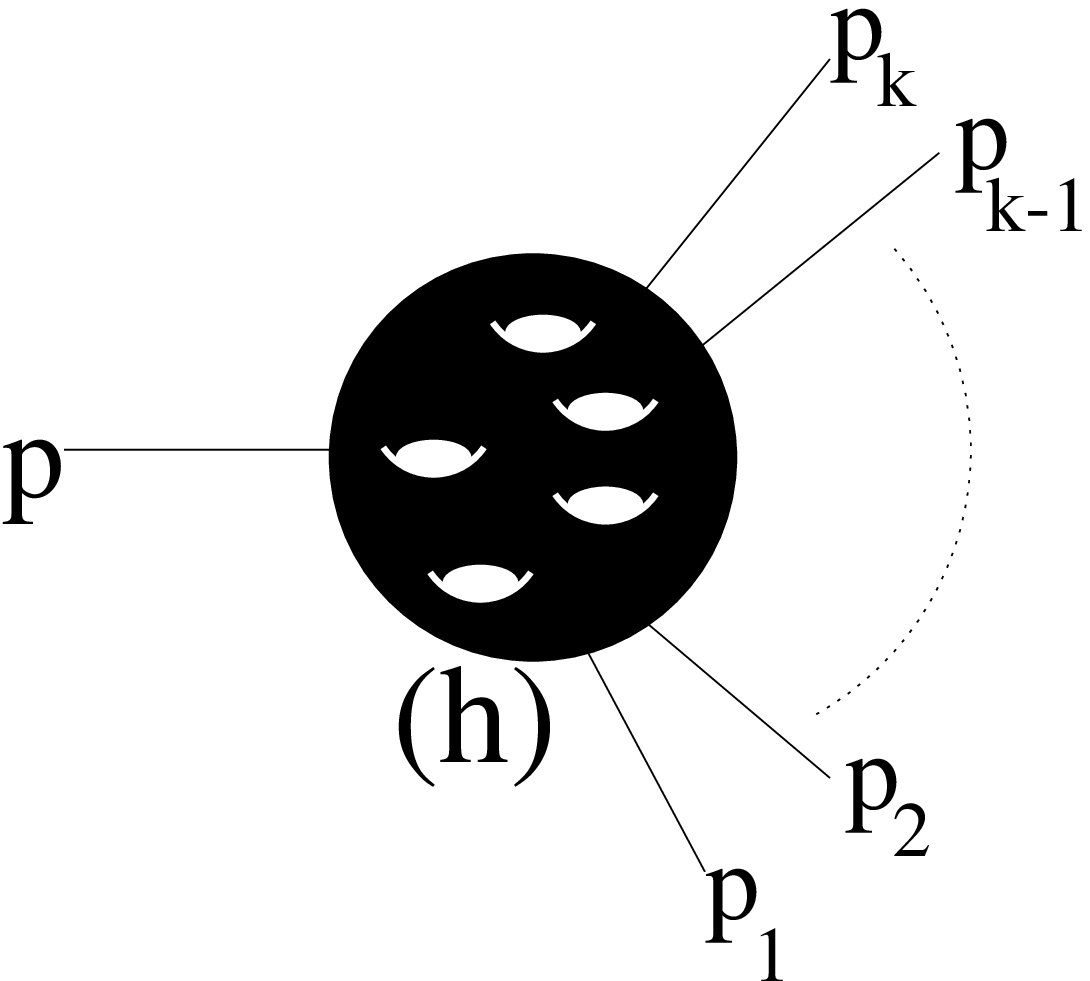}}
\end{array}
\eeq

\beq R_{k}^{i,(h)}(p,p_K)  := \begin{array}{l} {\epsfxsize
4cm\epsffile{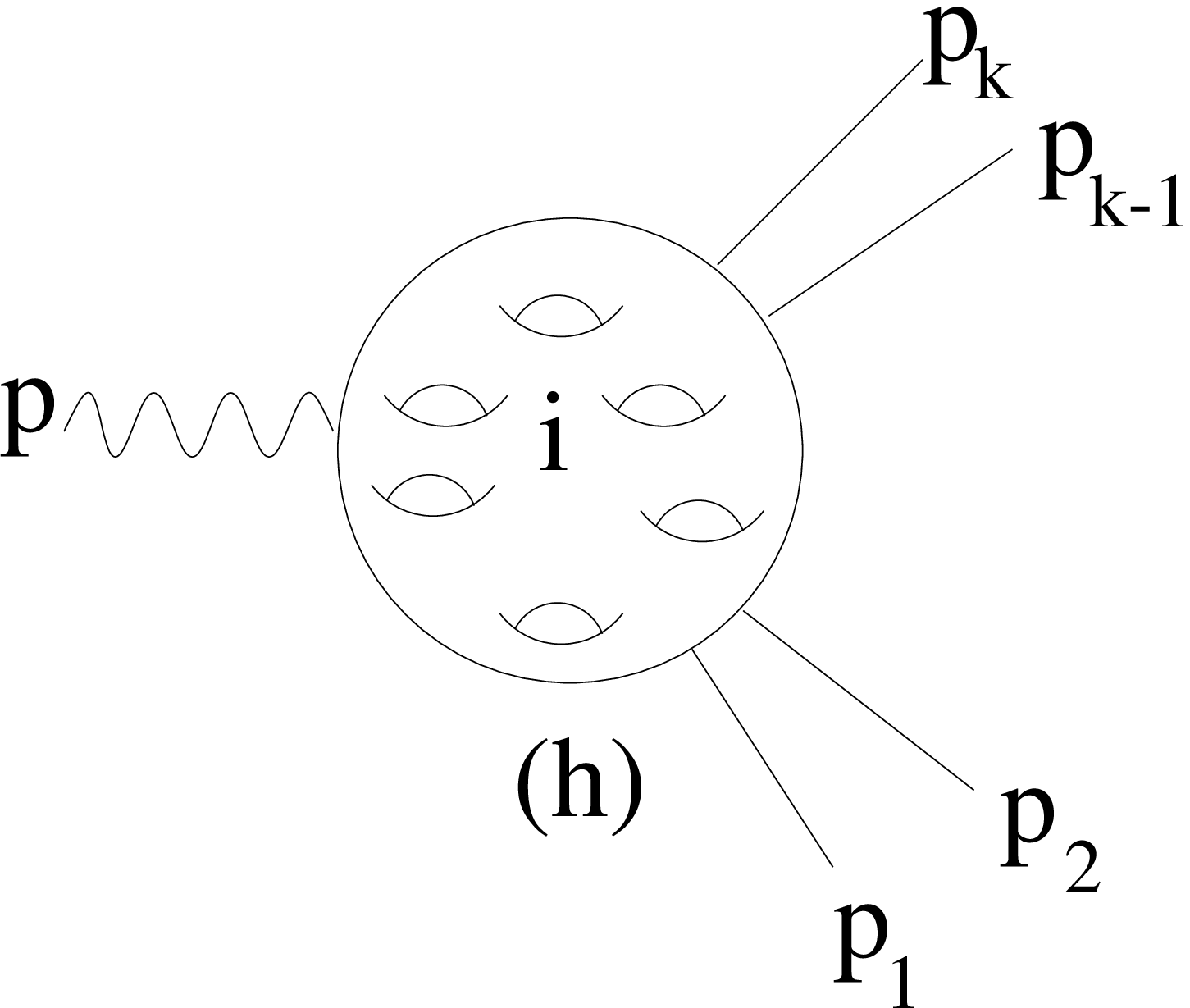}}
\end{array}
\eeq

Let us introduce also the following propagators and vertices:

\begin{center}
\begin{tabular}{|r|l|}\hline
non-arrowed propagator:&
$
\begin{array}{r}
{\epsfxsize 2.5cm\epsffile{W2.eps}}
\end{array}
:=W_{2}(p,q)
$
\cr\hline
arrowed propagator:&
$
\begin{array}{r}
{\epsfxsize 2.5cm\epsffile{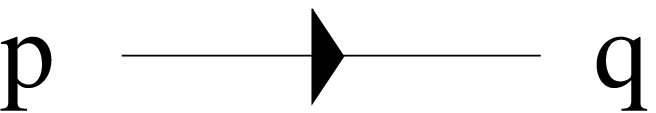}}
\end{array}
 :=dS_{q,o}(p)
$
\cr\hline
Residue cubic-vertex:&
$
\begin{array}{r}
{\epsfxsize 2.5cm\epsffile{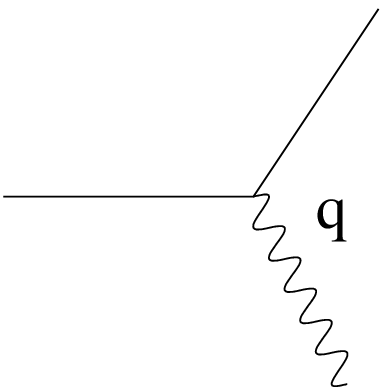}}
\end{array}
 := \sum_s \Res_{q\rightarrow a_s}
$
\cr\hline
colored cubic-vertices:&
$
\begin{array}{r}
{\epsfxsize 2cm\epsffile{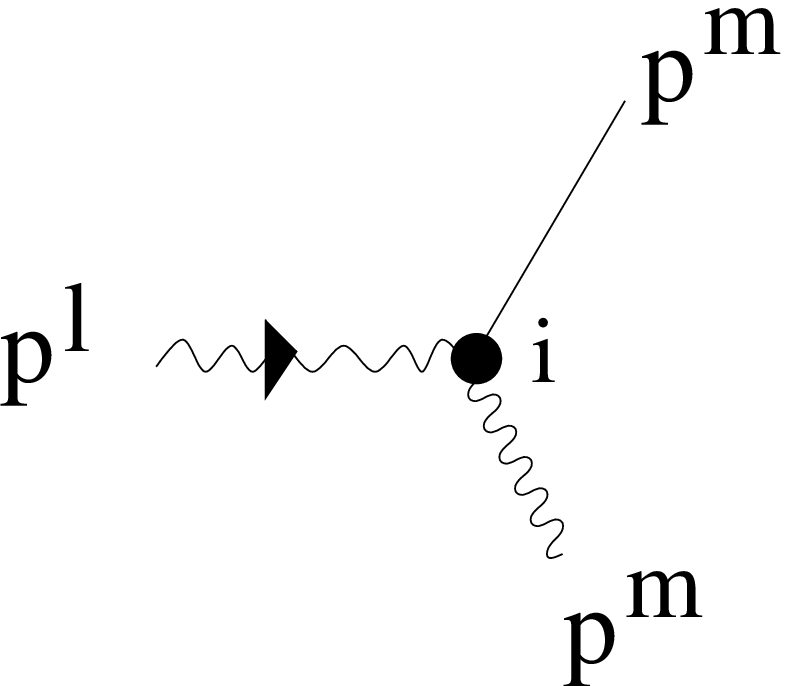}}
\end{array}
:={(1 - \delta_{l,m}) (1-\delta_{m,i}) (1-\delta_{i,l})\over (y(p^{i})-y(p^{l}))dx(p)}
$
\cr\hline
2-valent vertex:&
$
\begin{array}{r}
{\epsfxsize 2cm\epsffile{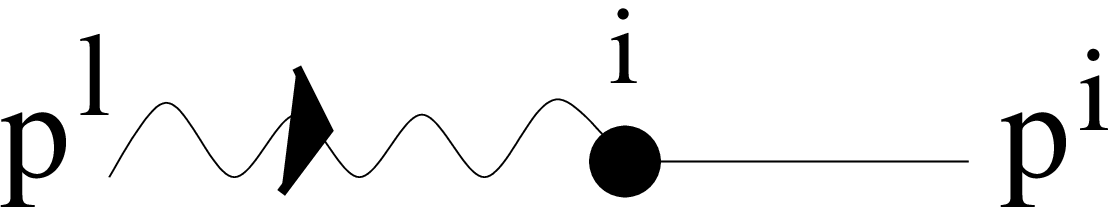}}
\end{array}
:= {1 \over (y(p^{i})-y(p^{l}))dx(p)} (1-\delta_{i,l})
$
\cr
\hline
\end{tabular}
\end{center}

One can now simply interpret the recursion relations \eq{soluce1}
and \eq{soluce2} in terms of diagrams.

The relation \eq{soluce1} reads:

\bea
\begin{array}{r}
{\epsfxsize 3cm\epsffile{Wh.eps}}
\end{array}
&=&\sum_{i=1}^{d_2} \sum_{m=0}^h \sum_{j=0, m+j\neq 0}^k
\sum_{J\in K_j}  \begin{array}{l} {\epsfxsize
4cm\epsffile{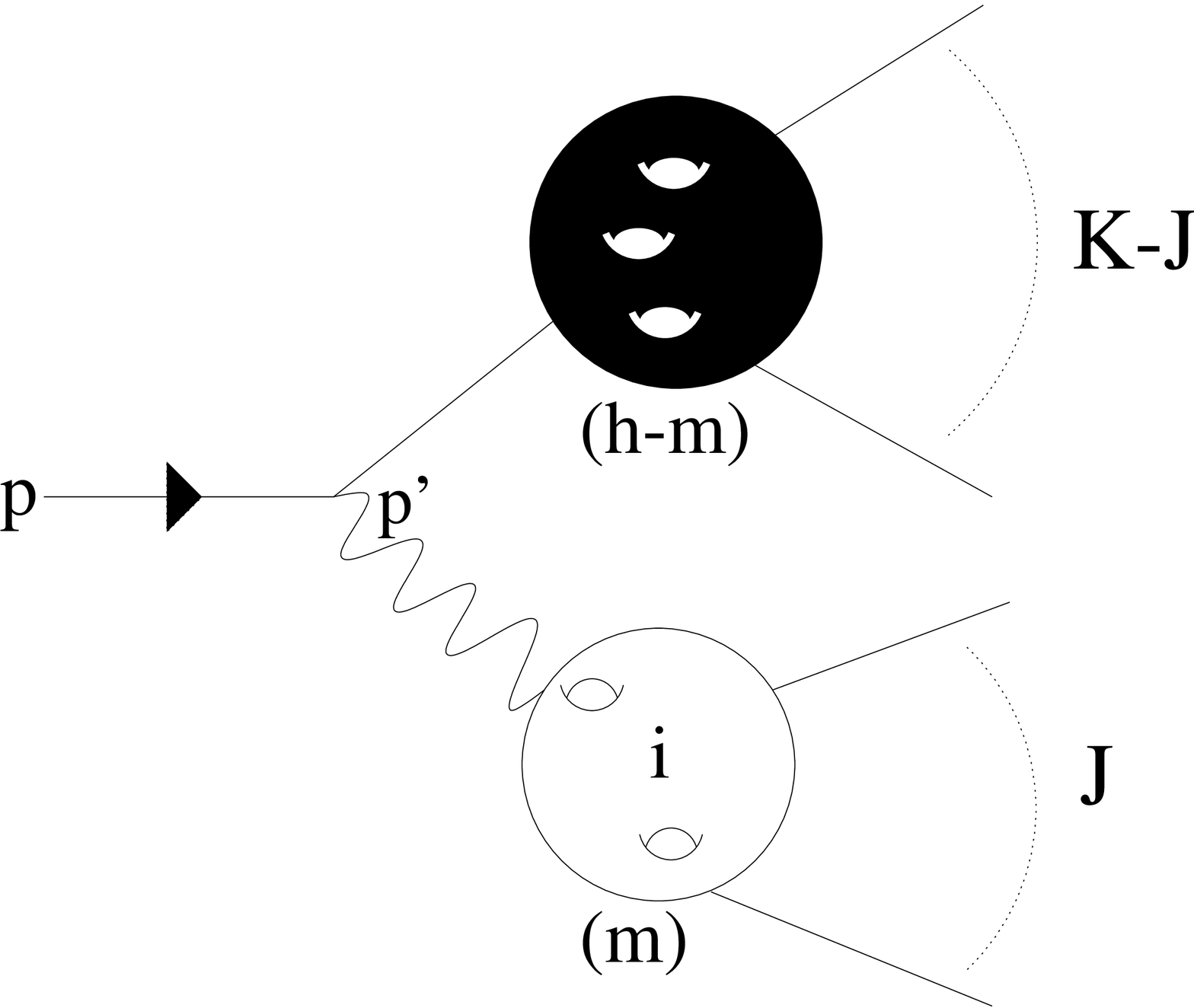}} \end{array} \cr && +
\sum_{i=1}^{d_2}\begin{array}{l} {\epsfxsize
5cm\epsffile{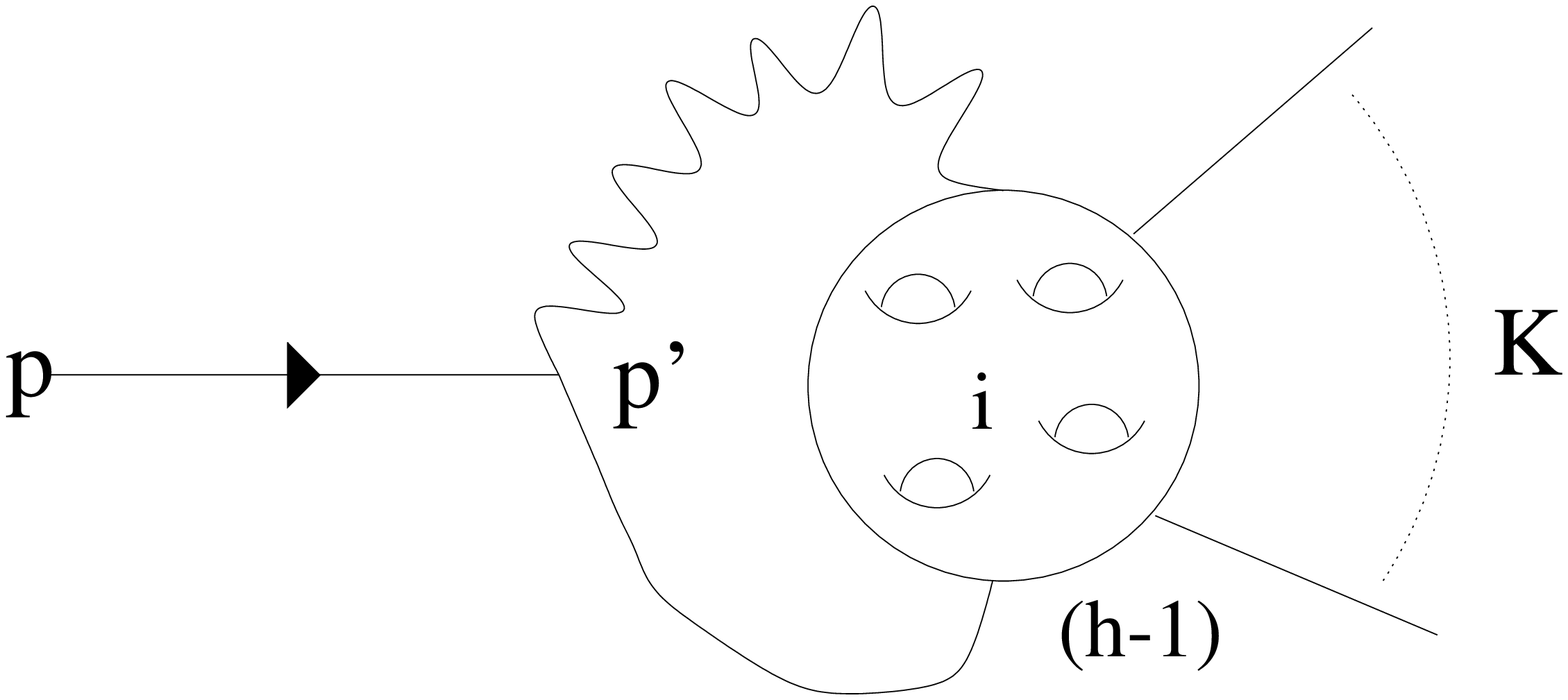}} \end{array} \cr \eea

And given lower order $R_{l}^{i,(m)}$'s and $W_{l}^{(m)}$'s, one can
obtain $R_{k}^{i,(h)}$ diagrammatically by writing \eq{soluce2}:

\beq
\begin{array}{rcl}
\begin{array}{r}
{\epsfxsize 3cm\epsffile{Rh.eps}}
\end{array}
&=& \sum_{m=0}^h \sum_{j=0, m+j\neq 0}^k \sum_{J\in K_j}
\sum_{l=0}^{d_2} \begin{array}{l} {\epsfxsize
3.2cm\epsffile{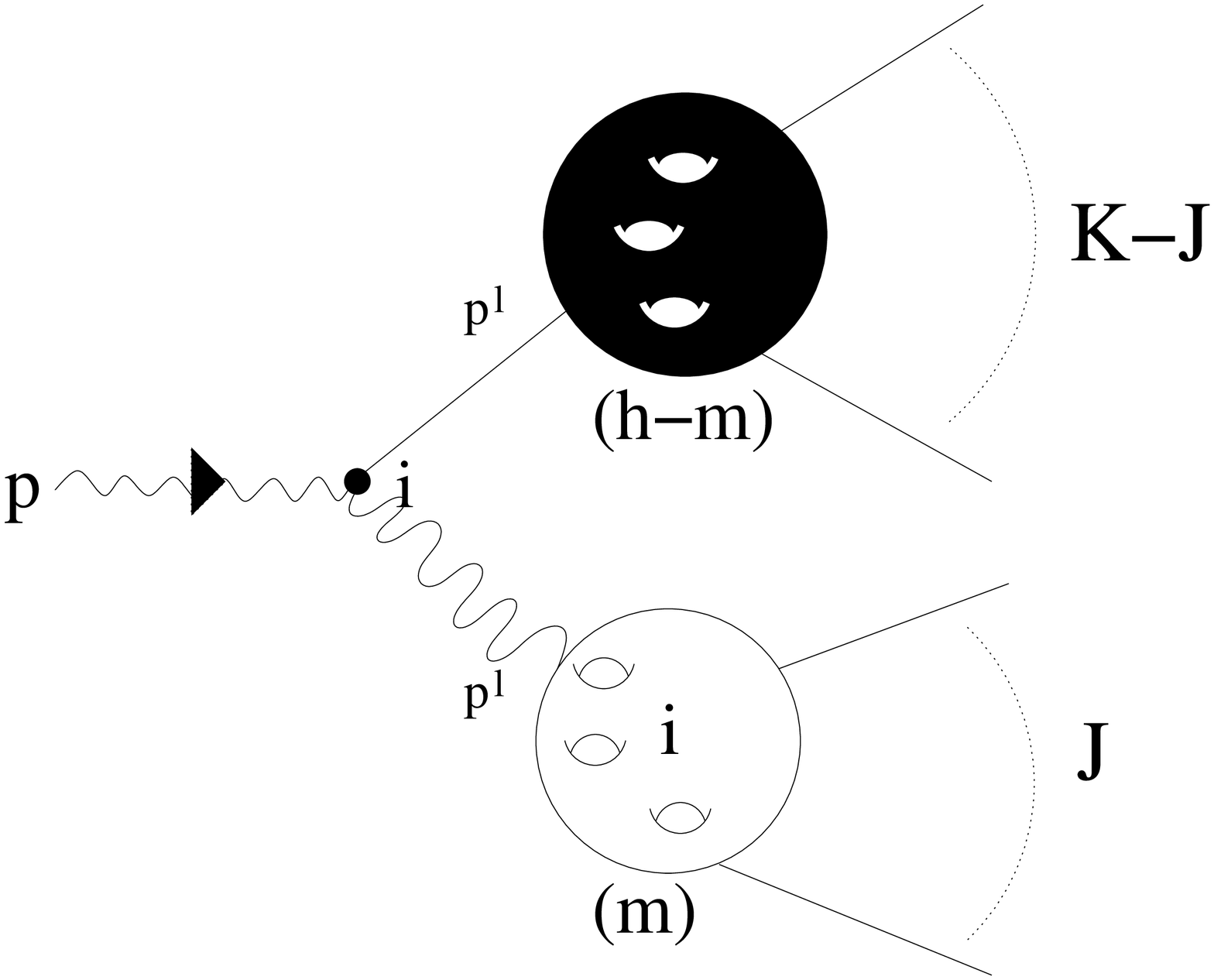}} \end{array}\cr &&+ \sum_{l=0}^{d_2}
\begin{array}{l} {\epsfxsize 5cm\epsffile{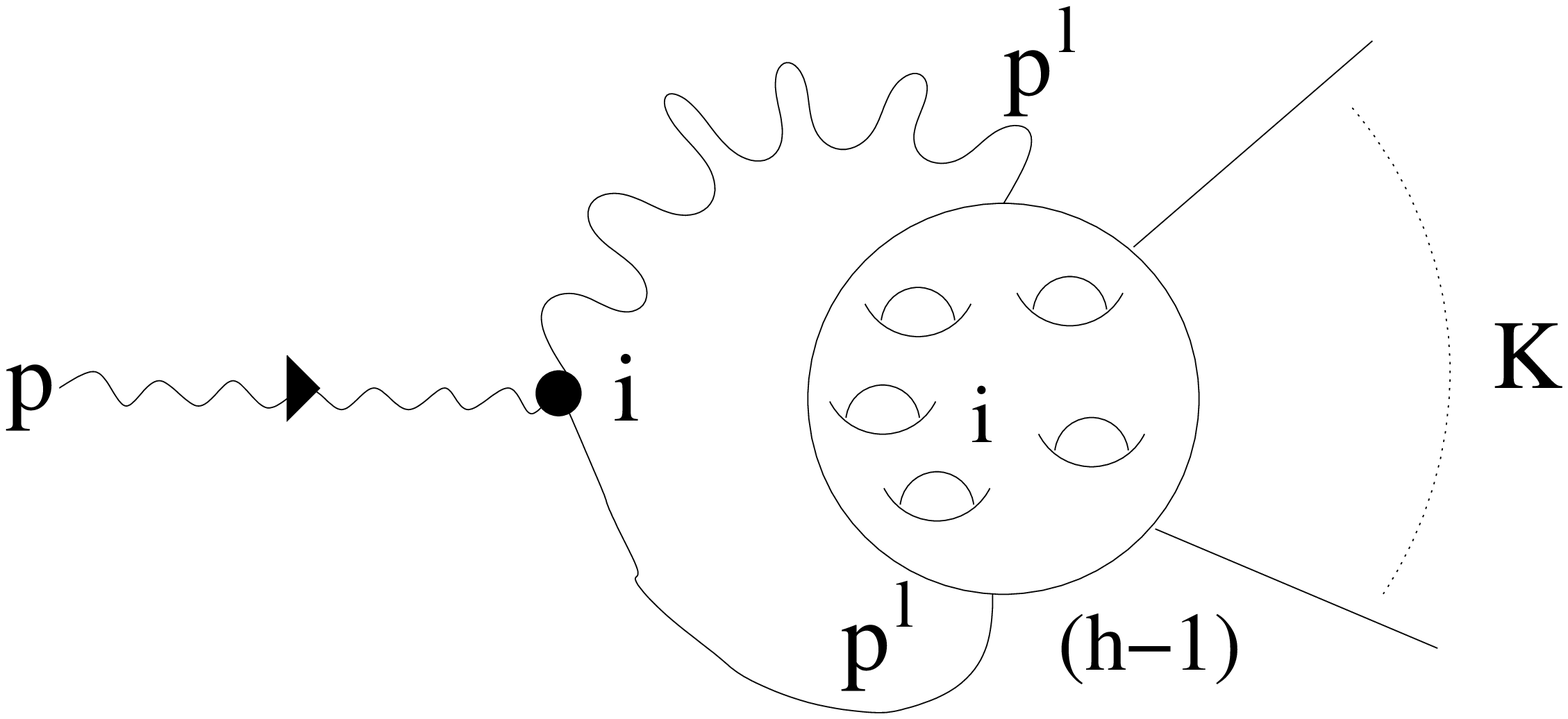}} \end{array}
\cr && \qquad +\begin{array}{l} {\epsfxsize
5cm\epsffile{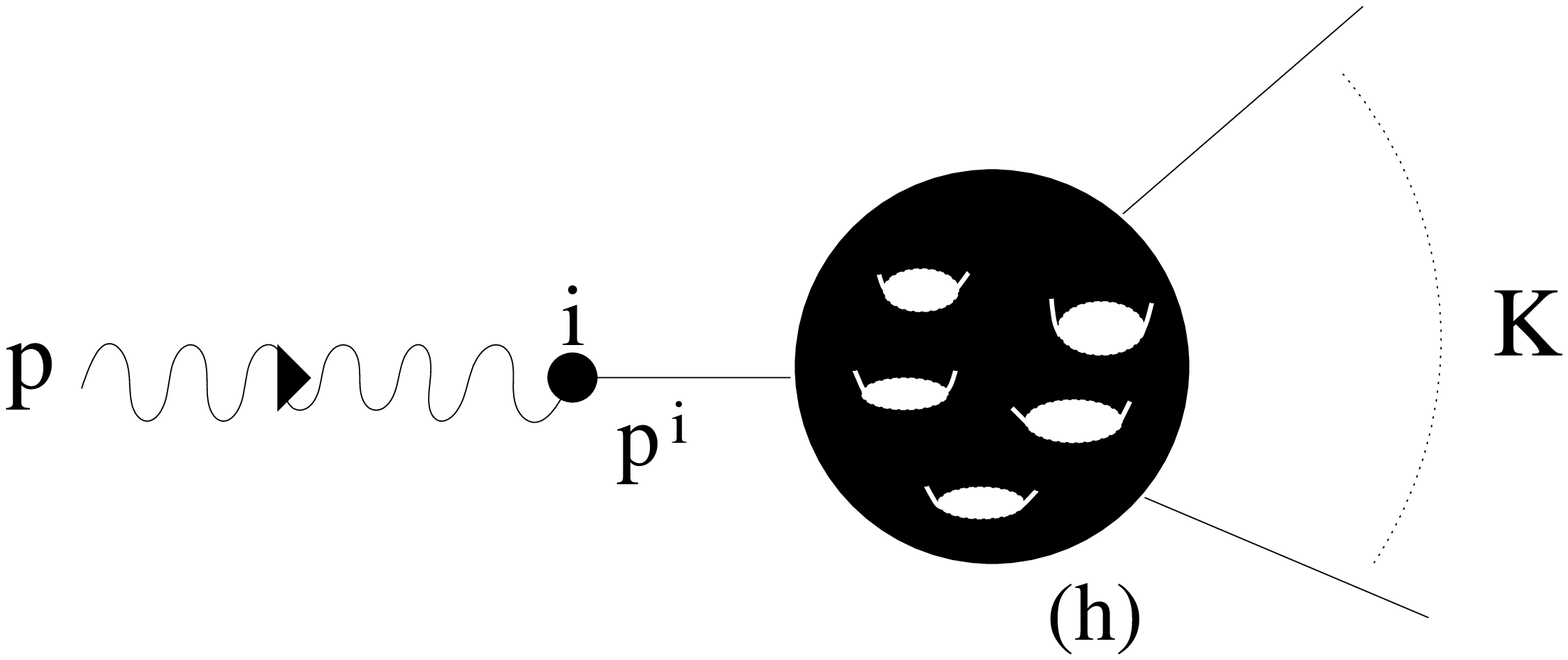}} \end{array} \cr
\end{array}
\eeq

From these diagrammatic relations, one can see that $W_{k+1}^{(h)}$ is obtained by {\em the summation
over all diagrams with 1 root, k leaves and $h$ loops}
following the rules:

{\em
\begin{itemize}

\item The vertices have valence 2 or 3; there are $2h+k-1$
trivalent vertices;

\item The edges, are arrowed or not, the arrowed edges are waved
or not;

\item The subgraph made of arrowed edges forms a skeleton tree (i.e. a tree whose vertices have valence up to 3);

\item from each trivalent vertex comes one waved and one
non-waved propagator;

\item two vertices linked with a waved propagator have different indices;

\item the k leaves are non-arrowed propagators finishing at
$p_j$'s (i.e. $B(.,p_j)$);

\item the root is an arrowed non waved propagator starting from $p$.

\end{itemize}
}

A practical way to draw these graphs is to draw every skeleton
tree of arrows, put $k$ non arrowed propagators as leaves,
close it with $h$ non arrowed propagators linking one vertex to one of its descendents
in order to obtain $h$ loops and then put waves so that from each trivalent
vertex comes one waved and one non-waved arrow with the possibility that every
waved arrow leads to a bivalent vertex.

{\bf Remarks:}
\begin{itemize}
\item The order for computing the residues is following the arrows backwards from leaves
to root.

\item $W_{k+1}$ is symmetric in its $k+1$ variables, although it is not obvious from this representation.

\item There is no symmetry factor arising in this representation unlike \cite{eynloop1mat}.

\end{itemize}

\subsection{Examples}

Let us briefly show some diagrams for
small $h$ and small $k$.

\subsubsection{Leading terms: tree level}

We begin by the leading terms of the first correlation functions,
i.e. for $h=0$.

$\bullet$ $k=3$:

\beq
\begin{array}{rcl}
W_{3}^{(0)}(p,p_1,p_2) &=& \sum_{i=1}^{d_2} \begin{array}{l} {\epsfxsize
5.5cm\epsffile{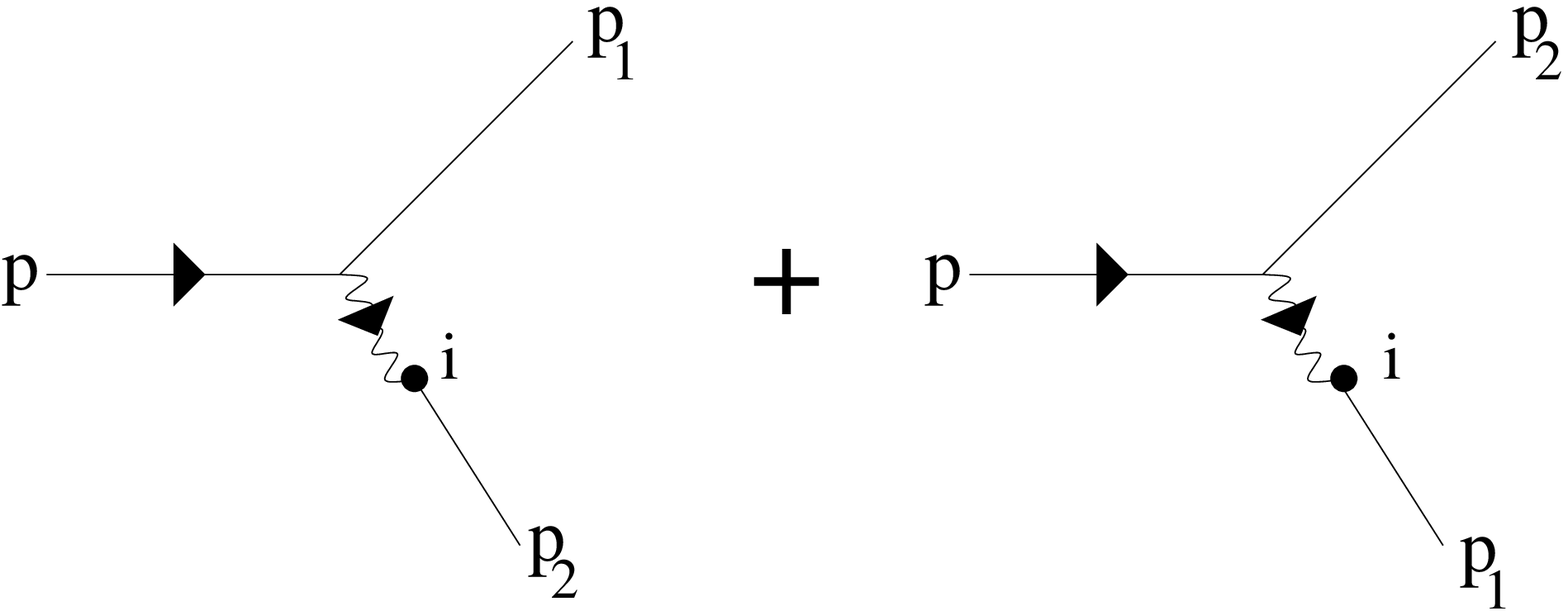}} \end{array}\cr &=& \sum_{i=1}^{d_2}\sum_s
\Res_{p' \rightarrow a_s} \left[ {B(p'^{i},p_1) B(p',p_2) \over
(y(p'^{i})-y(p')) dx(p')} + {B(p'^{i},p_2) B(p',p_1) \over
(y(p'^{i})-y(p')) dx(p')} \right] dS_{p',\a}(p)
\end{array}
\eeq

and

\beq
\begin{array}{rcl}
R_{2}^{i,(0)}(p,p_1,p_2) &=& \sum_{j=1}^{d_2} \begin{array}{l}
{\epsfxsize 6cm\epsffile{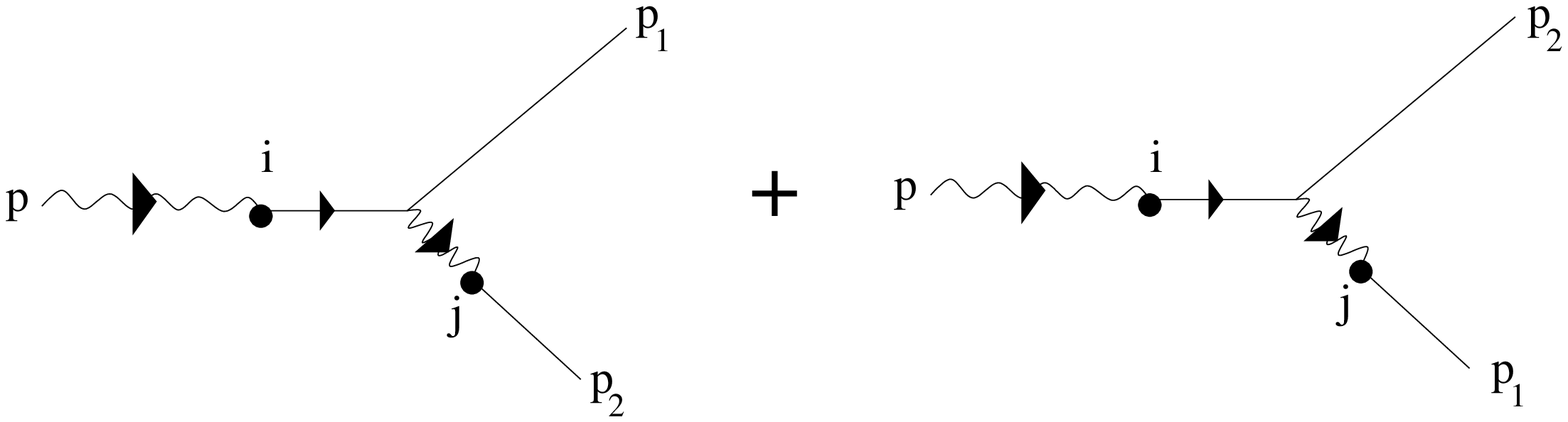}} \end{array} \cr && +
\sum_{j\neq i} \begin{array}{l} {\epsfxsize
7cm\epsffile{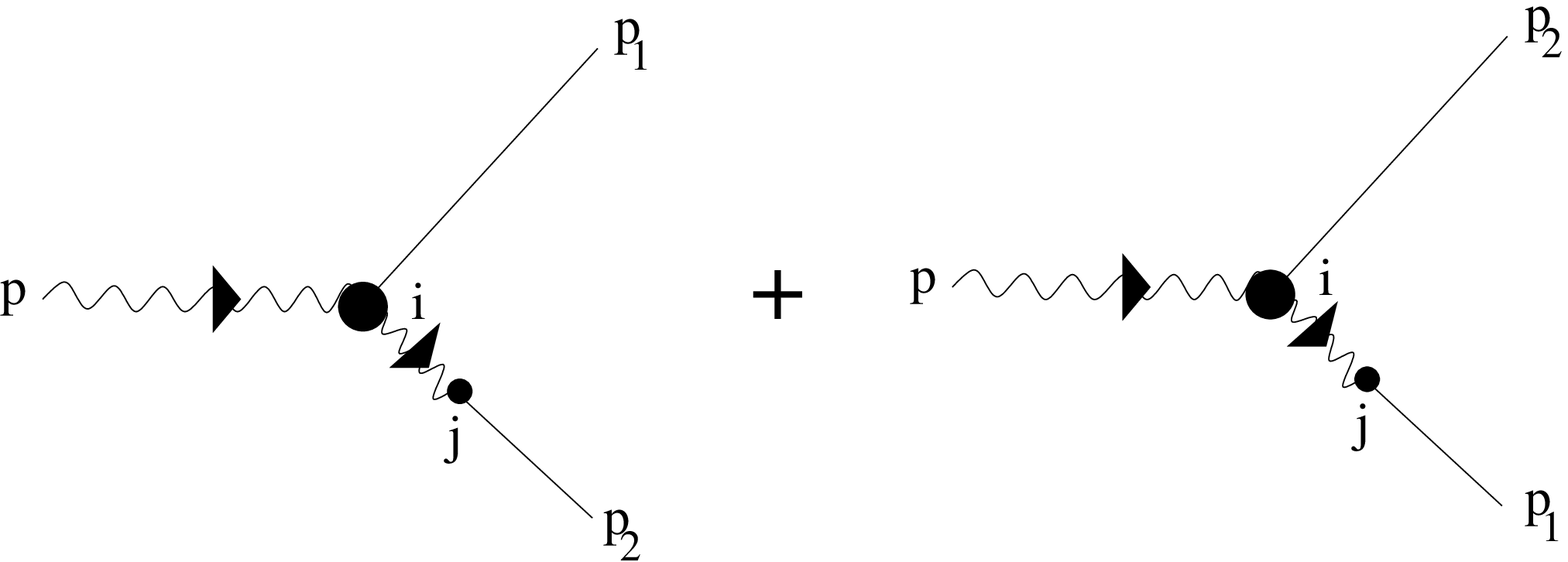}} \end{array} \cr
\end{array}
\eeq
\smallskip

Let us show that $W_3^{(0)}(p,p_1,p_2)$ is indeed symmetric in $p_1$, $p_2$ and $p_3$.

For every branch point $a$, let $\overline{q}$ be the only $q^i$ such that $dx(\overline{q}) \to 0$
when $q \to a$.

\beq
\begin{array}{rcl}
W_{3}^{(0)}(p,p_1,p_2) &=& \sum_{i=1}^{d_2} \sum_s
\Res_{q \rightarrow a_s} {B(q^{i},p_1) B(q,p_2)
+ B(q^{i},p_2) B(q,p_1) \over
(y(q^{i})-y(q)) dx(q)} dS_{q,\a}(p) \cr
&=& \sum_{i=1}^{d_2} \sum_s
\Res_{q \rightarrow a_s} \Res_{r \to q^i} {B(r,p_1) B(q,p_2)
+ B(r,p_2) B(q,p_1) \over
(y(r)-y(q)) (x(r)-x(q)) dx(q)} dS_{q,\a}(p) \cr
&=& \sum_s
\Res_{q \rightarrow a_s} \Res_{r \to \overline{q}} {B(r,p_1) B(q,p_2)
+ B(r,p_2) B(q,p_1) \over
(y(r)-y(q)) (x(r)-x(q)) dx(q)} dS_{q,\a}(p) \cr
&=& \sum_s
\Res_{q \rightarrow a_s} {B(q,p_1) B(\overline{q},p_2) dS_{q,\overline{q}}(p) \over (y(\overline{q})-y(q)) dx(q)} \cr
&=& - \sum_s
\Res_{q \rightarrow a_s} {B(q,p_1) B(q,p_2) dS_{q,\overline{q}}(p) \over (y(\overline{q})-y(q)) dx(q)} \cr
&=& \sum_s
\Res_{q \rightarrow a_s} {B(q,p_1) B(q,p_2) B(q,p) \over dx(q) dy(q)} \cr
\end{array}
\eeq

which is nothing but the formula found in \cite{Kri} and is a way of writing Rauch's variational formula.

$\bullet$ $k=4$: \beq
\begin{array}{rcl}
W_4^{(0)}(p,p_1,p_2,p_3) &=& \sum_{i=1}^{d_2} \sum_{j=1}^{d_2}
\begin{array}{l} {\epsfxsize 4cm\epsffile{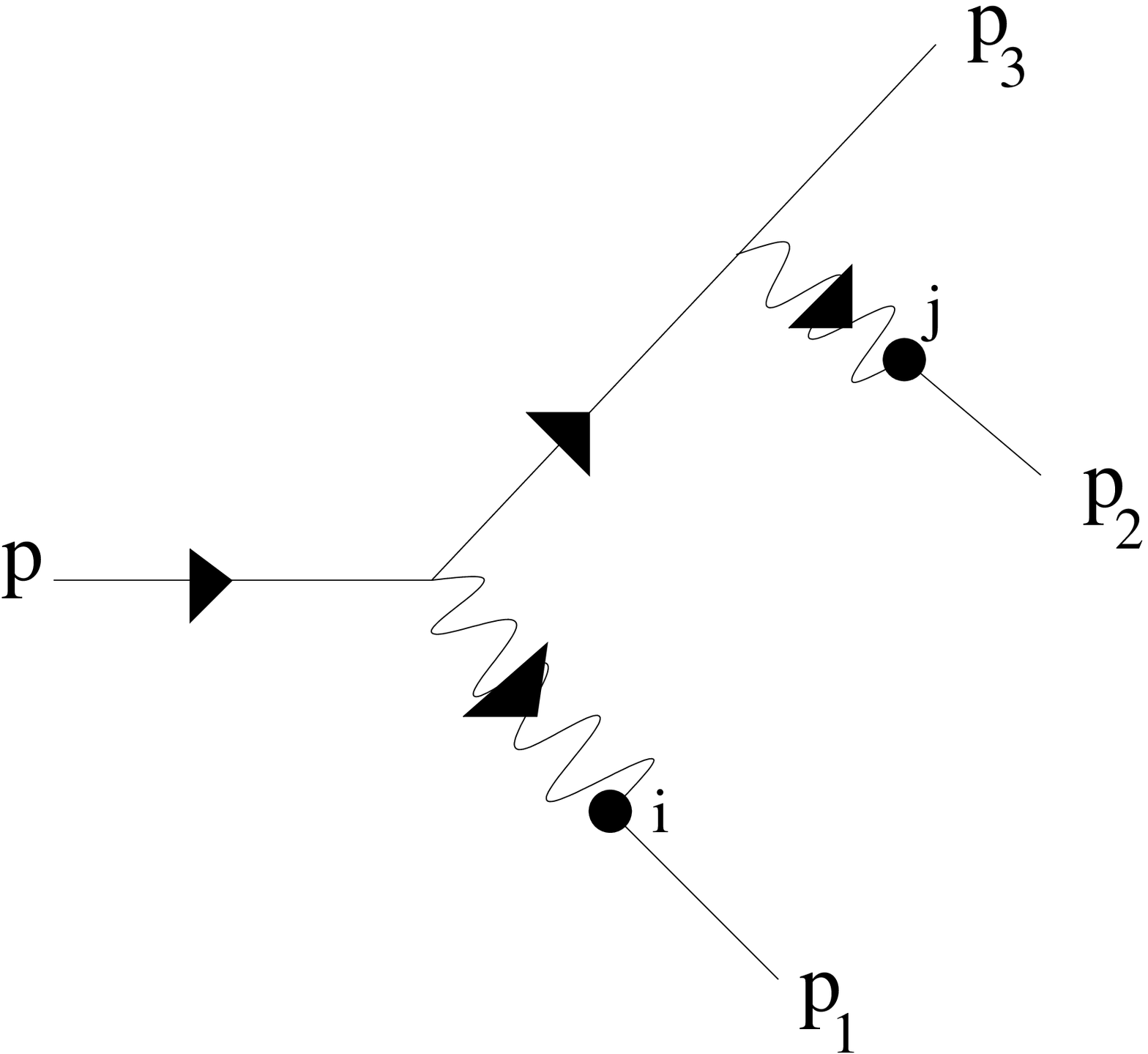}} \end{array}
\cr && + \sum_{i=1}^{d_2} \sum_{j=1}^{d_2} \begin{array}{l}
{\epsfxsize 4cm\epsffile{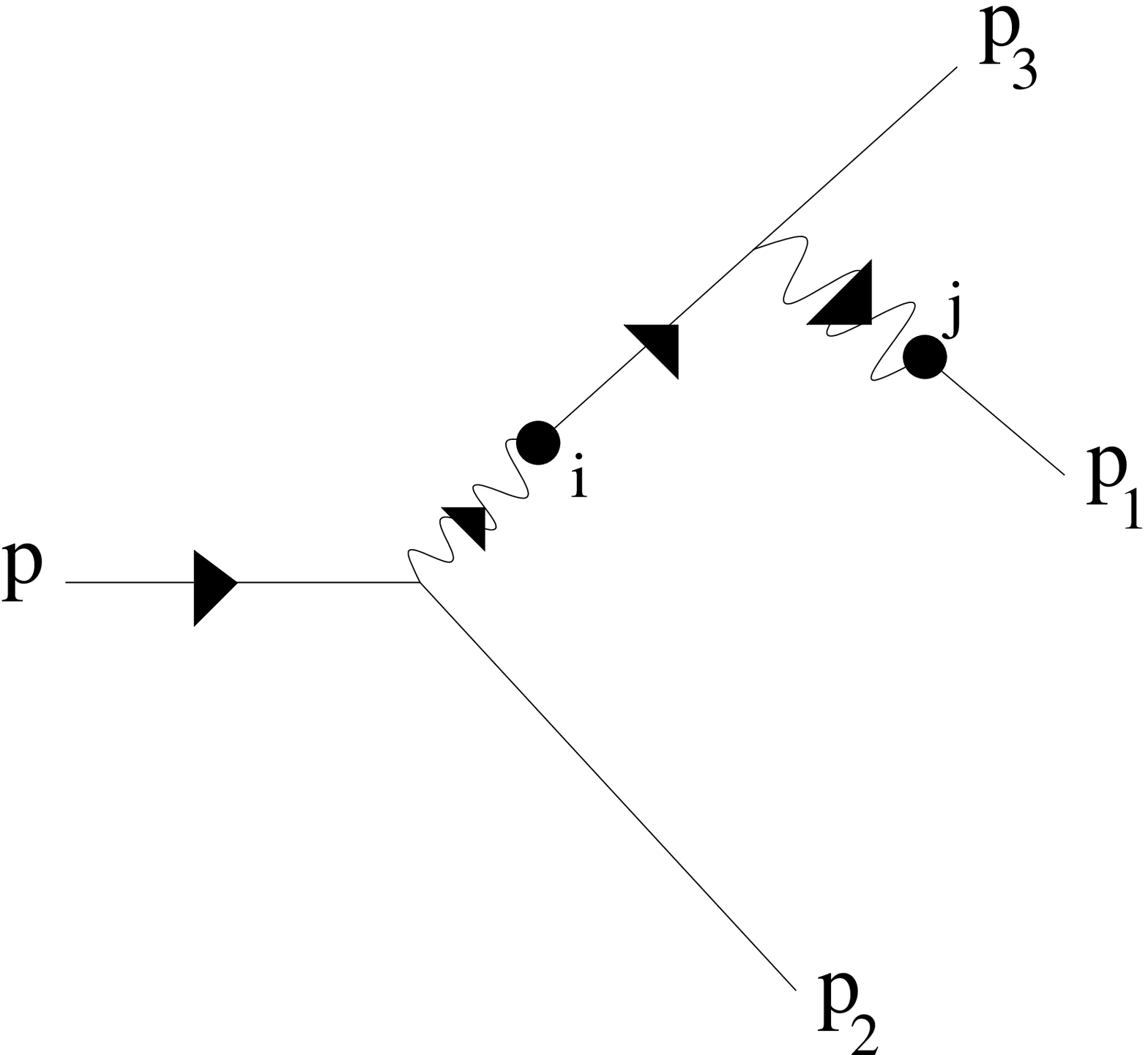}} \end{array} \cr && +
\sum_{i=1}^{d_2} \sum_{j\neq i =1}^{d_2} \begin{array}{l}
{\epsfxsize 4cm\epsffile{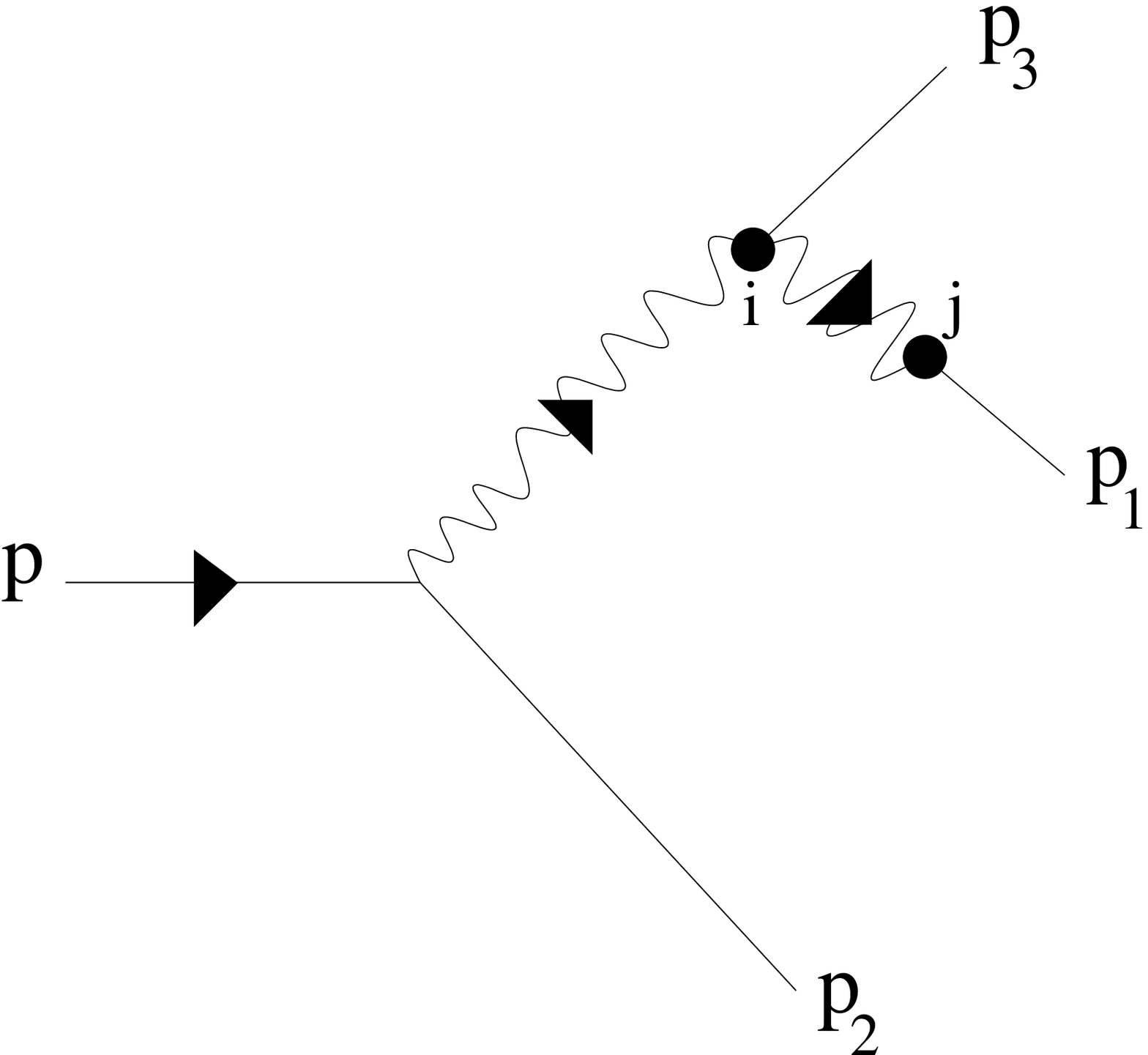}} \end{array} \cr && + (\;
\hbox{permutations of}\; \{p_1, p_2, p_3 \}\,)
\end{array}
\eeq

One has to consider all the permutations on the external legs.
Thus, $W_4^{(0)}$ is the sum over 18 different diagrams.

\subsubsection{Topological expansion: one and two loops level}

Consider now the first non planar examples beginning by the
simplest one, the one loop correction to the one point function.

$\bullet$ $k=1$ and $h=1$:

\bea W_{1}^{(1)}(x(p))dx(p) &=& \begin{array}{l} {\epsfxsize
3cm\epsffile{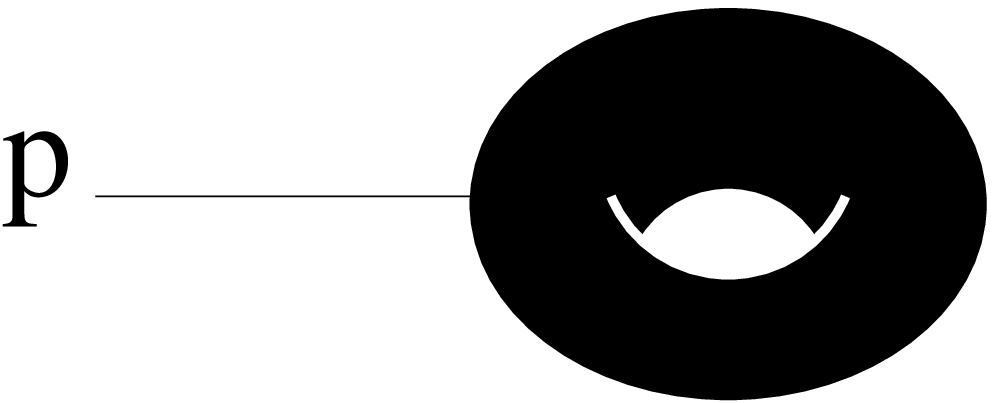}}
\end{array} \cr
& = & \sum_{i=1}^{d_2} \begin{array}{l} {\epsfxsize
3cm\epsffile{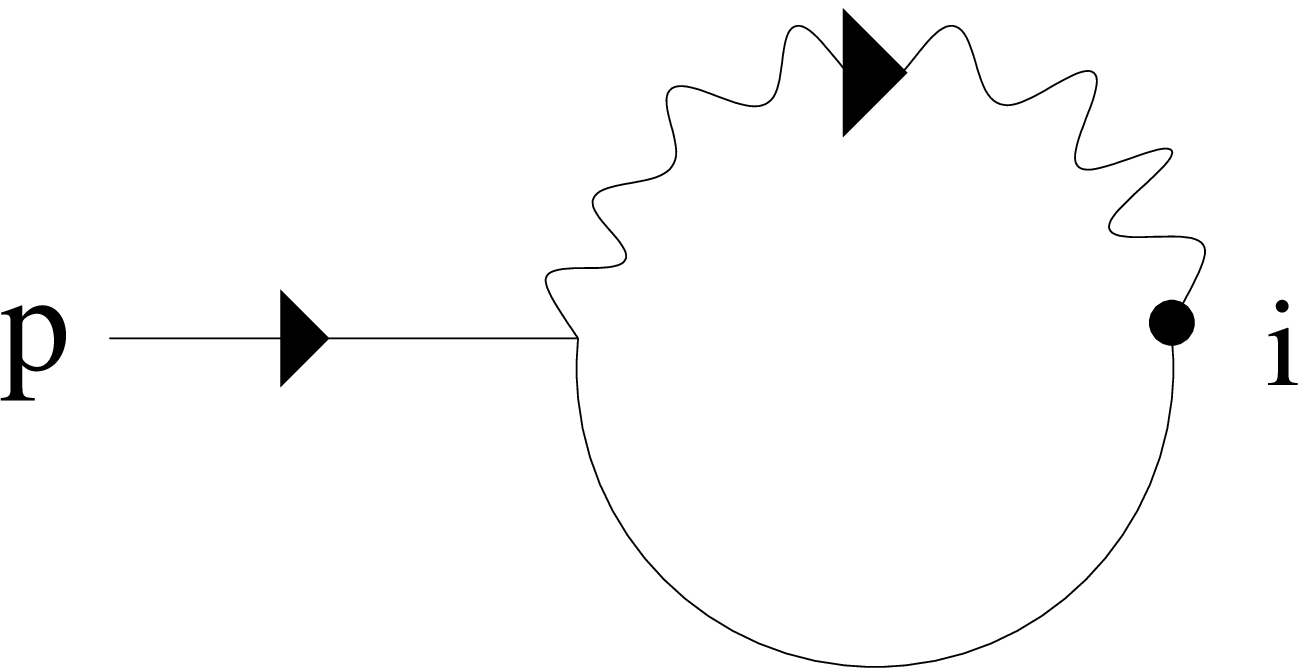}} \end{array} \cr &=& \sum_{i=1}^{d_2}
\sum_s \Res_{q\rightarrow a_s} dS_{q,o}(p) {B(q,q^{i}) \over
y(q^{i})-y(q)} \cr \eea

One can check that this is identical to the result of \cite{EKK}.

$\bullet$ $k=2$ and $h=1$:

\beq
\begin{array}{rl}
W_{2}^{(1)} = & \sum_{i=1}^{d_2} \sum_{j\in [1,d_2]-\{i\}} \left[
\begin{array}{l} {\epsfxsize 3cm\epsffile{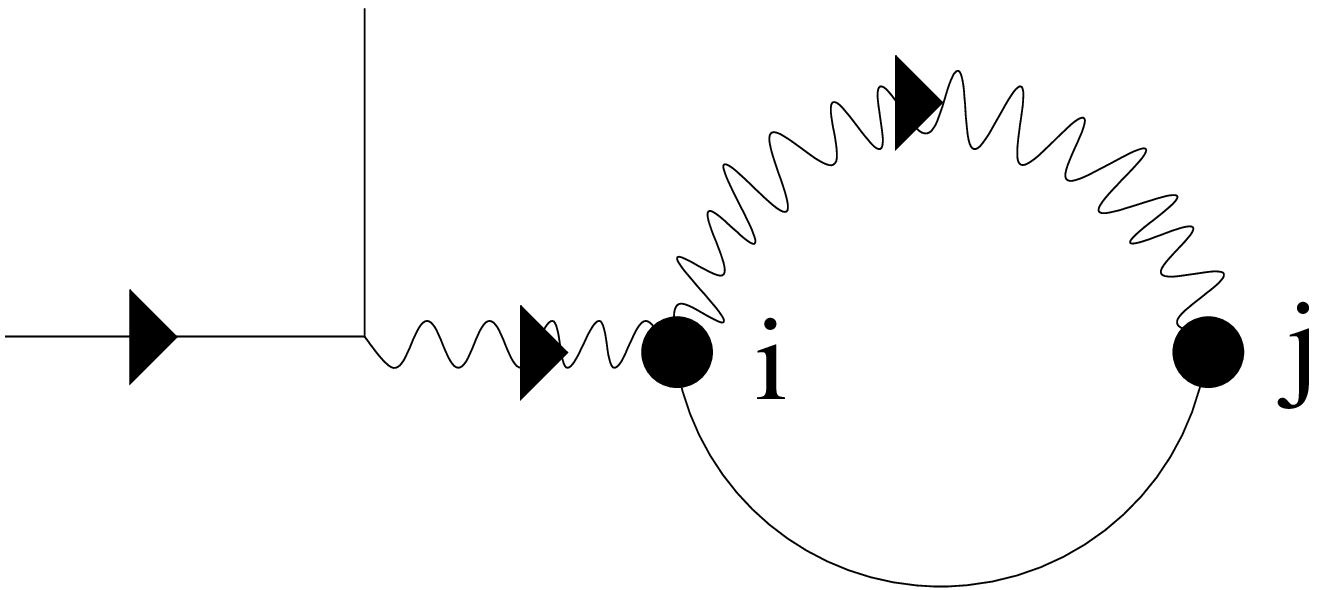}} \end{array} +
\begin{array}{l} {\epsfxsize 3cm\epsffile{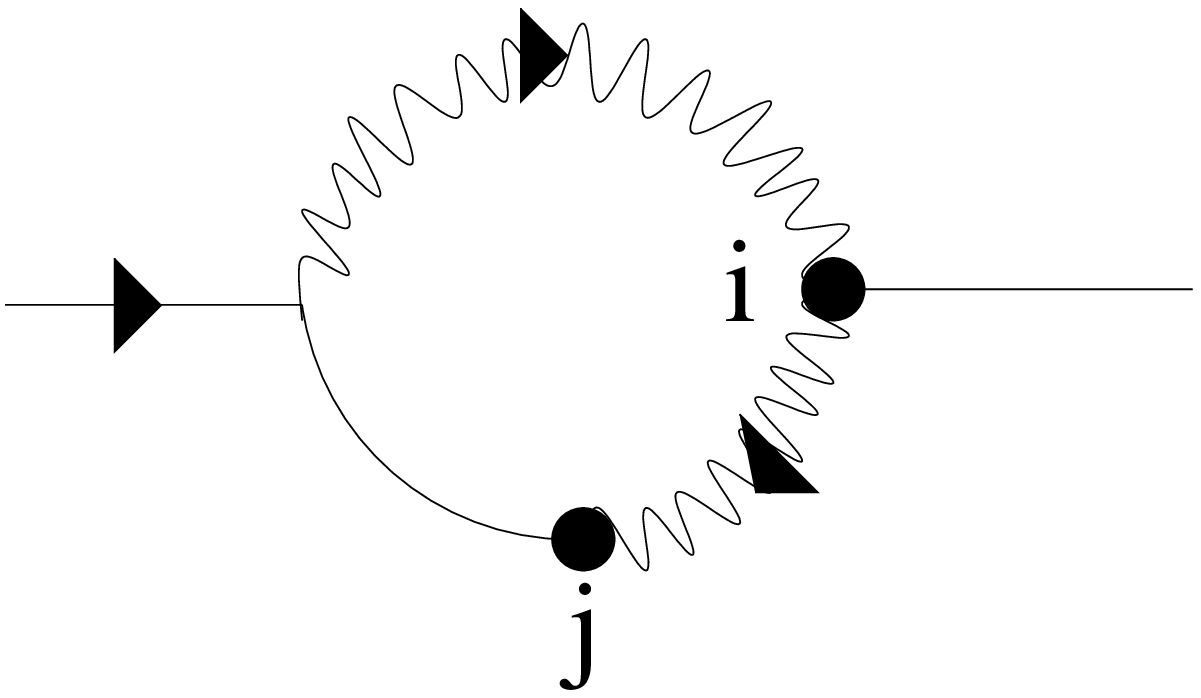}} \end{array}
\right. \cr & \left. \qquad + \begin{array}{l} {\epsfxsize
3cm\epsffile{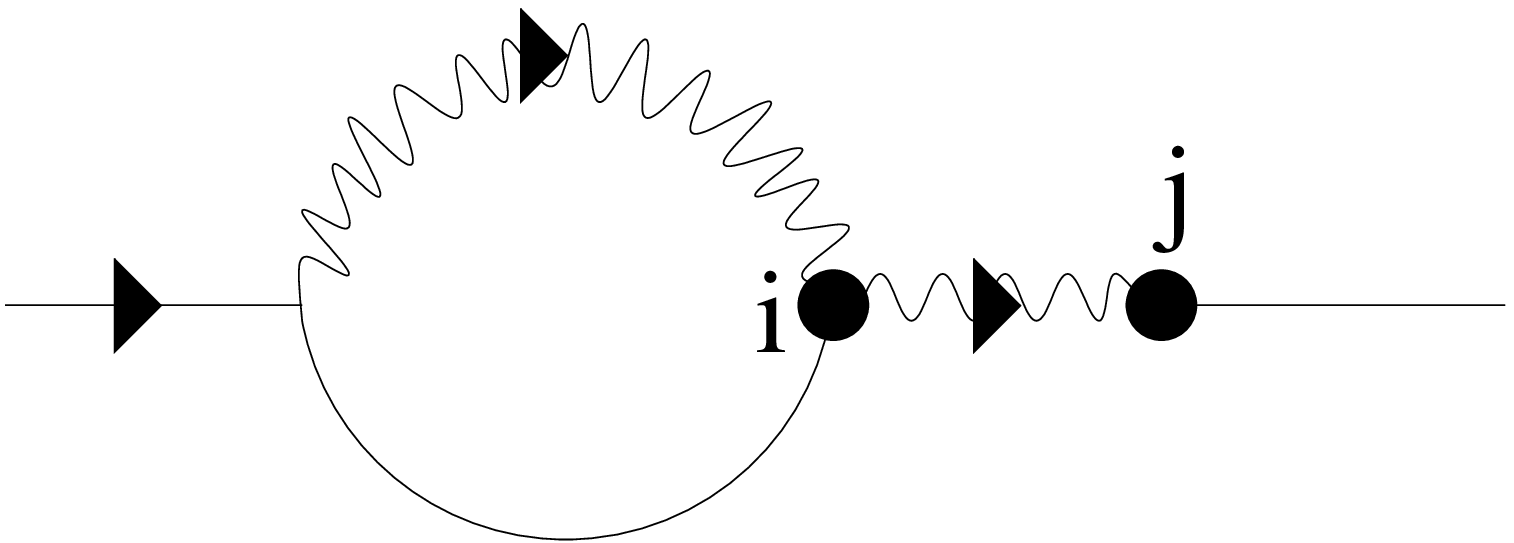}} \end{array}\right] \cr & +
\sum_{i=1}^{d_2} \sum_{j=1}^{d_2}\left[ \begin{array}{l}
{\epsfxsize 3cm\epsffile{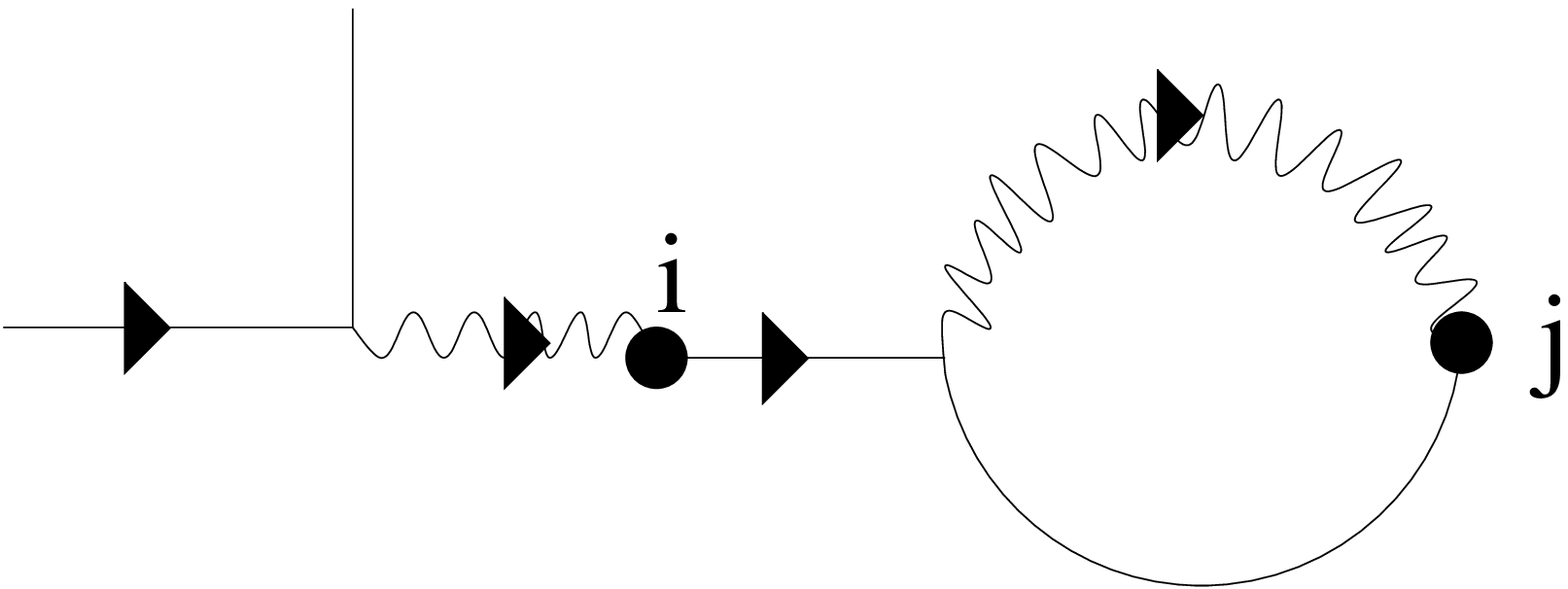}} \end{array} + \begin{array}{l}
{\epsfxsize 3cm\epsffile{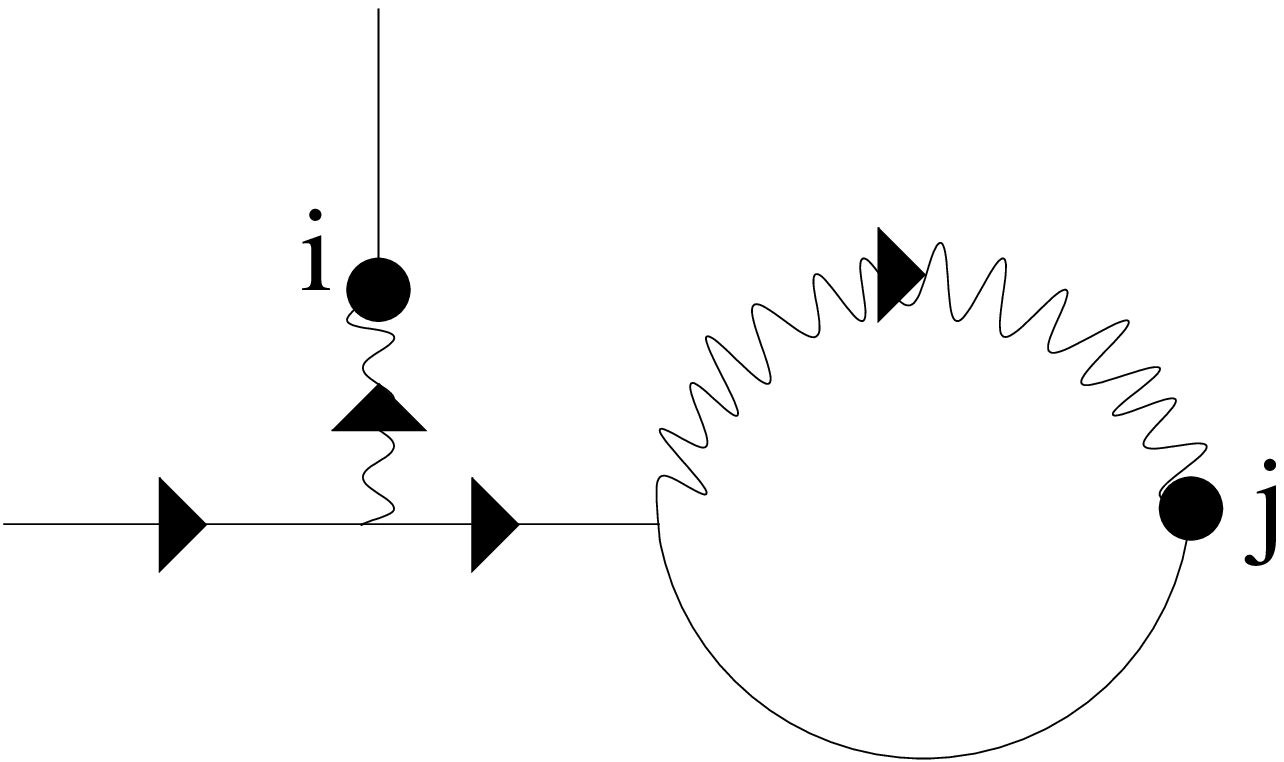}} \end{array}\right. \cr &
\qquad \left. + \begin{array}{l} {\epsfxsize
3cm\epsffile{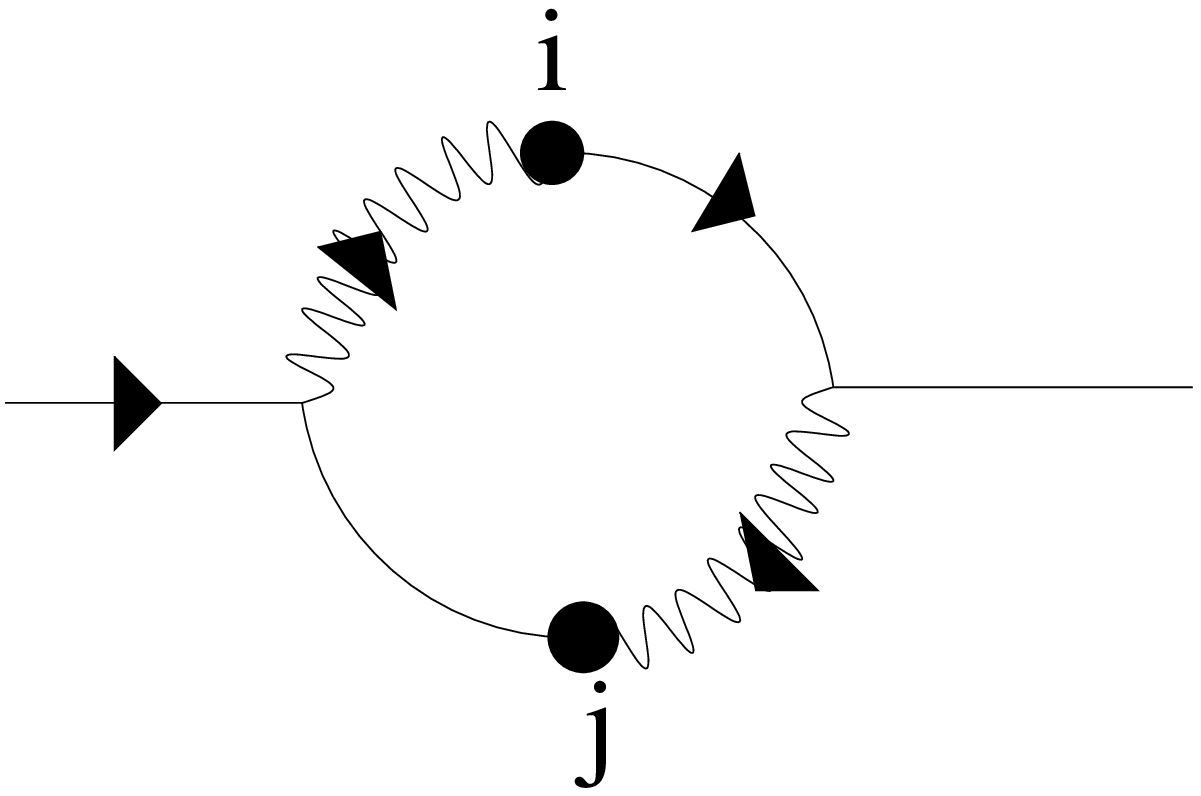}} \end{array} + \begin{array}{l} {\epsfxsize
3cm\epsffile{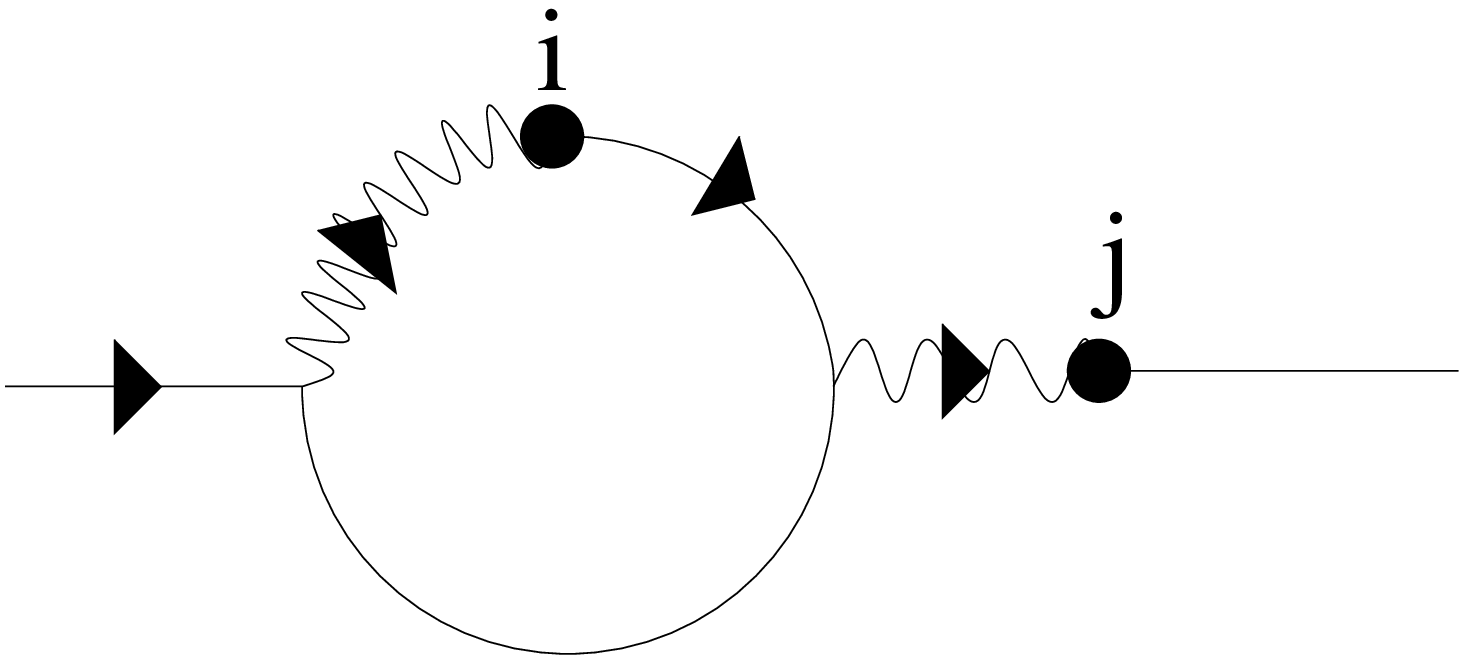}} \end{array}\right] \cr
\end{array}
\eeq

Analytically, this reads: \beq
\begin{array}{l}
W_2^{(1)}(p,p_1) = \cr \sum_{i=1}^{d_2} \sum_{j\in [1,d_2]-\{i\}}
\sum_s \Res_{p' \rightarrow a_s} {dS_{p',o}(p) \over
(y(p'^{i})-y(p')) (y(p'^{i})-y(p'^{j})) dx^2(p')} \cr
 \quad \left[ B(p',p_1) B(p'^{i},p'^{j}) + B(p'^{i},p_1) B(p',p'^{j}) + B(p',p'^{i}) B(p_1,p'^{j}) \right] \cr
 + \sum_{i=1}^{d_2} \sum_{j=1}^{d_2} \sum_{s,t} \Res_{p' \rightarrow a_s} \Res_{p'' \rightarrow a_t} {dS_{p',o}(p) \over (y(p'^{i})-y(p')) (y(p''^{j})-y(p'')) dx(p') dx(p'')} \cr
 \quad \left[ B(p',p_1) B(p'',p''^{j}) dS_{p'',o}(p'^{i}) + B(p'^{i},p_1) B(p'',p''^{j}) dS_{p'',o}(p') \right. \cr
 \qquad \left. + B(p'',p') B(p_1,p''^{j}) dS_{p'',o}(p'^{i}) + B(p_1,p'') B(p',p''^{j}) dS_{p'',o}(p'^{i}) \right] \cr
\end{array}
\eeq

$\bullet$ $k=1$ and $h=2$:

\bea W_{1}^{(2)} = & \sum_{i=1}^{d_2} \sum_{j=1}^{d_2}
\sum_{k=1}^{d_2} \left[ \begin{array}{l} {\epsfxsize
3cm\epsffile{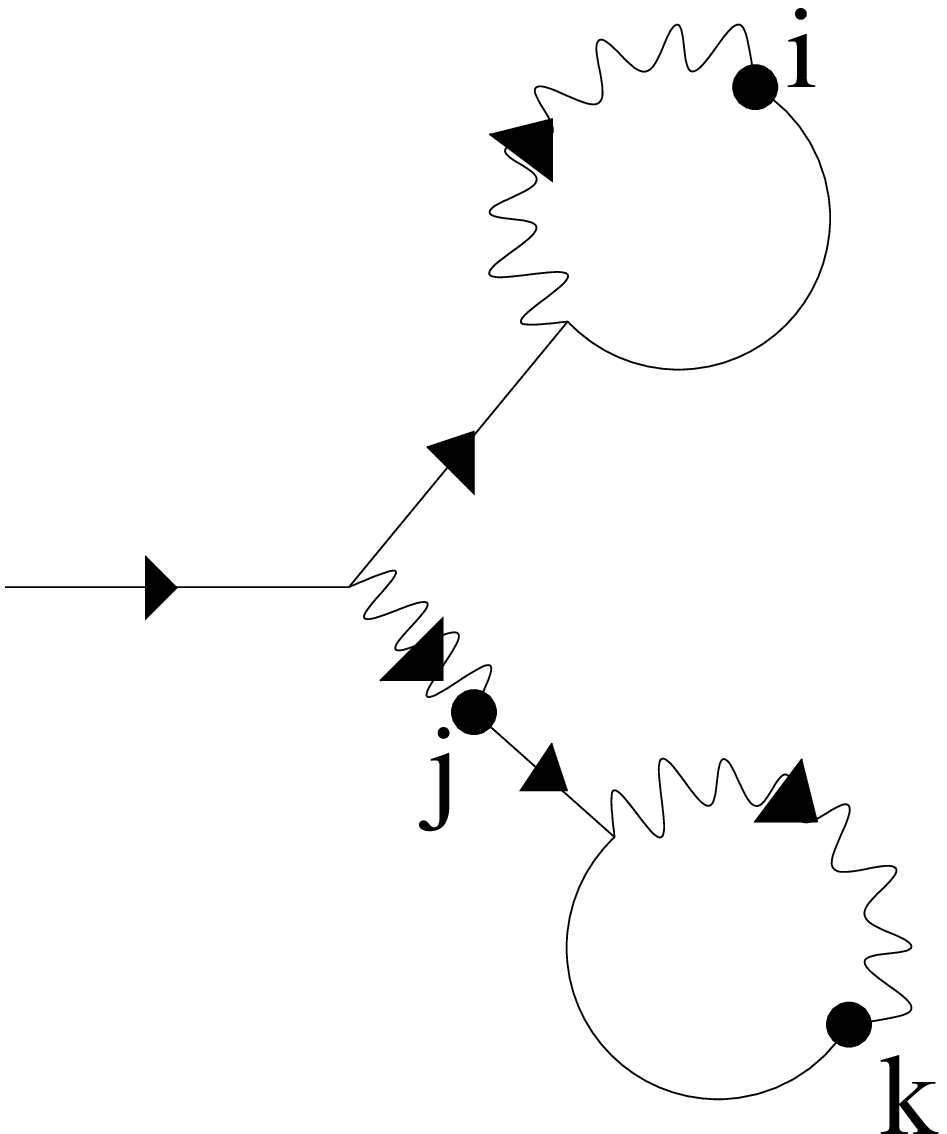}} \end{array} + \begin{array}{l} {\epsfxsize
3cm\epsffile{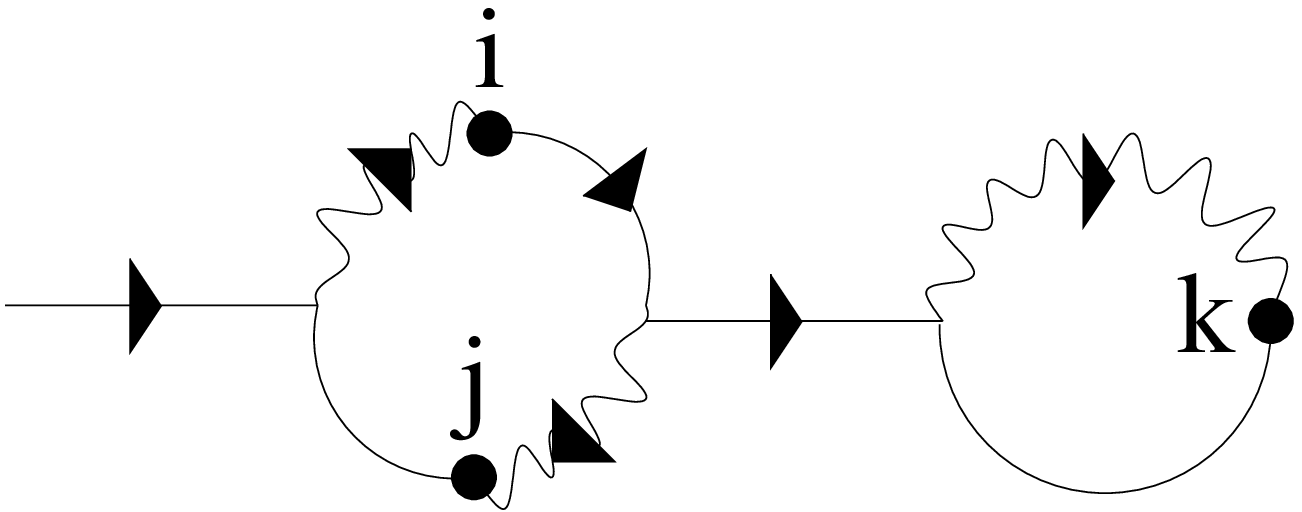}} \end{array} \right. \cr & \qquad \left. +
\begin{array}{l} {\epsfxsize 3cm\epsffile{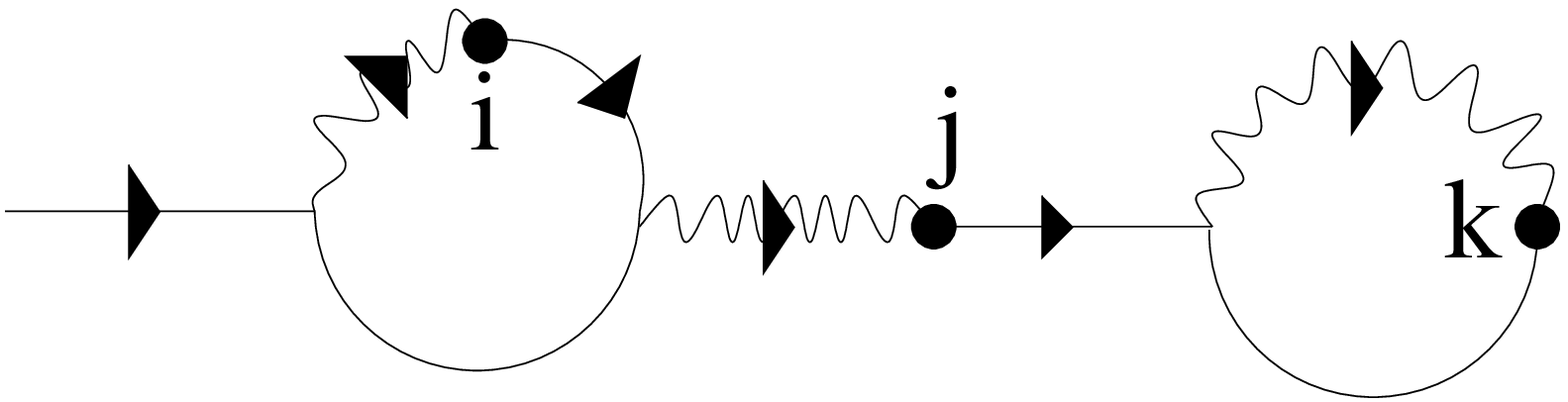}} \end{array} +
\begin{array}{l} {\epsfxsize 3cm\epsffile{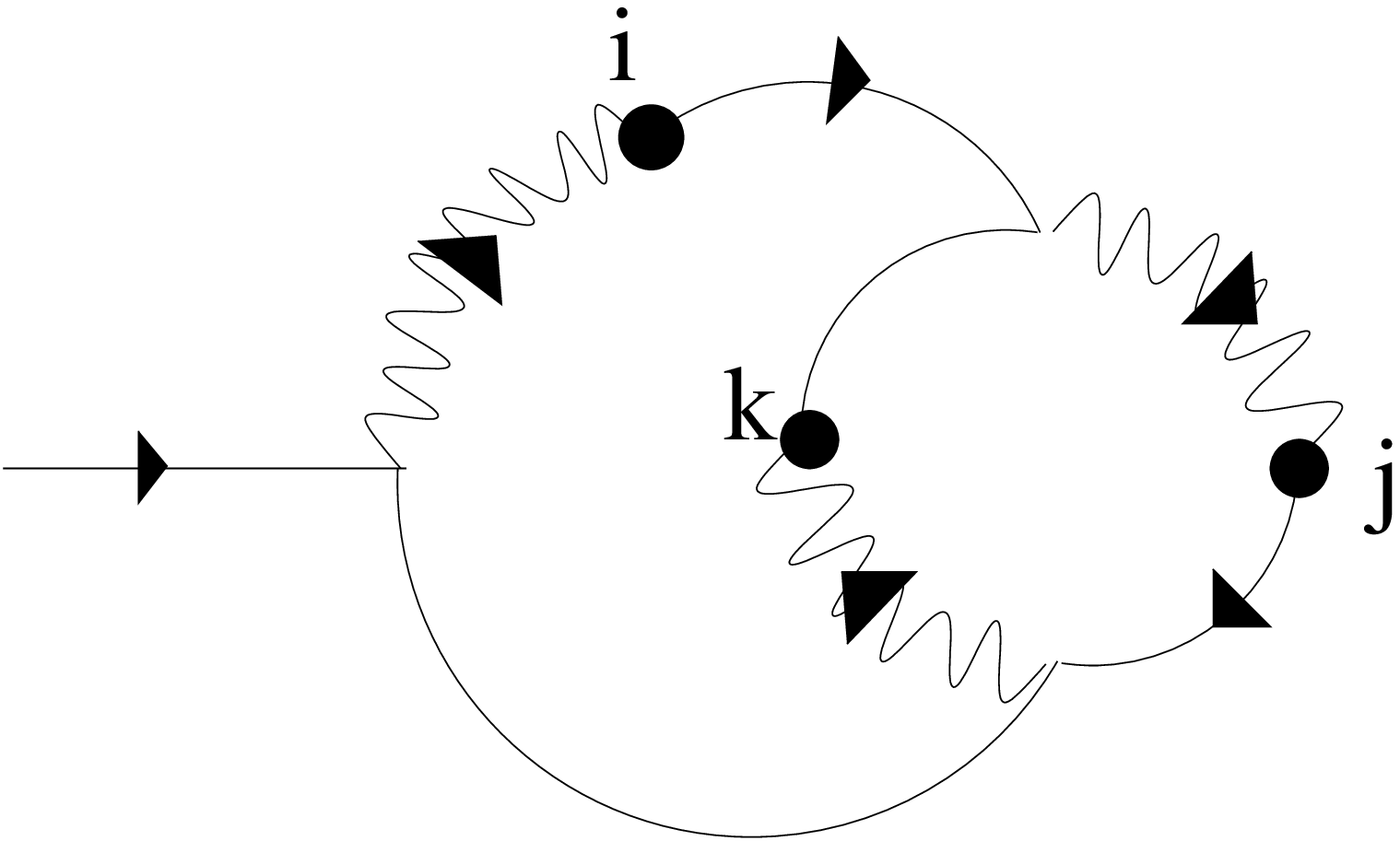}} \end{array}
+ \begin{array}{l} {\epsfxsize 3cm\epsffile{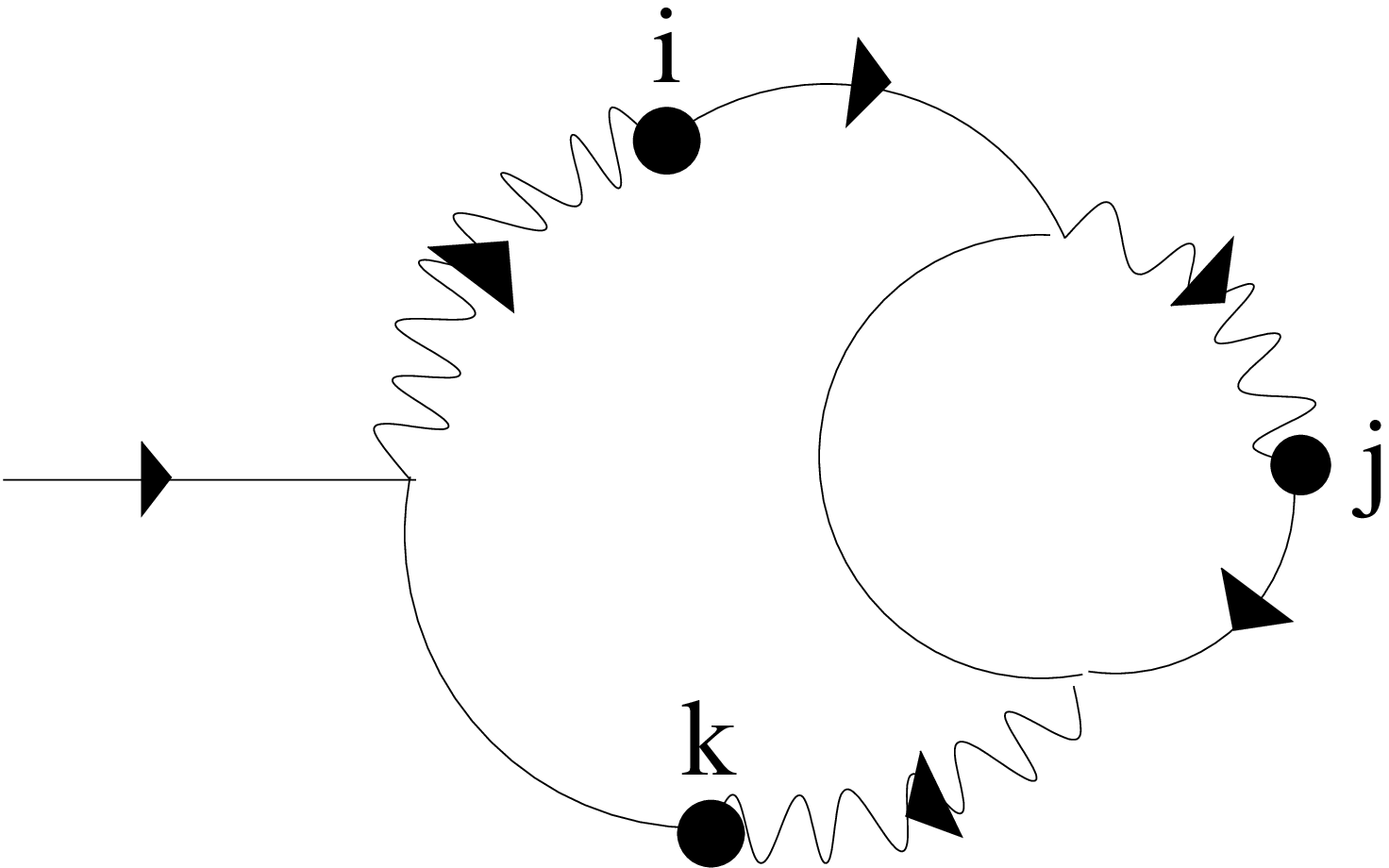}}
\end{array} \right] \cr & + \sum_{i=1}^{d_2} \sum_{j\in
[1,d_2]-\{i\}} \sum_{k=1}^{d_2} \left[ \begin{array}{l}
{\epsfxsize 3cm\epsffile{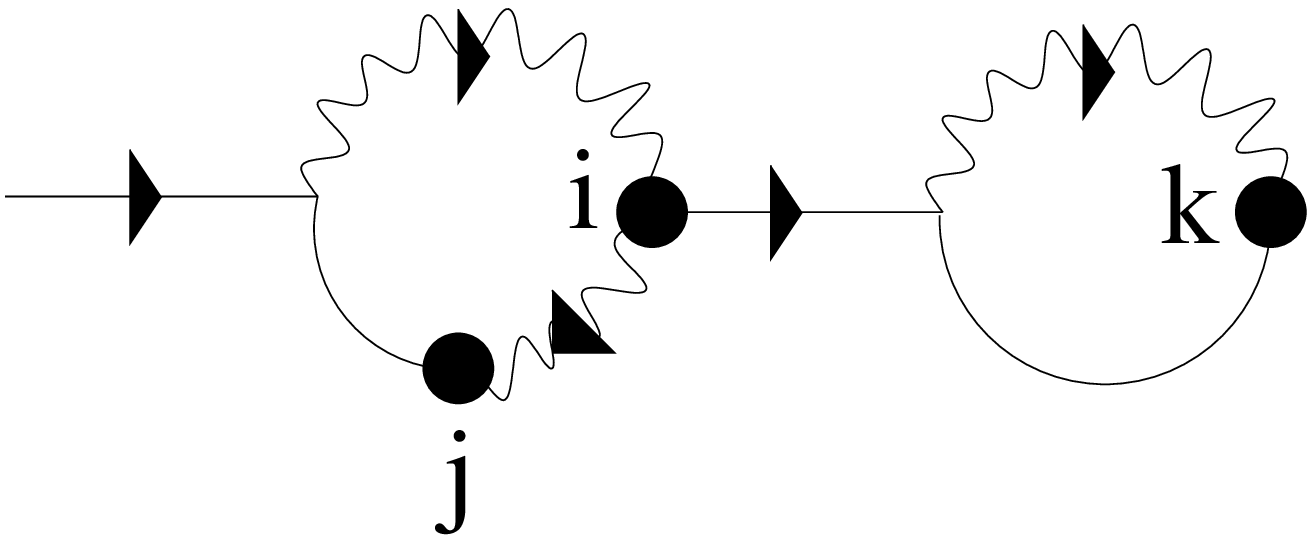}} \end{array} + \begin{array}{l}
{\epsfxsize 3cm\epsffile{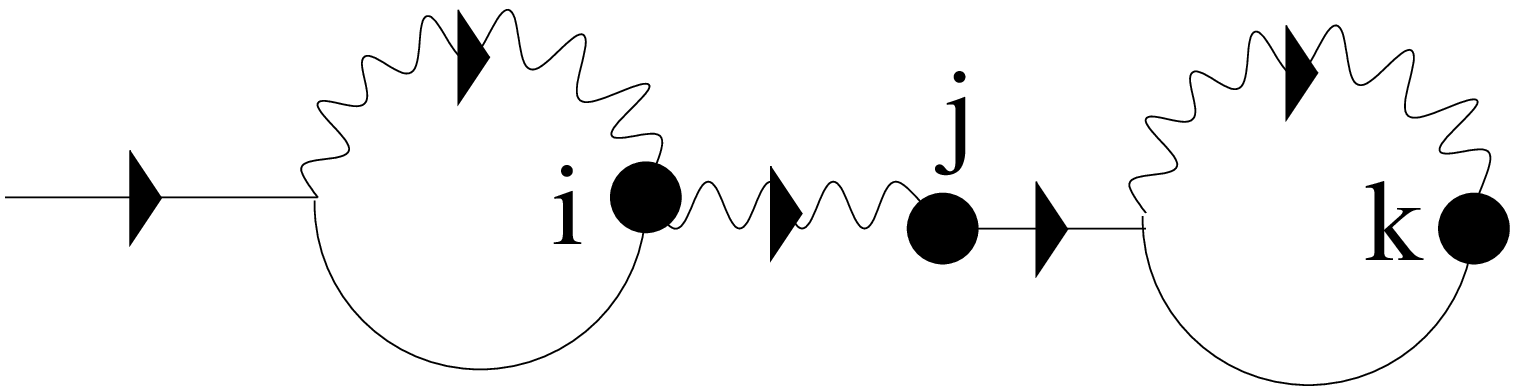}} \end{array} \right. \cr &
\qquad \left. + \begin{array}{l} {\epsfxsize
3cm\epsffile{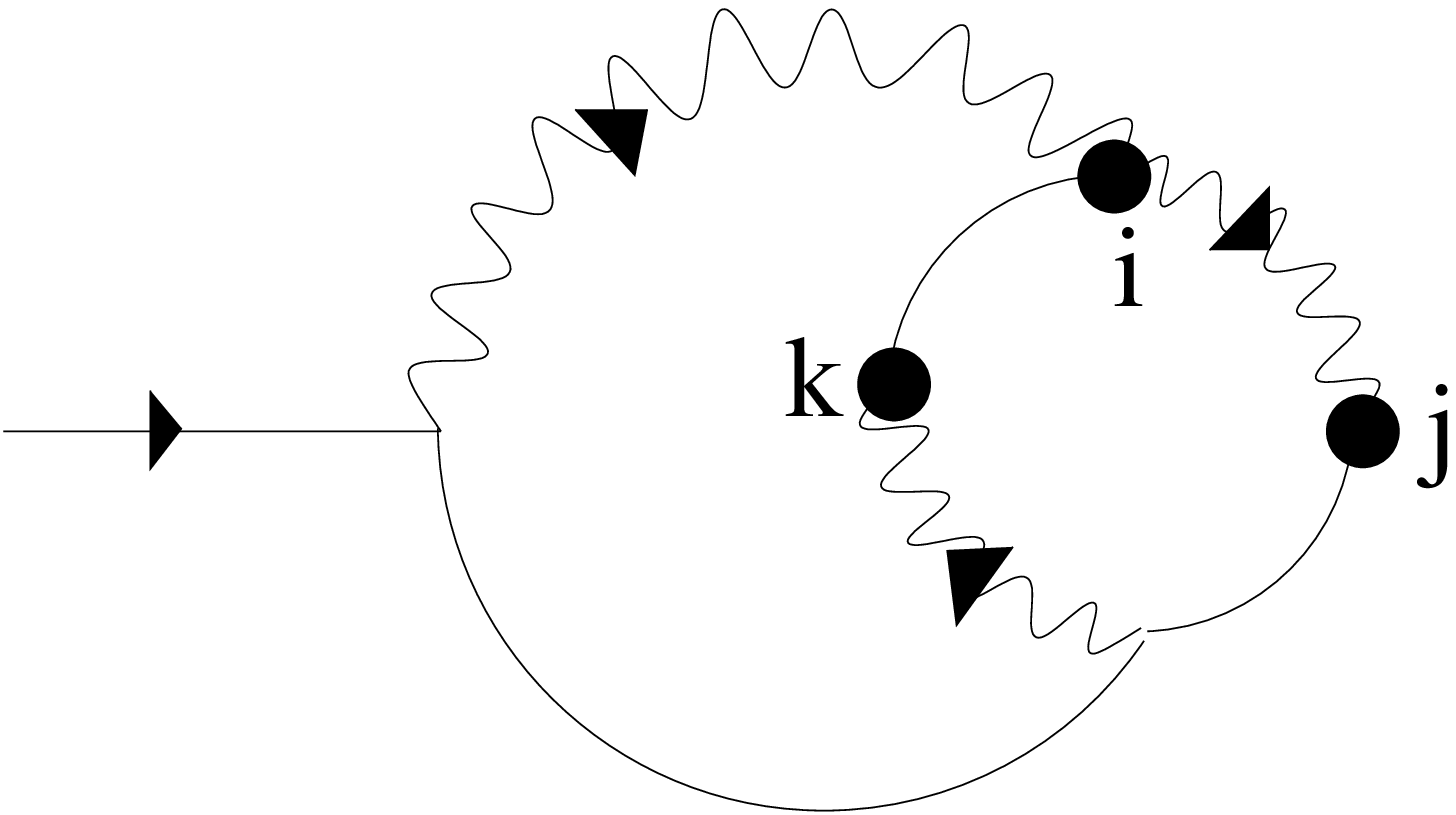}} \end{array} + \begin{array}{l} {\epsfxsize
3cm\epsffile{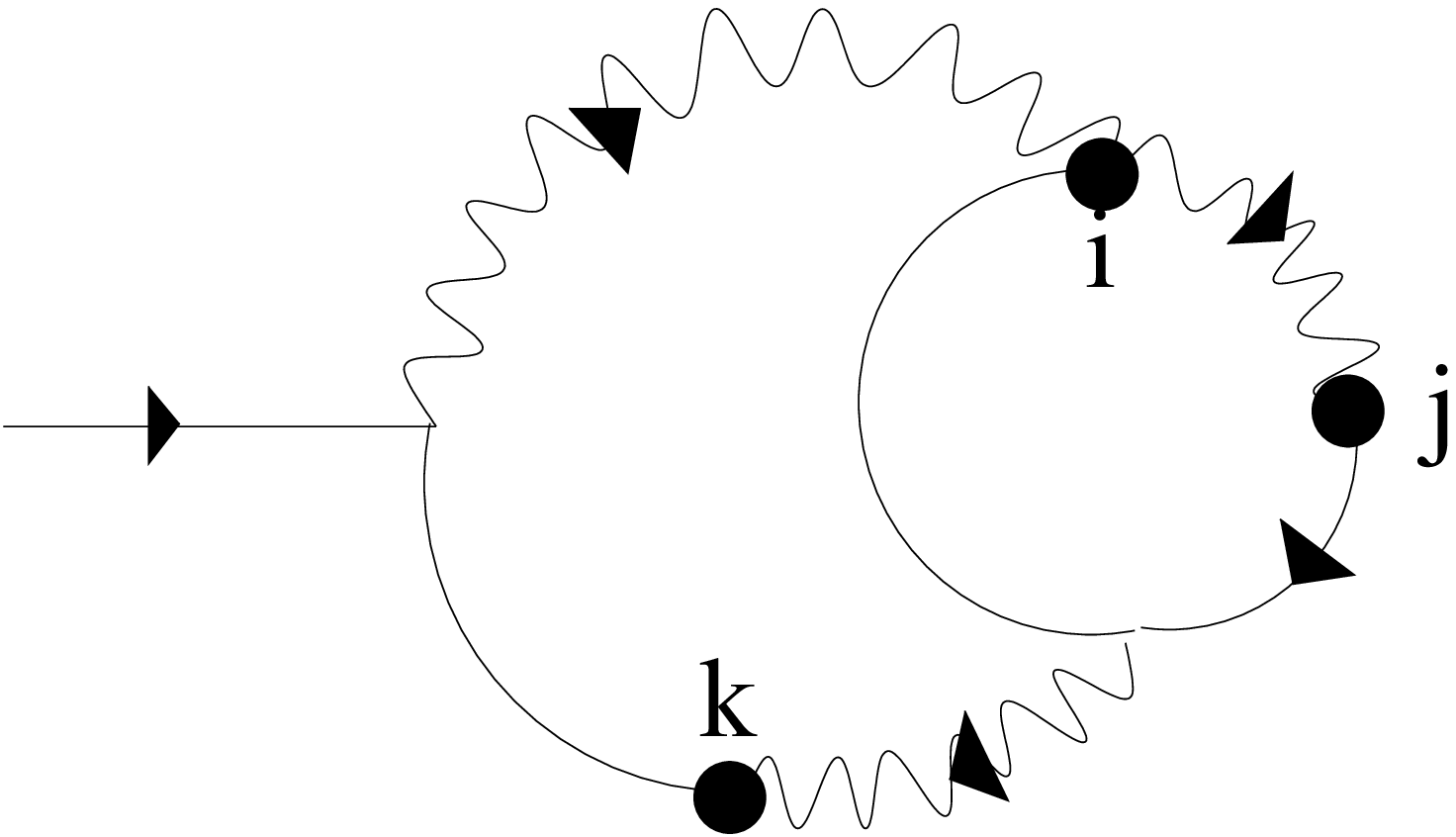}} \end{array}\right] \cr & +
\sum_{i=1}^{d_2} \sum_{j=1}^{d_2} \sum_{k\in [1,d_2]-\{j\}} \left[
\begin{array}{l} {\epsfxsize 3cm\epsffile{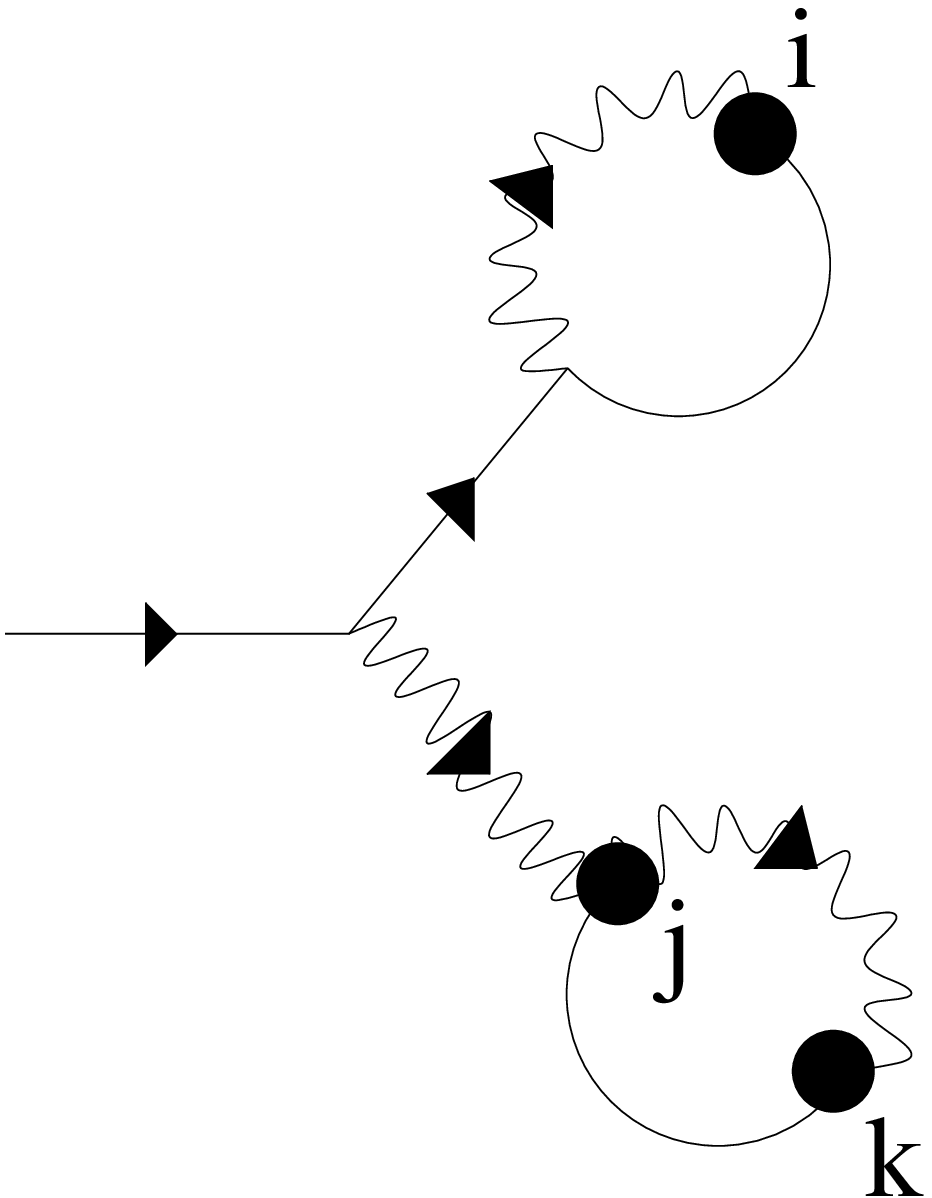}} \end{array} +
\begin{array}{l} {\epsfxsize 3cm\epsffile{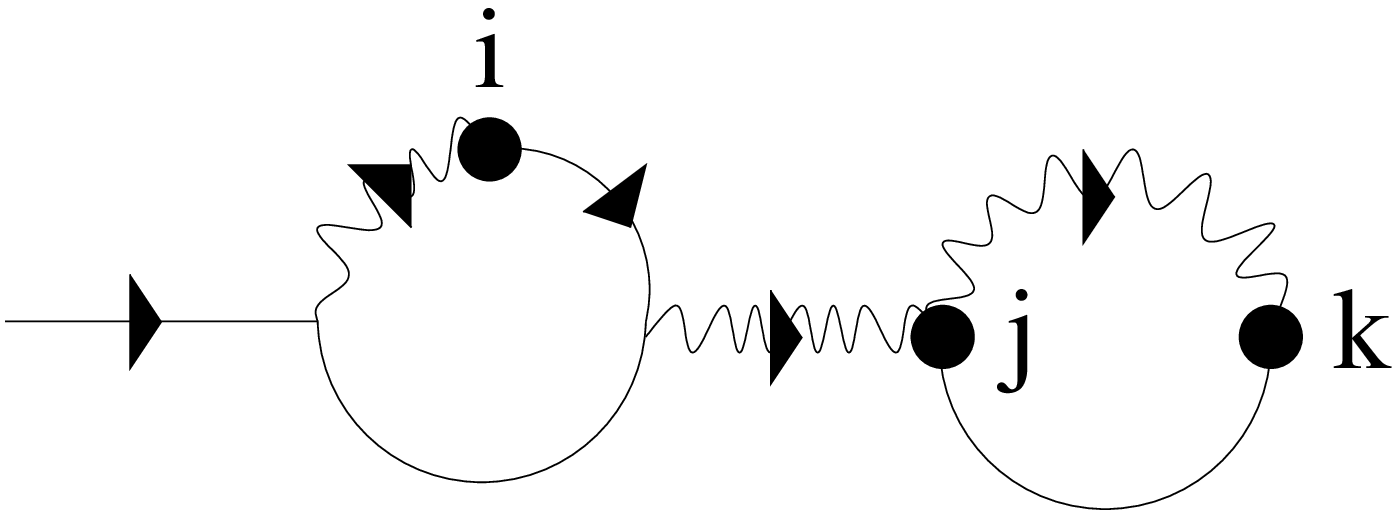}}
\end{array}\right. \cr & \qquad \left. + \begin{array}{l}
{\epsfxsize 3cm\epsffile{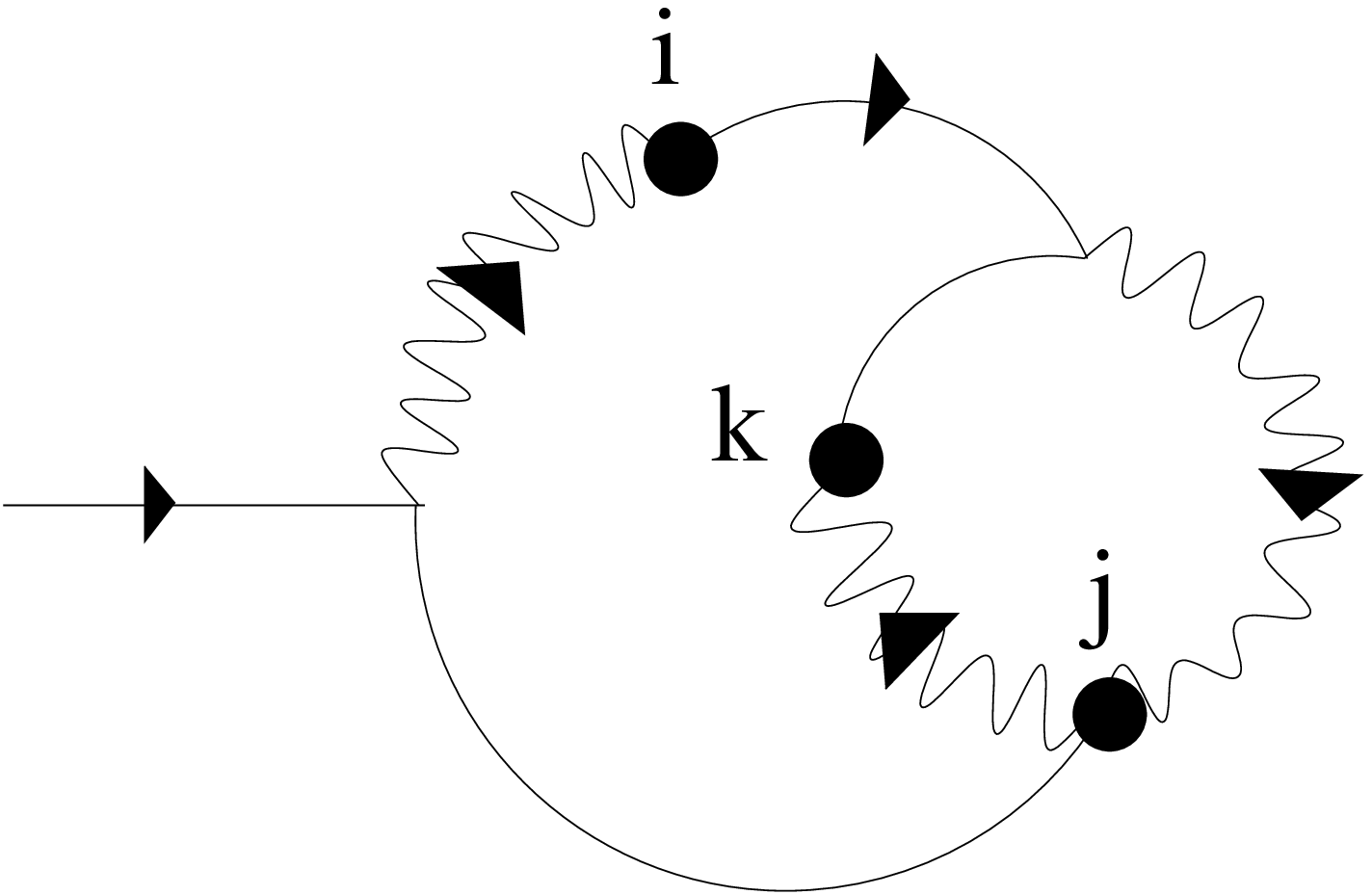}} \end{array} +
\begin{array}{l} {\epsfxsize 3cm\epsffile{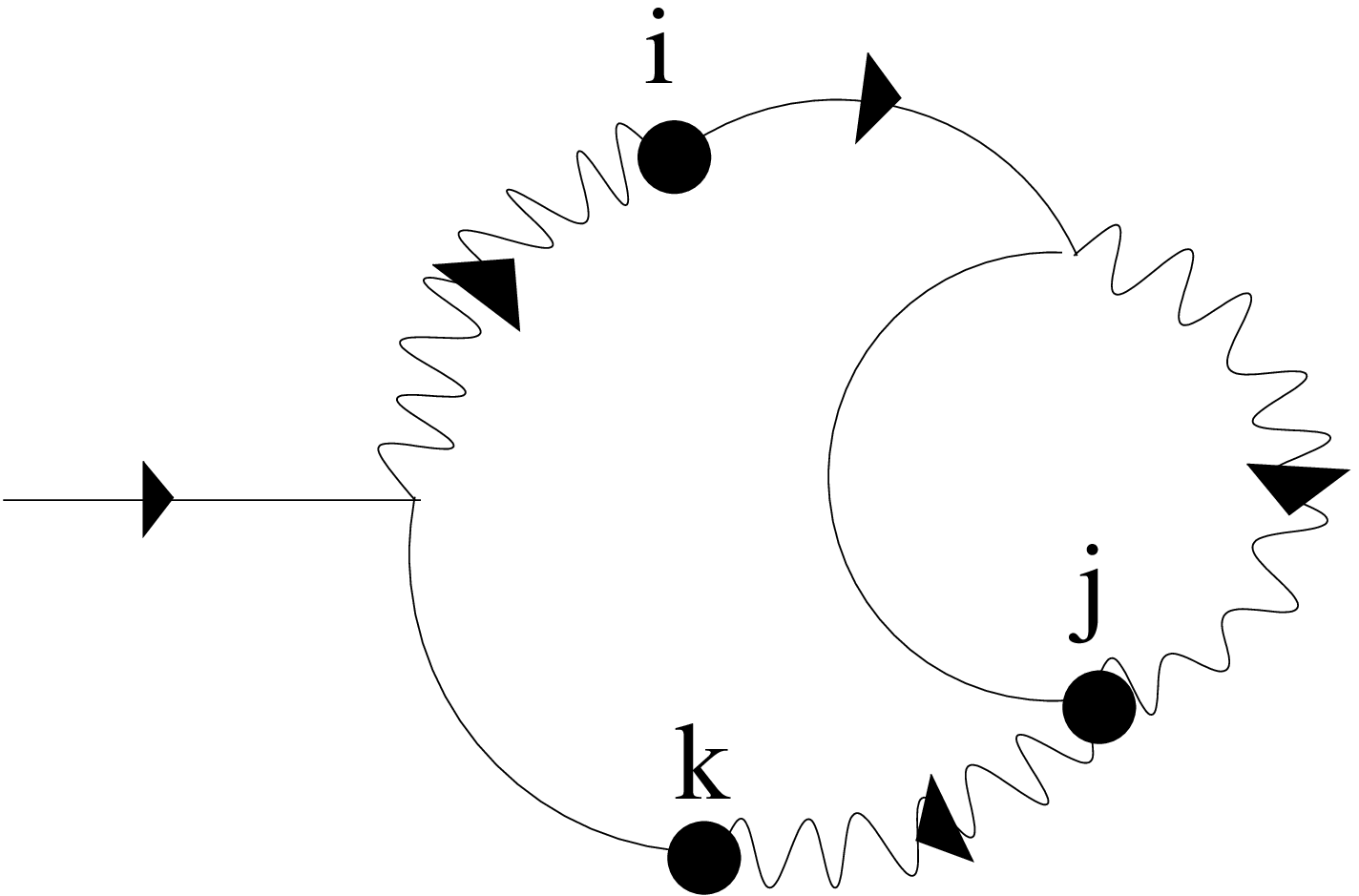}}
\end{array}\right] \cr & + \sum_{i=1}^{d_2} \sum_{j\in
[1,d_2]-\{i\}} \sum_{k\in [1,d_2]-\{j\}} \left[ \begin{array}{l}
{\epsfxsize 3cm\epsffile{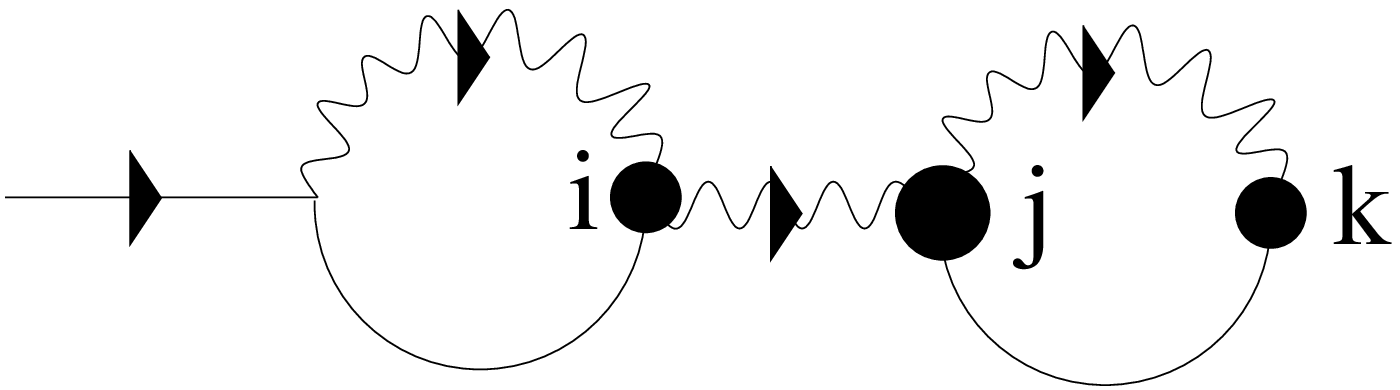}} \end{array} \right. \cr &
\qquad \left. + \begin{array}{l} {\epsfxsize
3cm\epsffile{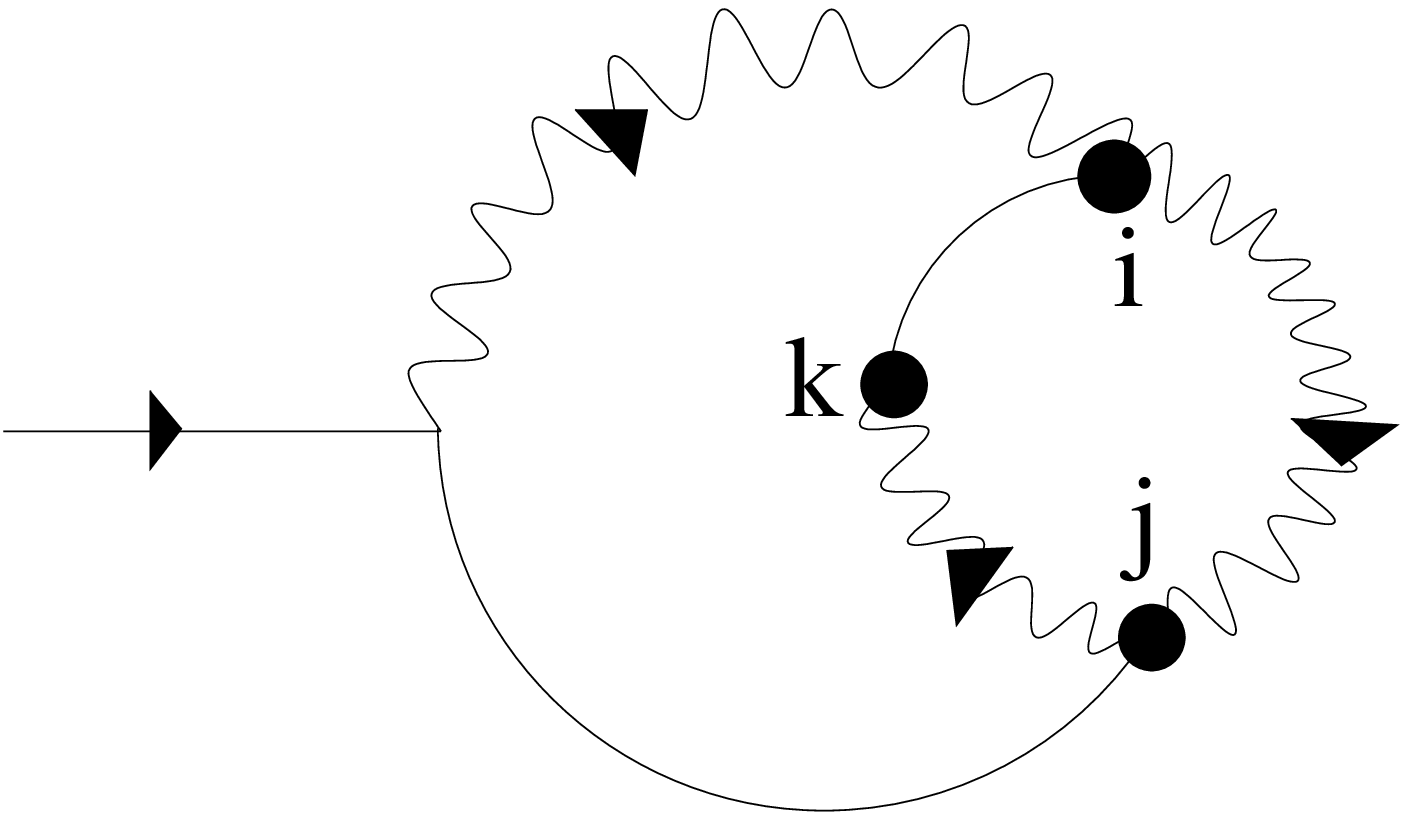}} \end{array} + \begin{array}{l}
{\epsfxsize 3cm\epsffile{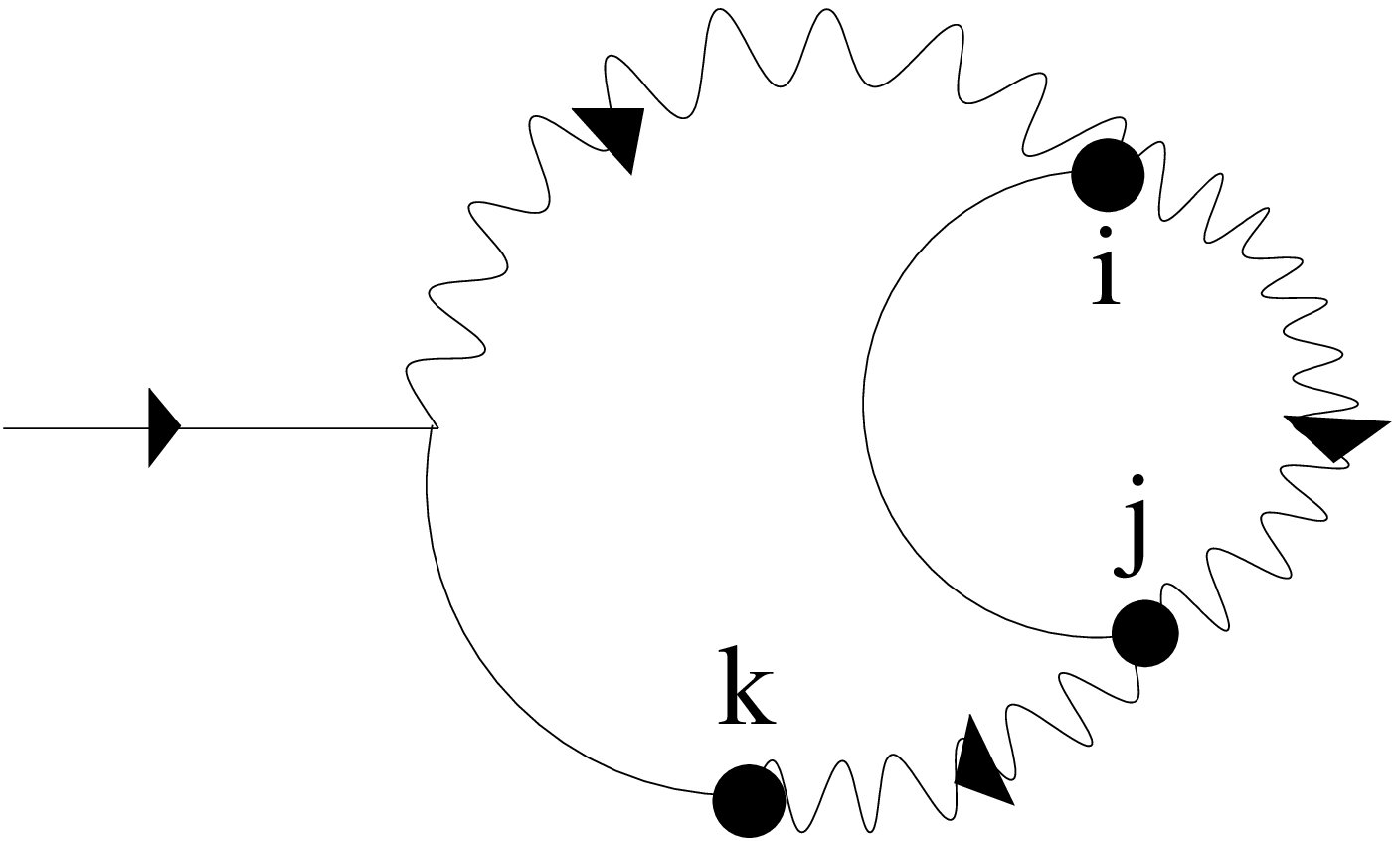}} \end{array} \right] \cr \eea

\newsection{An effective non cubic theory}
The Feynman-like graphs described up to now correspond to cubic
vertices only, but the price to pay is the introduction of auxiliary functions $R_k^{i,(h)}$.
Nevertheless, in order to study some problems, this property is not needed
and one may prefer an effective diagrammatic representation for only $W_k^{(h)}$
but vertices with valence up to $d_2-1$. This section is dedicated to
building such a diagrammatic representation. It
consists in resumming the linked waved vertices into one
multivalent vertex: \beq
\begin{array}{r}
{\epsfxsize 4cm\epsffile{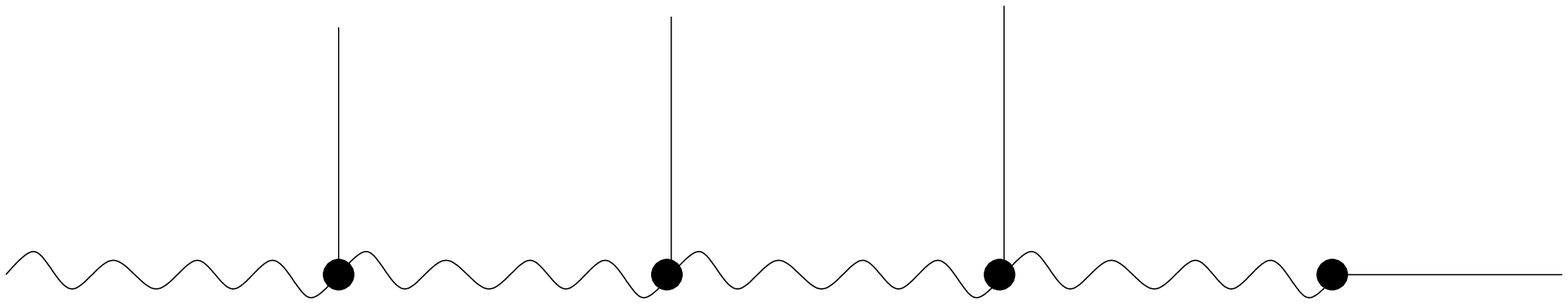}}
\end{array}
\sim
\begin{array}{r}
{\epsfxsize 2.5cm\epsffile{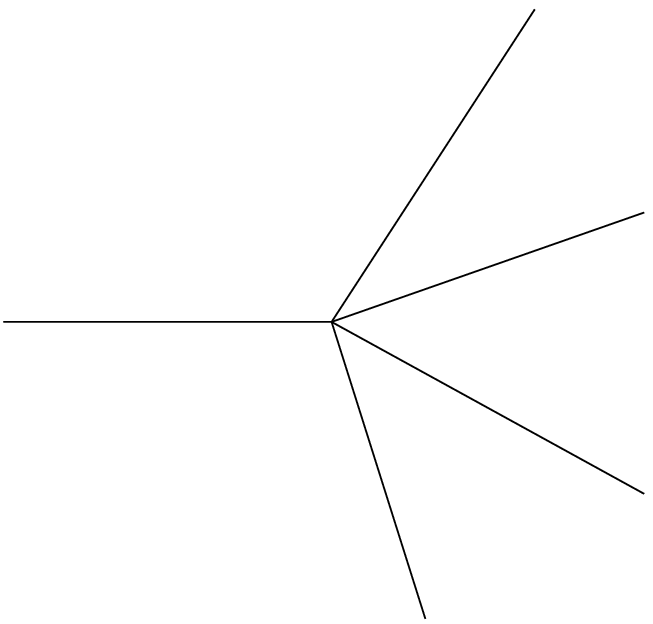}}
\end{array}
\eeq

\subsection{Leading order: Genus 0}
We have already written the equations necessary to
define this effective theory. Let us consider \eq{recW} and
\eq{UWW}: \beq\label{recWeff} W_{k+1}(p,{\bfp}_K) = -\sum_s
\Res_{p'\to a_s} \sum_{j=1}^{k-1} \sum_{J\in K_j} \frac{1}{dx'}
{{U}_{j}(p',y(p');\bfp_J)\over  E_y(x(p'),y(p')) }\,
{W}_{k-j+1}(p',\bfp_{K-J}) dS_{p',\a}(p) \eeq

\beq U_{k}(p,y;\bfp_K) = {E(x(p),y) dx(p)\over
y-y(p)}\sum_{r=1}^{d_2} \sum_{K_1\cup\dots\cup K_r=K}
\sum_{j_1\neq j_2\neq \dots \neq j_r=1}^{d_2} \prod_{t=1}^r
{W_{|K_t|+1}(p^{j_t},\bfp_{K_t})\over (y-y(p^{j_t}))\,dx(p)}
\eeq

This second equation taken for $y=y(p)$ reads:
\beq
{U_{k}(p,y(p);\bfp_K) \over E_y(x(p),y(p)) dx(p)} =
\sum_{r=1}^{d_2} \sum_{K_1\cup\dots\cup K_r=K} \sum_{j_1\neq
j_2\neq \dots \neq j_r=1}^{d_2} \prod_{t=1}^r
{W_{|K_t|+1}(p^{j_t},\bfp_{K_t})\over
(y(p)-y(p^{j_t}))\,dx(p)}
\eeq

Introduce it in \eq{recWeff} and get a closed recursive formula
for the $W_k$'s: \beq\encadremath{\label{recWeff2}
\begin{array}{rcl}
W_{k+1}(p,{\bfp}_K) &=& - \sum_s \Res_{p'\to a_s} \sum_{r=1}^{d_2}
\sum_{K_0\cup K_1\cup\dots\cup K_r=K} \sum_{j_1\neq j_2\neq \dots
\neq j_r=1}^{d_2}  \cr && {W}_{|K_0|+1}(p',\bfp_{K_0})
\prod_{t=1}^r {W_{|K_t|+1}(p'^{j_t},\bfp_{K_t})\over
(y(p')-y(p'^{j_t}))\,dx(p')} dS_{p',\a}(p)
\end{array}
}\eeq

Let us introduce the following Feynman rules:
\begin{center}
\begin{tabular}{|r|l|}\hline
non-arrowed propagator:& $
\begin{array}{r}
{\epsfxsize 2.5cm\epsffile{W2.eps}}
\end{array}
:=W_{2}(p,q) $ \cr\hline
arrowed propagator:& $
\begin{array}{r}
{\epsfxsize 2.5cm\epsffile{arrowedpropagator.eps}}
\end{array}
 :=dS_{q,o}(p)
$ \cr\hline
\begin{tabular}{c}
r+2 - vertex\cr ($1\leq r \leq d_2$)\cr with one marked \cr edge:
\end{tabular}&
$
\begin{array}{r}
{\epsfxsize 3.8cm\epsffile{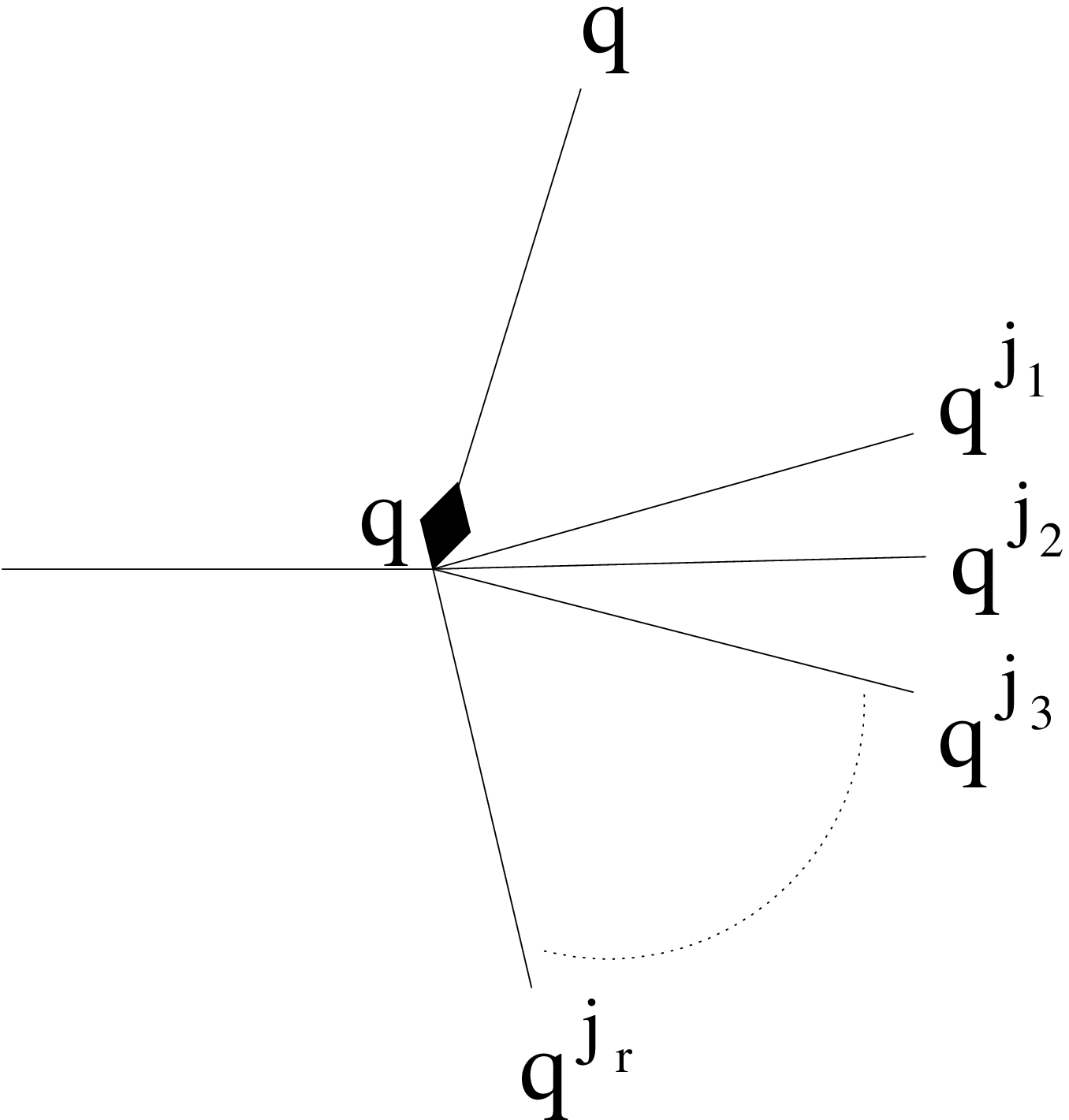}}
\end{array}
 := \begin{array}{l}
 - \sum_s \sum_{j_1\neq \dots \neq j_r \neq 0} \Res_{q\rightarrow a_s} \cr
 \prod_{t=1}^r {1 \over (y(q) - y(q^{j_t})) dx(q)}\cr
 \end{array}
$ \cr
\hline
\end{tabular}
\end{center}

Remark that one leg of the multiple vertex is marked: on this leg,
there is no summation over the different sheets.
\bigskip

Using these rules, one can diagrammatically write the recursive
relation as follows: \beq
\begin{array}{r}
{\epsfxsize 4.5cm\epsffile{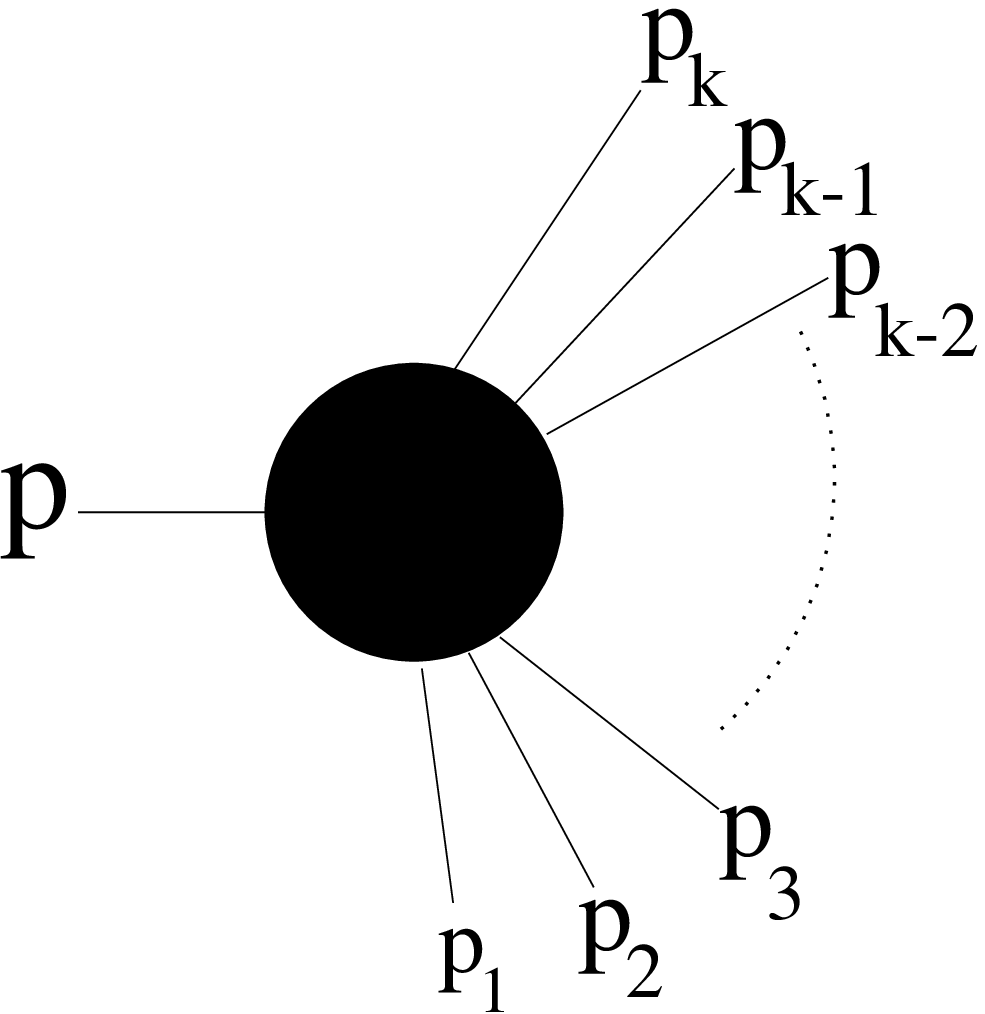}}
\end{array}
= \sum_{r=1}^{d_2} \sum_{K_0\cup K_1\cup\dots\cup K_r=K}
\begin{array}{l}
{\epsfxsize 5.5cm\epsffile{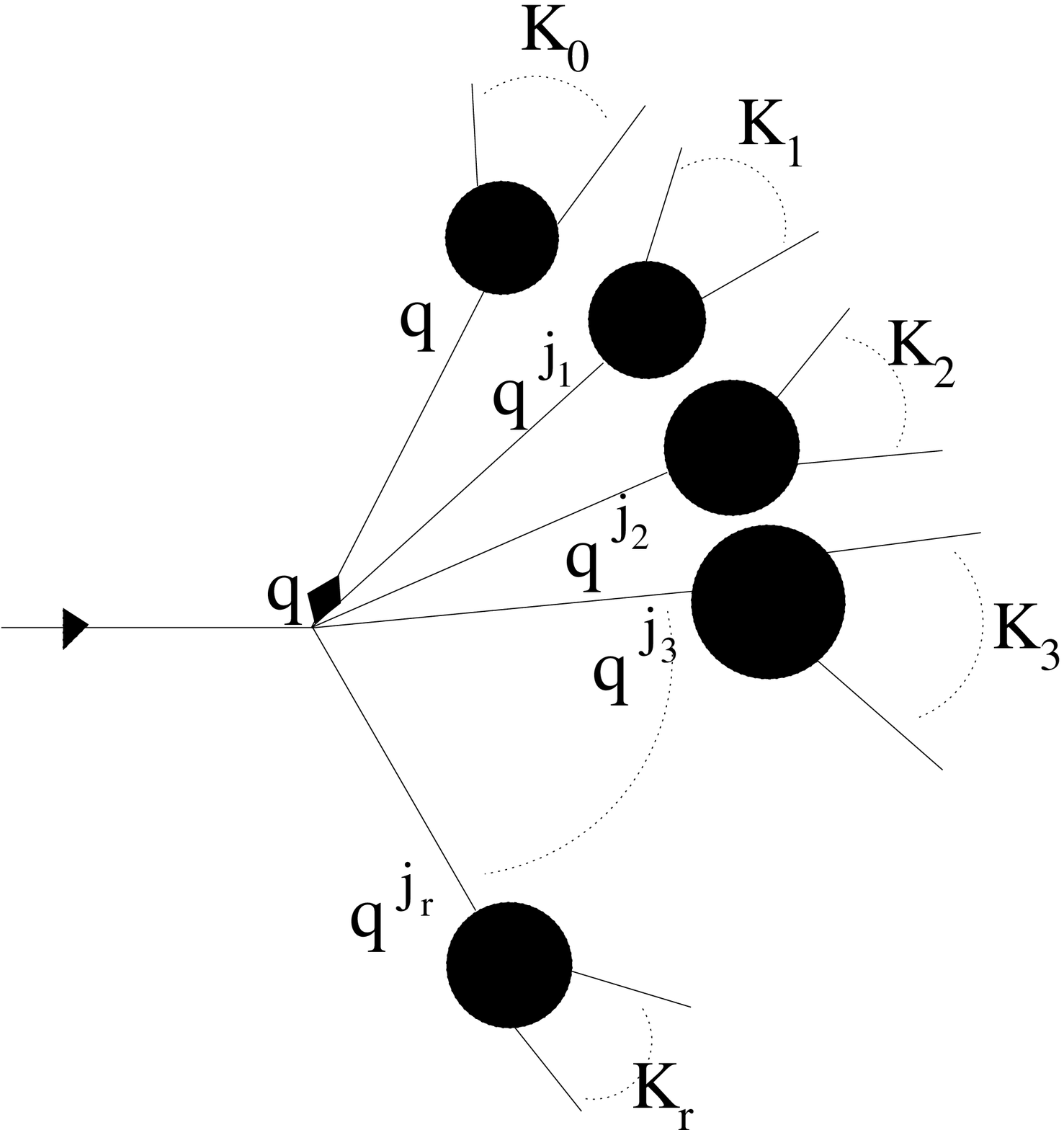}}
\end{array}
\eeq

From this relation, one can see that $W_{k+1}(p,{\bfp}_K)$ is
obtained as the {\em summation over all trees with $k+1$ external
legs} and following the rules: {\em \begin{itemize} \item The
vertices have valence r+2 such as $1 \leq r \leq min(k-1,d_2)$;
\item The edges are arrowed; \item One of the legs of each vertex
is marked \item The k leaves are non arrowed propagators
ending at $p_j$'s;
\item The root is an arrowed propagator starting from $p$.
\end{itemize}}

The drawbacks of these effective rules induced by the existence of
multivalent vertices is balanced by the simplicity of the
vertices and the absence of different propagators.

\subsection{Any genus h}
Let us now study the extension of this theory to any genus.

Once again, the fundamental equations have already been written.
Let us recall to mind \eq{soluce1} and
\eq{eqUW}: \beq\label{Wkh3}
\begin{array}{l}
W_{k+1}^{(h)}(p,\bfp_K) = \cr
 - \sum_{m=0}^h \sum_{j=0, m+j \neq 0}^{k}  \sum_s \Res_{p'\to a_s} {{U}_{j}^{(m)}(p',y(p');\bfp_J) \over E_y(x(p'),y(p'))} {W}_{k-j+1}^{(h-m)}(p',\bfp_{K-J}) dS_{p',\alpha}(p)\cr
 - \sum_s \Res_{p'\to a_s} {{U}_{k+1}^{(h-1)}(p',y(p');p',\bfp_K) \over E_y(x(p'),y(p'))} dS_{p',\alpha}(p)
\end{array}
\eeq

and, for $i\neq 0$:

\beq\label{eqUW3}
\begin{array}{l}
U_{k}^{(h)}(p,y(p^{i});\bfp_K)= \cr {E_y(x,y(p^{i})) \over
y(p^{i})-y(p)} \sum_{r=1}^{min(d_2,k+h)} \sum_{ K_1 \bigcup
\dots \bigcup K_r = K}  \sum_{h_{\alpha} = 0}^h
\sum_{k_\alpha=|K_\alpha|}^{k+h} \sum_{j_{\alpha,\beta} \neq
j_{\alpha',\beta'} \in [1,d_2]-\{i\}} {1 \over \Omega} \cr
{W_{k_1+1}^{(h_1)}(p^{i}, \bfp_{K_1} ,p^{j_{1,1}}, \dots
,p^{(j_{1,k_1-|K_1|})}) \left(\prod_{\alpha=2}^{r}
W_{k_{\alpha}+1}^{(h_{\alpha})}(p^{j_{\alpha,0}},\bfp_{K_{\alpha}}
,p^{j_{\alpha,1}}, \dots
,p^{j_{\alpha,k_\alpha-|K_\alpha|}})\right) \over
dx(p)^{r-k-1+\sum k_\alpha} \prod_{\alpha,\beta}
y(p^{i})-y(p^{j_{\alpha,\beta})}}\cr
\end{array}
\eeq

In order to introduce this second formula inside the first one,
one has to use the interpolation formula to consider the case
where $i = 0$ : \beq
\begin{array}{l}
{U_{l}^{(m)}(p,y(p);\bfp_L) \over E_y(x(p),y(p))} = \cr
 - \sum_{r=1}^{min(d_2,l+m)} \sum_{ L_1 \bigcup \dots \bigcup L_r = L} \sum_{m_{\alpha} = 0}^m \sum_{l_{\alpha}=|L_\alpha|}^{l+m}
\sum_{j_1 \neq \dots \neq j_r \in [1,d_2]} {1 \over \Omega} \cr
{W_{l_1+1}^{(m_1)}(p^{j_{1,0}}, \bfp_{L_1} ,p^{j_{1,1}}, \dots
,p^{j_{1,l_1-|L_1|}})  \prod_{\alpha=2}^{r}
W_{l_{\alpha}+1}^{(m_{\alpha})}(p^{j_{\alpha,0}},\bfp_{L_{\alpha}}
,p^{j_{\alpha,1}}, \dots ,p^{j_{\alpha,l_\alpha-|L_\alpha|}})
\over dx(p)^{r-l-1+\sum l_\alpha} (y(p^{j_{1,0}})-y(p))
\prod_{\alpha,\beta}
(y(p^{j_{1,0}})-y(p^{j_{\alpha,\beta})}}\cr
\end{array}
\eeq

Recursively, it is easy to check that it can be written:
\bea\label{Ulm} &&{U_{l}^{(m)}(p,y(p);\bfp_L) \over E_y(x(p),y(p))
dx(p)} = \cr &&\sum_{r=1}^{min(d_2,l+m)} \sum_{ L_1 \bigcup \dots
\bigcup L_r = L} \sum_{m_{\alpha} = 0}^m
\sum_{l_{\alpha}=|L_\alpha|}^{l+m} \sum_{j_{\alpha,\beta} \neq
j_{\alpha',\beta'} \in [1,d_2]} {1 \over \Omega '}  \cr &&\quad
\prod_{\alpha=1}^{r}
{W_{l_{\alpha}+1}^{(m_{\alpha})}(p^{j_{\alpha,0}},\bfp_{L_{\alpha}}
,p^{j_{\alpha,1}}, \dots ,p^{j_{\alpha,l_\alpha-|L_\alpha|}})
\over  dx(p)^{l_\alpha-|L_\alpha|+1} \prod_{\beta =
0}^{l_\alpha-|L_\alpha|}   (y(p)-y(p^{j_{\alpha,\beta})})}\cr
\eea

where $\Omega '$ is some other symmetry factor depending only on
the same parameters as $\Omega$.

One is now able to write an explicit recursion formula for the
$W_k^{(h)}$'s that can be graphically represented with the Feynman
rules introduced in this section. The introduction of \eq{Ulm} in
\eq{Wkh3} gives:

\bea\label{Wkh4} &&W_{k+1}^{(h)}(p,\bfp_K) = \cr && - \sum_s
\Res_{p'\to a_s} \sum_{r=1}^{d_2} \sum_{ K_0 \bigcup K_1 \bigcup
\dots \bigcup K_r = K} \sum_{h_{\alpha} = 0}^h
\sum_{k_{\alpha}=|K_\alpha|}^{k+h} \sum_{j_{\alpha,\beta} \neq
j_{\alpha',\beta'} \in [1,d_2]} {1 \over \Omega '} \cr && \quad
dS_{p',o}(p) W_{|K_0|+1}^{(h_0)}(p',\bfp_{K_0}) \prod_{\alpha=1}^r
{W_{k_{\alpha}+1}^{(h_{\alpha})}(p'^{j_{\alpha,0}},\bfp_{K_{\alpha}}
,p'^{j_{\alpha,1}}, \dots
,p'^{j_{\alpha,k_\alpha-|K_\alpha|}}) \over
dx(p')^{k_\alpha-|K_\alpha|+1} \prod_{\beta =
0}^{k_\alpha-|K_\alpha|}   (y(p')-y(p'^{j_{\alpha,\beta})})}\cr
&& - \sum_s \Res_{p'\to a_s}
{{U}_{k+1}^{(h-1)}(p',y(p');p',\bfp_K) \over E_y(x(p'),y(p'))}
dS_{p',\alpha}(p) \cr \eea

That is to say: \bea
\begin{array}{r}
{\epsfxsize 4cm\epsffile{Wh.eps}}
\end{array}
& = & \sum_{r=1}^{d_2} \sum_{h_\alpha} \sum_{K_0\cup
K_1\cup\dots\cup K_r=K} {1 \over \Omega'}
\begin{array}{l}
{\epsfxsize 5.5cm\epsffile{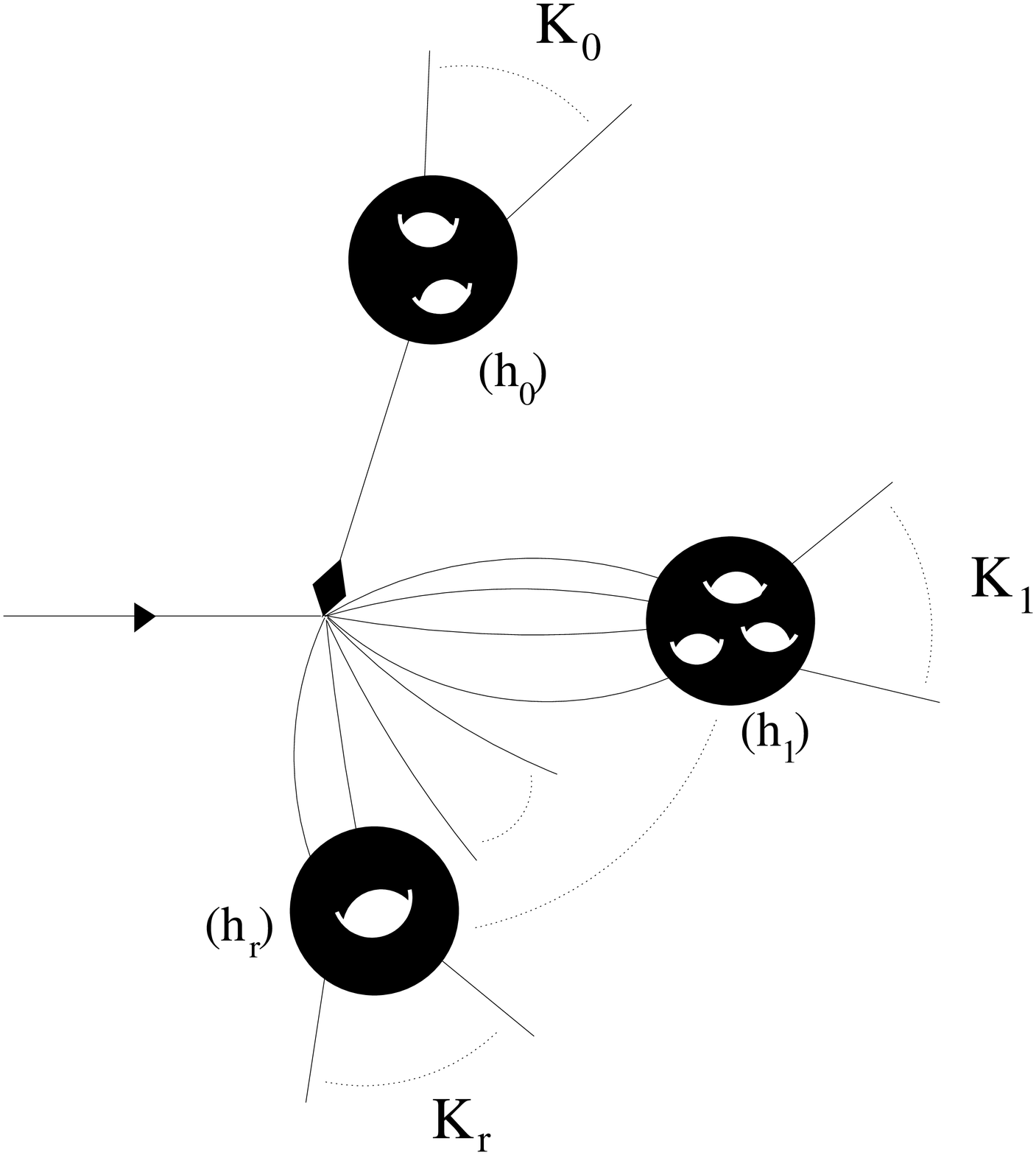}}
\end{array} \cr
&& + \sum_{r=1}^{d_2} \sum_{h_\alpha} \sum_{K_1\cup\dots\cup
K_r=K} {1 \over \Omega'}
\begin{array}{l}
{\epsfxsize 5.5cm\epsffile{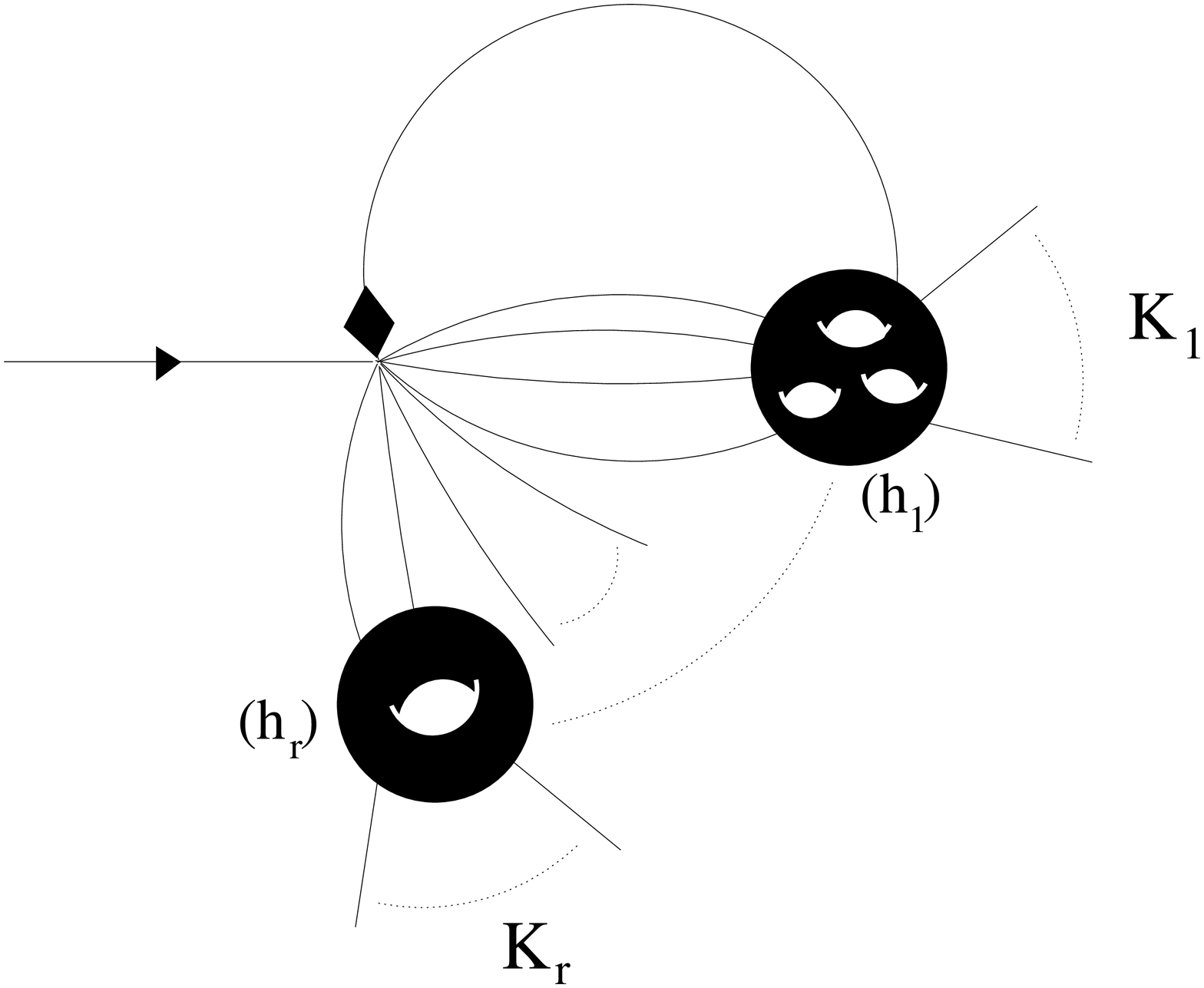}}
\end{array}
\eea

Remark that we have splitted the diagrams in the RHS in order to
reproduce the recursion relation. Nevertheless, the first term in
the RHS is nothing else but a particular case of the second term
where the marked leg of the vertex is left alone inside one of the
$W$'s.

\bigskip

Hence, the $h$-th order expansion term of the correlation function
$W_{k+1}^{(h)}$ is obtained as the {\em summation over all Feynman
diagrams with $k+1$ external legs and $h$ loops} following the same
rules as exposed in the genus 0 case, i.e.: {\em \begin{itemize}
\item The vertices have valence r+2 such as $1 \leq r \leq d_2$;
\item The edges are arrowed or not; \item One of the legs of each
vertex is marked; \item The subgraph made of arrowed edges forms a
skeleton tree; \item The k leaves are non arrowed propagators
ending at $p_j$'s;
\item The root is an arrowed propagator starting from $p$;
\item a non arrowed edge links a vertex to one of its descendants along the tree.
\end{itemize}}

\subsection{Examples}

Let us review some simple examples of this description.

\beq W_3^{(0)}(p,p_1,p_2) = \begin{array}{l} {\epsfxsize
10cm\epsffile{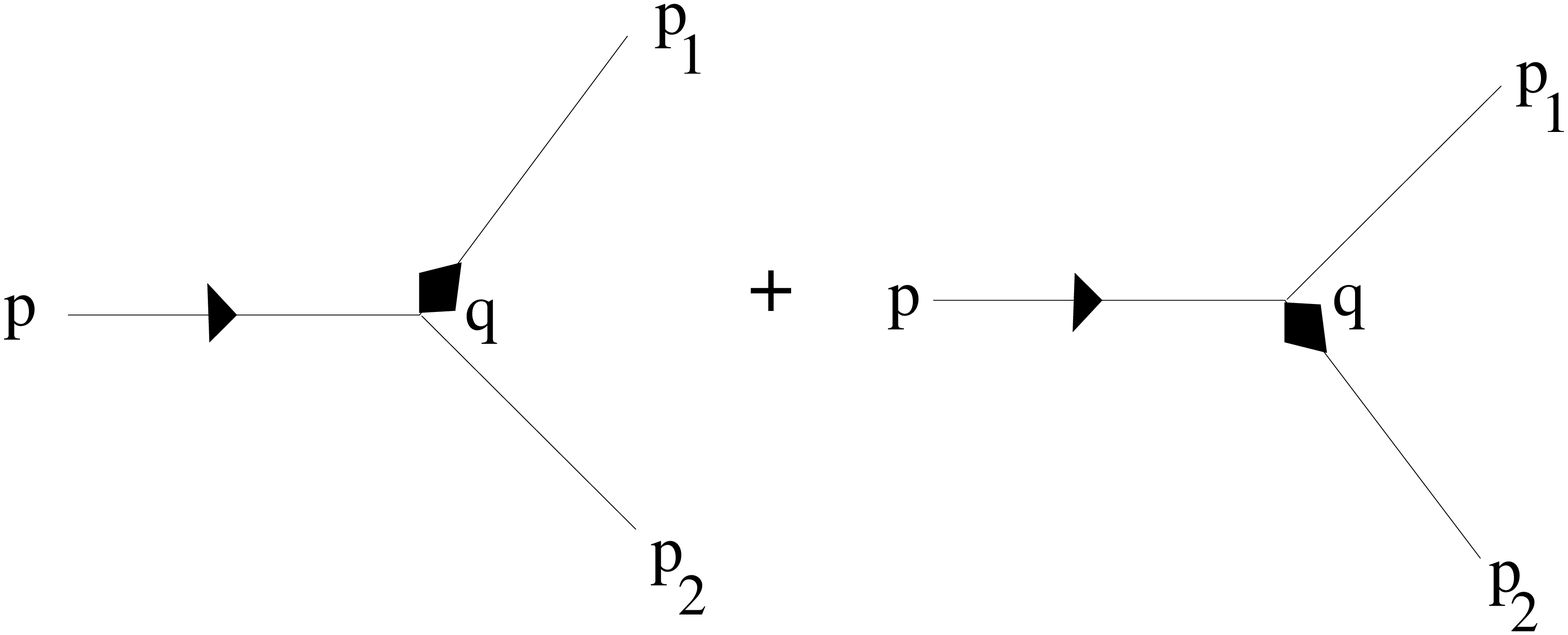}}
\end{array}
\eeq

Analytically, this reads: \beq
\begin{array}{l}
W_3^{(0)}(p,p_1,p_2) = \cr \sum_{i=1}^{d_2}\sum_s \Res_{q \rightarrow
a_s} \left[ B(q^{i},p_1) B(q,p_2)  + B(q^{i},p_2) B(q,p_1)
\right] {dS_{q,\a}(p) \over (y(q^{i})-y(q)) dx(q)}
\end{array}
\eeq

\bigskip

\bea W_1^{(1)}(p) &=& \begin{array}{l} {\epsfxsize
6cm\epsffile{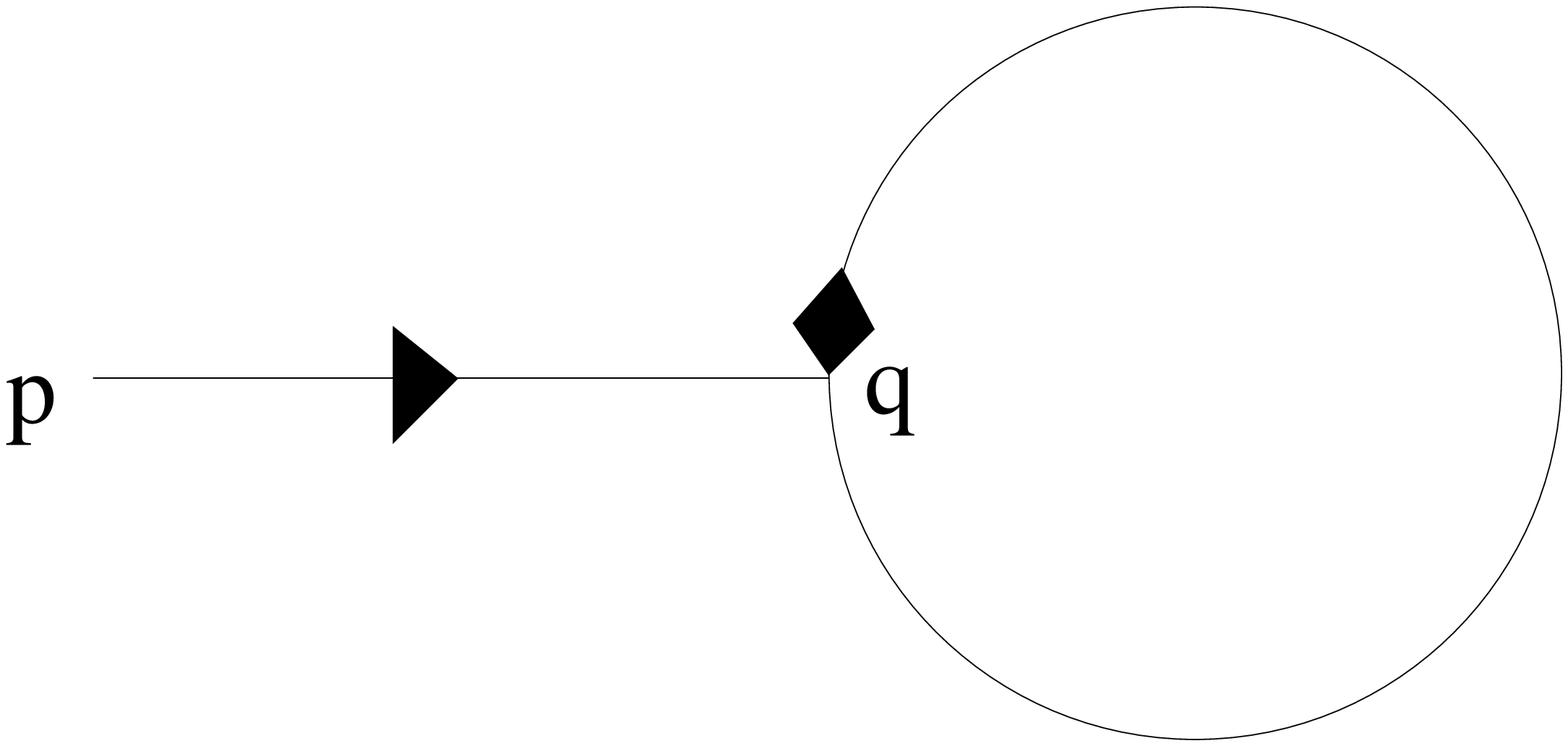}}
\end{array}\cr
&=& \sum_s \sum_{i=1}^{d_2} \Res_{q\rightarrow a_s} dS_{q,o}(p)
{B(q,q^{i}) \over (y(q^{i})-y(q)) dx(q)} \eea

\bigskip

\bea && W_2^{(1)}(p,p_1) =\cr && \begin{array}{l} {\epsfxsize
6cm\epsffile{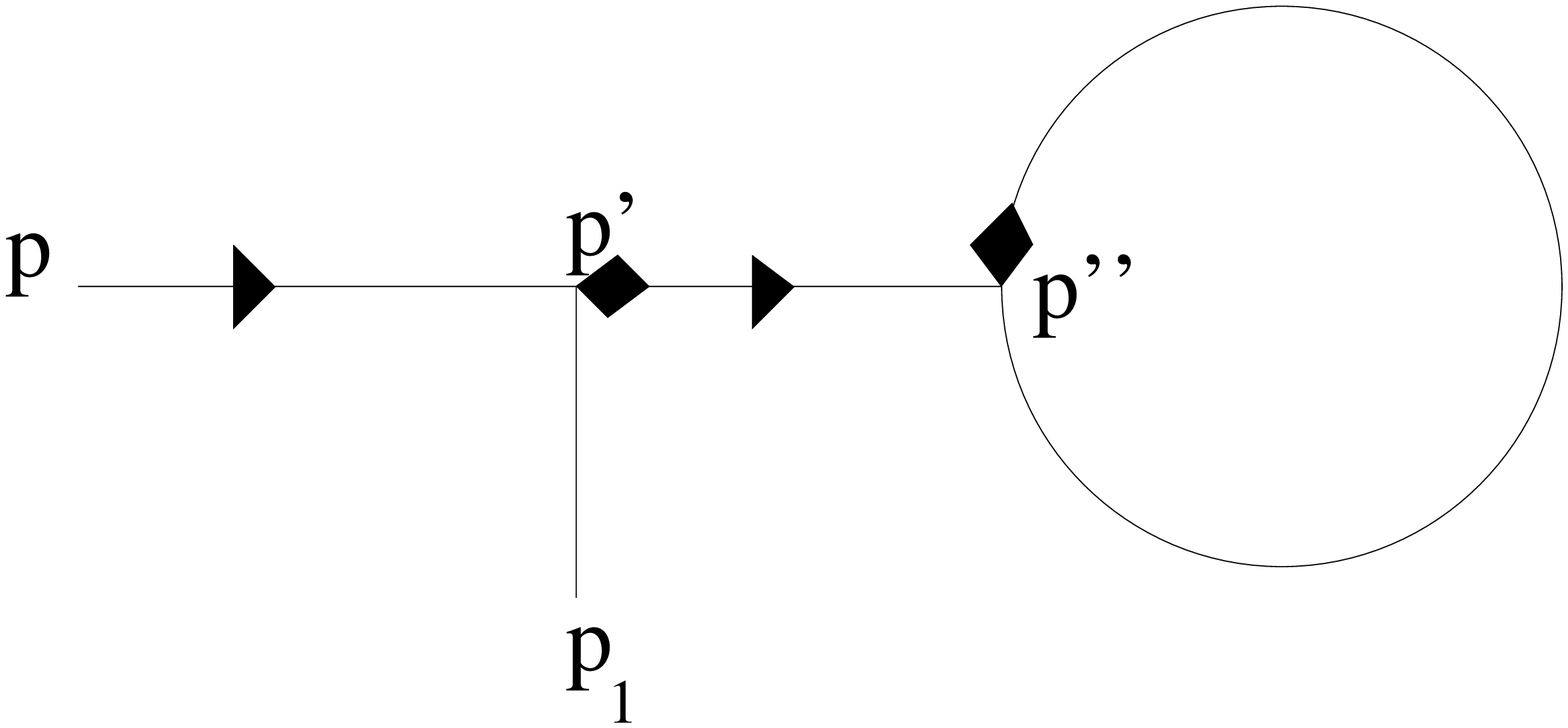}}
\end{array}
+
\begin{array}{l}
{\epsfxsize 6cm\epsffile{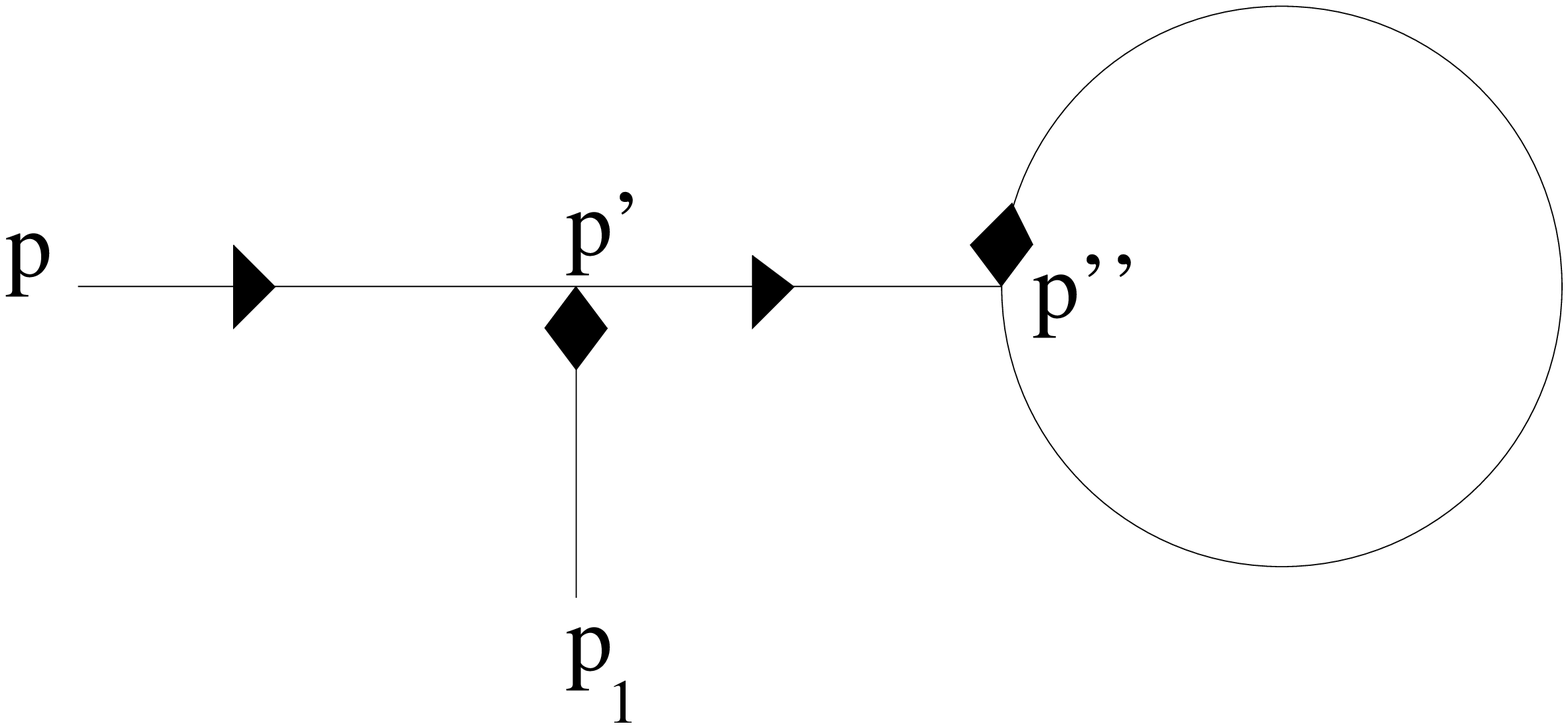}}
\end{array} \cr
&& +{1 \over 2} \begin{array}{l} {\epsfxsize
6cm\epsffile{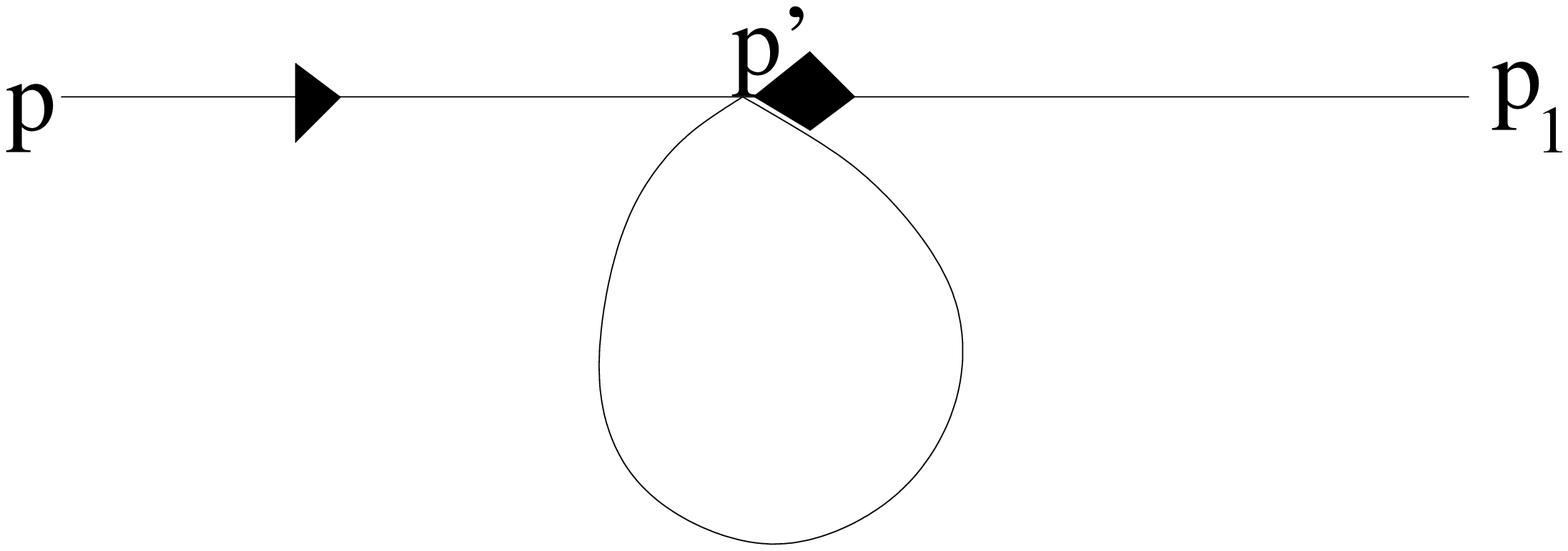}}
\end{array}
+
\begin{array}{l}
{\epsfxsize 6cm\epsffile{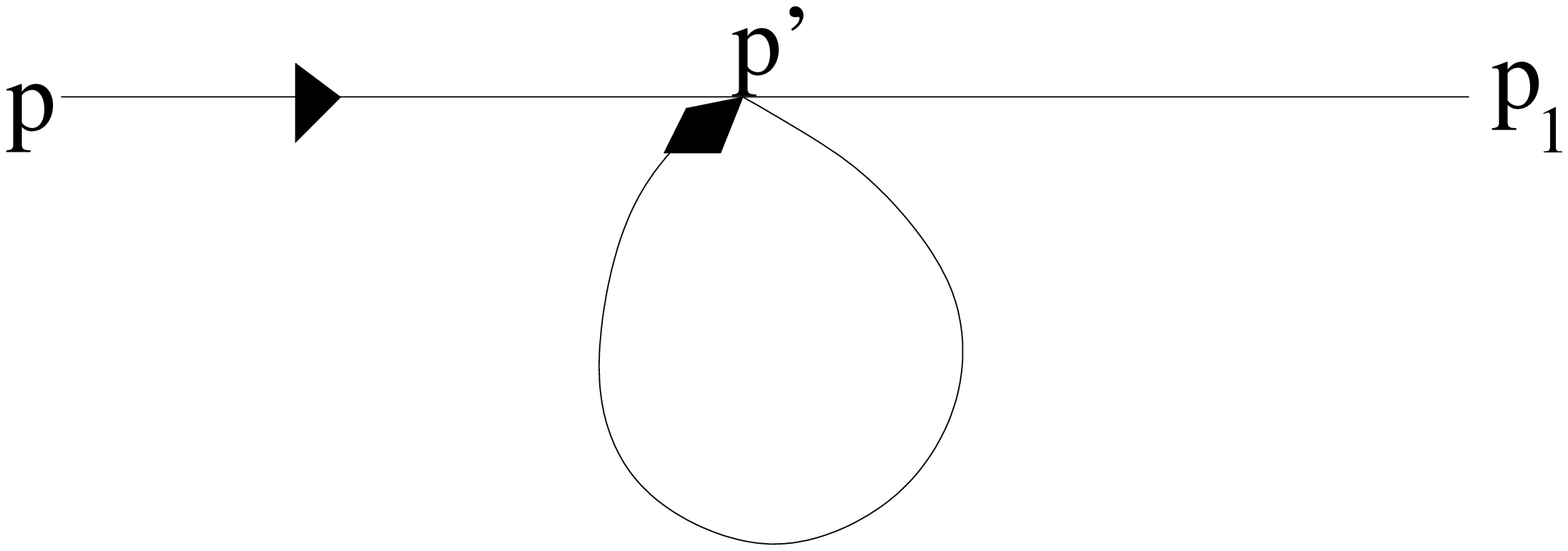}}
\end{array}\cr
&& + \begin{array}{l} {\epsfxsize 6cm\epsffile{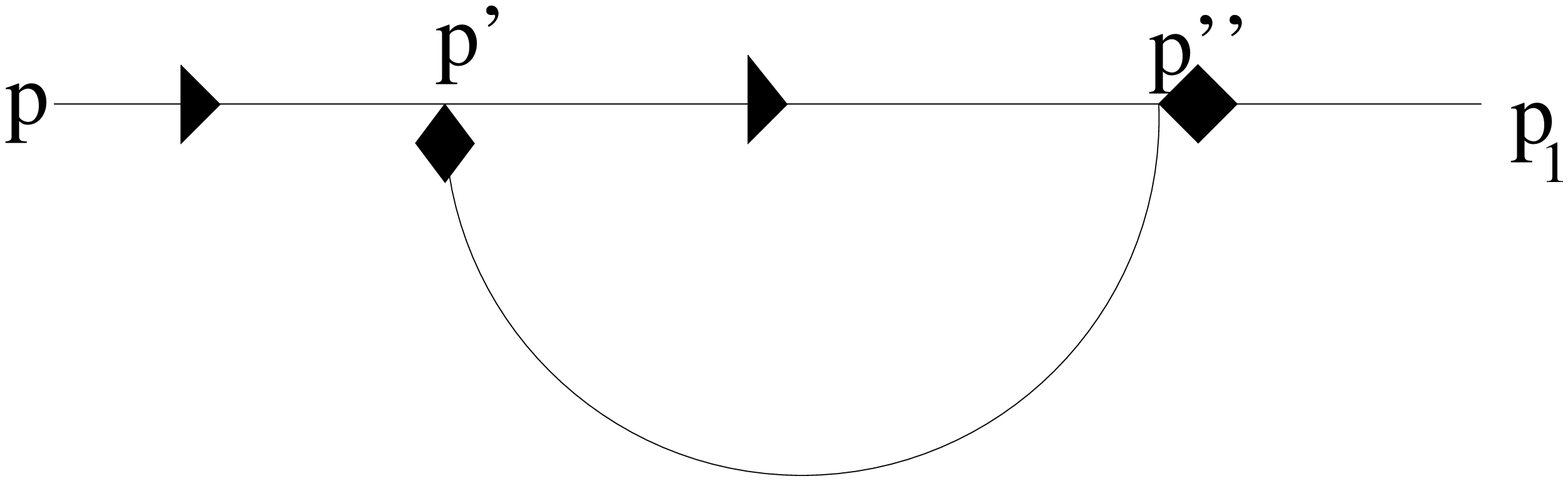}}
\end{array}
+\begin{array}{l} {\epsfxsize 6cm\epsffile{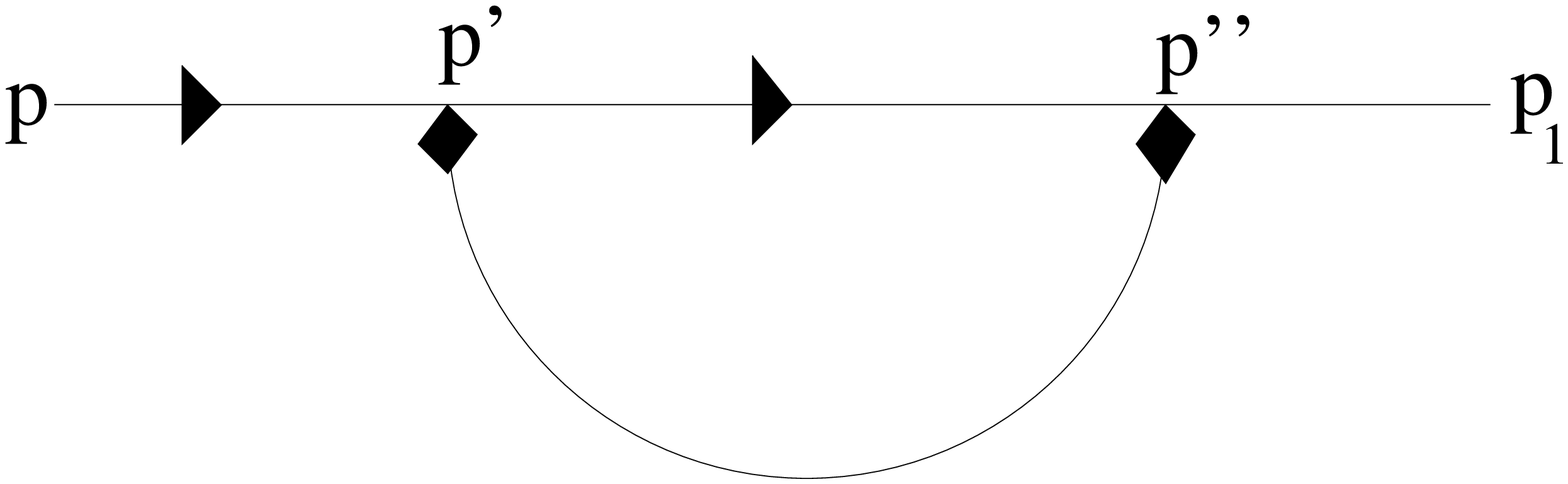}}
\end{array}
\eea

\newsection{The gaussian case: the 1-matrix model limit.}
In this section, we are interested in the special case where
$d_2=1$, i.e. one has a gaussian potential in $M_2$. This
situation is very important because it links our results to the
1-matrix model studied in \cite{eynloop1mat}. Indeed, when one of
the potentials is gaussian -- $V_2$ for example --, the
integration over one of the variables -- $M_2$ in this case -- is
gaussian and can be straightforwardly performed without giving any
contribution to the formal expansion. Then, the 2-matrix model
with one gaussian potential $V_2(y)= {g_2 \over 2} y^2$ is equivalent to the 1-matrix model
with a potential $V = V_1 - {x^2 \over 2 g_2}$. We check in this
part that our results coincide with the ones obtained directly
from the 1-matrix model in \cite{eynloop1mat}. Actually, it is a
good way to better understand the structure obtained.

In this case, the Riemann surface is an hyperelliptical surface with
only two $x$-sheets. The equation $x(p)=x$ has only two solutions.
Let us call them $p$ and $\overline{p}$, i.e. $p^{0} =p$ and
$p^{1}=\overline{p}$. They obey the following relations: \beq
x(p)= x(\overline{p}) \and y(p) = - y(\overline{p}) \eeq

The algebraic equation generating the Riemann surface reads: \beq
E(x(p),y(r))= - g_2 (y(r)-y(p)) (y(r)-y(\overline{p})) = - g_2
(y(r)^2 - y(p)^2) \eeq

One can also remark that: \beq U_{k}(p,y;\bfp_K) = g_2
W_{k+1}(p,\bfp_K) \eeq

That is to say: \beq R_k^{0}(p,\bfp_K) = {U_{k}(p,y(p);\bfp_K)
\over E_y(x(p),y(p)) dx(p)} = - {W_{k+1}(p,\bfp_K) \over 2 y(p) dx(p)} \eeq

So that: \beq R_k^{0}(\overline{p},\bfp_K) = R_k^{1}(p,\bfp_K) =
{W_{k+1}(p,\bfp_K) \over 2 y(p) dx(p)} \eeq

\subsection*{Diagrammatic rules.}
One can now study how the diagrammatic rules
introduced earlier  behave in this limit.

\begin{itemize}
\item {\bf The cubic rules}

Because $V_2$ is gaussian, the Feynman rules become:

\begin{center}
\begin{tabular}{|r|l|}\hline
non-arrowed propagator:& $
\begin{array}{r}
{\epsfxsize 2.5cm\epsffile{W2.eps}}
\end{array}
:=W_{2}(p,q) $ \cr\hline
arrowed propagator:& $
\begin{array}{r}
{\epsfxsize 2.5cm\epsffile{arrowedpropagator.eps}}
\end{array}
 :=dS_{q,o}(p)
$ \cr\hline
Residue cubic-vertex:& $
\begin{array}{r}
{\epsfxsize 2.5cm\epsffile{ncvertex.eps}}
\end{array}
 := \sum_s Res_{q\rightarrow a_s}
$ \cr\hline
simple vertex:& $
\begin{array}{r}
{\epsfxsize 2cm\epsffile{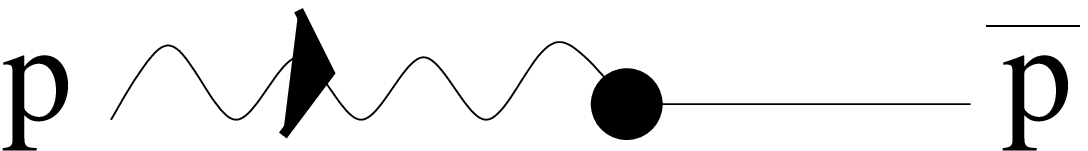}}
\end{array}
:= -{1 \over 2 y(p) dx(p)} $ \cr
\hline
\end{tabular}
\end{center}

The last component of the Feynman diagrams, the colored
cubic-vertex, implies three different $x$-sheets. Because there
exists only two such sheets in the gaussian case, this vertex
vanishes: \beq
\begin{array}{r}
{\epsfxsize 3.5cm\epsffile{cvertex.eps}}
\end{array}
\equiv  0 \eeq

Considered that the bivalent and trivalent vertices only appear
together, one can merge them into one whose value is equal to $-
\sum_s \Res_{q\rightarrow a_s} {1 \over 2 y(q) dx(q)}$, and one
recovers \cite{eynloop1mat}: \beq
\begin{array}{r}
{\epsfxsize 3cm\epsffile{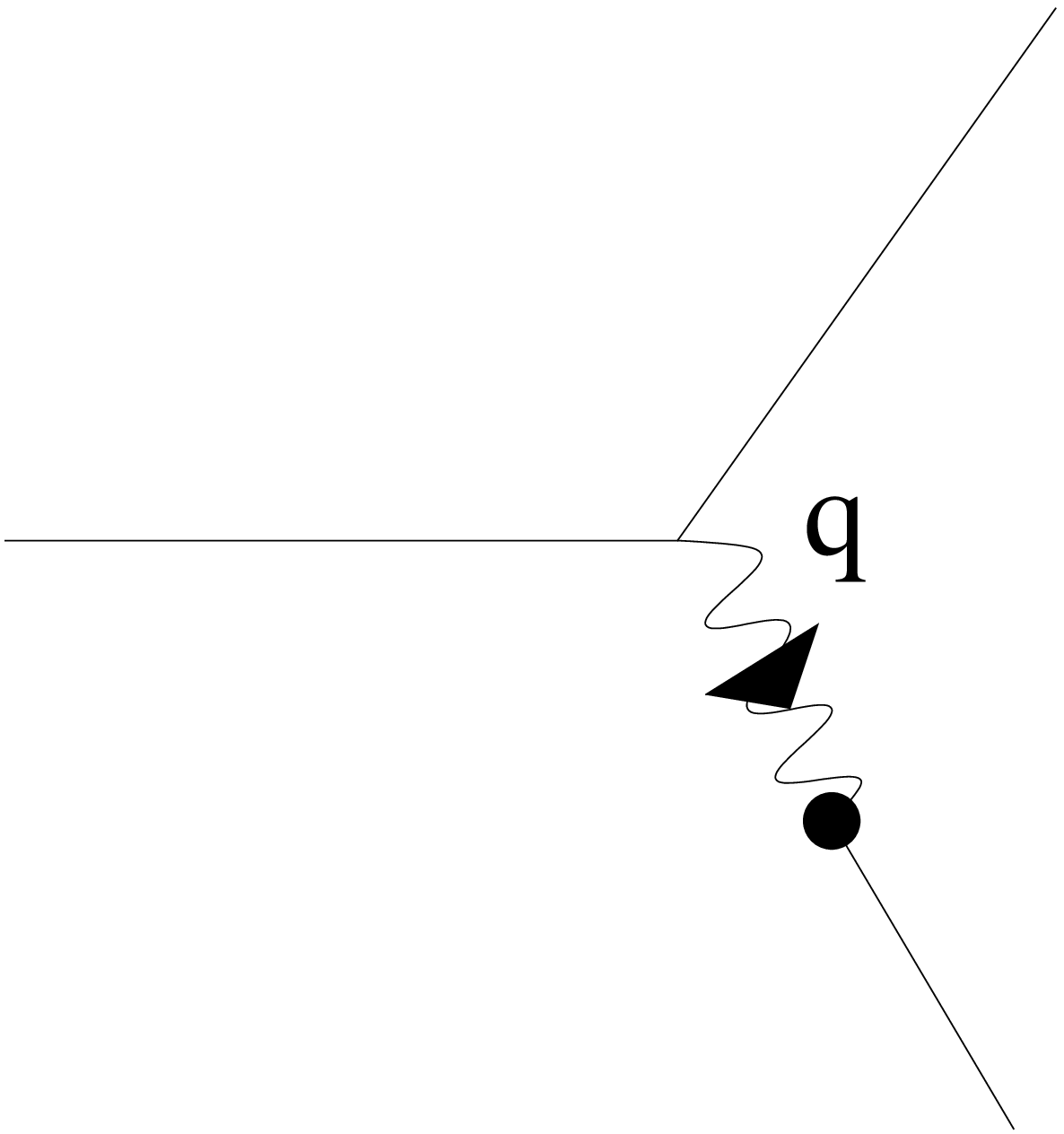}}
\end{array}
\rightarrow
\begin{array}{r}
{\epsfxsize 3cm\epsffile{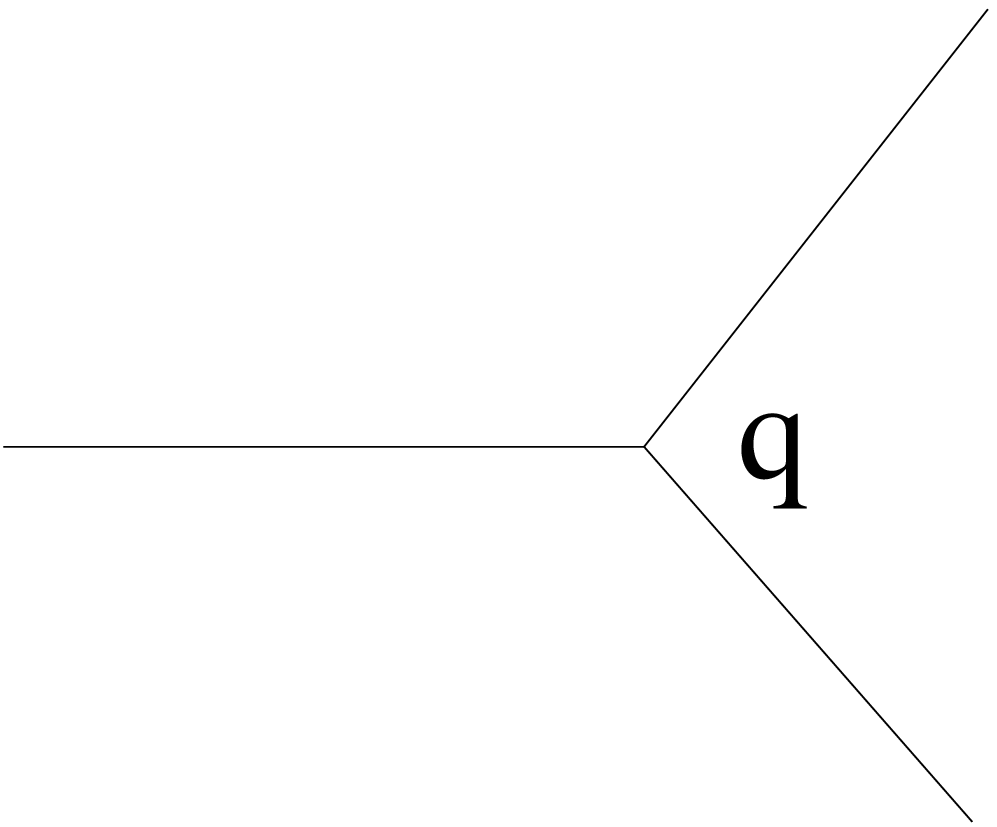}}
\end{array}
\eeq

\item {\bf The effective theory}

The effect of the gaussian limit on the effective theory is to
make it cubic. One obtains the following rules:
\begin{center}
\begin{tabular}{|r|l|}\hline
non-arrowed propagator:& $
\begin{array}{r}
{\epsfxsize 2.5cm\epsffile{W2.eps}}
\end{array}
:=W_{2}(p,q) $ \cr\hline
arrowed propagator:& $
\begin{array}{r}
{\epsfxsize 2.5cm\epsffile{arrowedpropagator.eps}}
\end{array}
 :=dS_{q,o}(p)
$ \cr\hline
\begin{tabular}{c}
cubic vertex \cr (only for r=1):
\end{tabular}&
$
\begin{array}{r}
{\epsfxsize 3.8cm\epsffile{multiplevertex.eps}}
\end{array}
 := - \sum_s  \Res_{q\rightarrow a_s} {1 \over 2 y(q) dx(q)}
$ \cr
\hline
\end{tabular}
\end{center}

\end{itemize}

Hence, the two theories turn into only one cubic theory in this
limit which is the one derived in \cite{eynloop1mat}. Indeed, the
corresponding recursive relation appears to be: \bea
W_{k+1}^{(h)}(p,p_K) &=& - \sum_l \Res_{q \to a_l}
{W_{k+2}^{(h-1)}(q,q,p_K) dS_{q,o}(p) \over 2 y(q) dx(q)} \cr && -
\sum_{m=0}^h \sum_{j=0, j+m \neq 0}^k \sum_l \Res_{q \to a_l}
{W_{j+1}^{(m)}(q,p_J) W_{k-j+1}^{(h-m)}(q,p_{K-J}) dS_{q,o}(p)
\over 2 y(q) dx(q)} \eea

{\bf Remark:}

Diagrammatically, this limit can be easily interpreted. Starting
from the general cubic theory, in order, to obtain the 1-matrix
model graphs from the 2-matrix model ones, one only has to take
the length of the waved propagators to 0. In this case, the
graphs containing at least one colored vertex vanish.

Everything works as if the waved propagators of the 2-matrix
model were unstable particles which decay into stable ones
represented by non-waved propagators. Then the 1-matrix
limit is obtained by taking the life time of these particles to 0.
\vs

One shall also note that there is no symmetry factor in the
2-matrix model graphs of the cubic theory whereas there are not
well understood ones in the 1-matrix case. The derivation of the
1-matrix model as a limit exhibits how these factors arise. They
come from the same contribution given by different diagrams in
this limit. This observation exhibits how the 2-matrix model seems
more fundamental.

\eop

\newsection{Conclusion}
In this article, we have generalized the diagrammatic technique of
\cite{eynloop1mat} to compute all non-mixed correlation functions
of the 2-matrix model, to all orders in the topological expansion.

The result can be represented diagrammatically, with some cubic
Feynman rules, which are just convenient notations for writing
residues on an algebraic curve and it is not clear whether there exists a
field theory giving rise to these graphs or not.

This shows that the method discovered in \cite{eynloop1mat} is
very universal, i.e. it works for all algebraic curves, not only
hyper elliptical curves.

The future prospects of that work are to find the diagrammatic
rules for computing the free energy to all order in the
topological expansion, and also all mixed correlation functions
(using the result of \cite{EOtrmixte}). Another possible extension
is to work out the multimatrix model, i.e.  the chain of matrices
as in \cite{eynmultimat}, and in particular the limit of matrix
quantum mechanics. We believe that this technique could apply to
many other integrable models.

Another question, is to understand the limit of critical points,
i.e. when some branch points and double points start to coalesce.
It seems that the diagrammatic technique should just reduce to
consider only residues at branch points which become critical. One
may expect to recover some relation with the Kontsevich integral,
in relationship with KP integrable hierarchies.

\subsection*{Acknowledgments}
The authors want to thank L.Chekhov, I. Kostov, V. Kazakov for
stimulating discussions. This work was partly supported by the
european network Enigma (MRTN-CT-2004-5652).

\eop

\setcounter{section}{0}

\appendix{Needed tools of algebraic geometry}

We review here some definitions and properties all along this article.

\bigskip
{\bf Behaviors at $\infty$.} We see from \eq{defY}, that at large
$x$, we have $y\sim V'_1(x)-{1\over x} +O(1/x^2)$ in the
$x$-physical sheet. (resp. at large $y$, we have $x\sim
V'_2(y)-{1\over y} +O(1/y^2)$ in the $y$-physical sheet). This
means that the functions $x(p)$ and $y(p)$ have two poles,
$\infty_+$ and $\infty_-$ on ${\cal E}$. The function $x(p)$ has a
simple pole at $\infty_+$ and a pole of degree $d_2$ at
$\infty_-$, while the function $y(p)$ has a simple pole at
$\infty_-$ and a pole of degree $d_1$ at $\infty_+$. We have:
\beq\label{largey} y(p)\mathop\sim_{p\to\infty_+} V'_1(x(p)) -
{1\over x(p)} + O(1/x(p)^2) \eeq \beq\label{largex}
x(p)\mathop\sim_{p\to\infty_-} V'_2(y(p)) - {1\over y(p)} +
O(1/y(p)^2) \eeq In particular: \beq \Res_{\infty_+} y\, dx =
\Res_{\infty_-} x\, dy = 1 \eeq

\bigskip
{\bf Genus and cycles.} The curve ${\cal E}$ is a compact
Riemann surface with a finite genus $g\leq d_1 d_2 -1$. If $g=0$,
${\cal E}$ is simply connected, and if $g\neq 0$, there exist $2g$
linearly independent irreducible cycles on ${\cal E}$, such that
by removing those $2g$ cycles we get a simply connected domain. It
is possible to choose canonically the $2g$ cycles as ${\cal A}_i$,
${\cal B}_i$, $i=1,\dots, g$, such that: \beq {\cal A}_i\cap {\cal
A}_j=0 \virg {\cal B}_i\cap {\cal B}_j=0 \virg {\cal A}_i\cap
{\cal B}_j=\delta_{ij} \eeq

\bigskip
{\bf Branch points.} The $x$-branch points $a_i$, $i=1,\dots,
d_2+1+2g$, are the zeroes of the differential $dx$, respectively,
the $y$-branch points $b_i$, $i=1,\dots, d_1+1+2g$, are the zeroes
of $dy$. We assume here, that all branch points are simple and
distinct, i.e. that the potentials are not critical. Notice also,
that $E_y(x(p),y(p))$ vanishes (simple zeroes) at the branch points
(it vanishes in other points too).

\bigskip
{\bf Bergmann kernel.} On the Riemann surface ${\cal E}$, there
exists a unique Abelian bilinear differential $B(p,q)$, with one
double pole at $p=q$, such that: \beq
B(p,q)\mathop\sim_{p\to q} {dx(p)dx(q)\over (x(p)-x(q))^2}+{\rm
finite} \quad {\rm and} \quad \forall i \,\,\,\oint_{{p\in{\cal
A}_i}} B(p,q) = 0 \eeq It is symmetric: \beq B(p,q)=B(q,p) \eeq
Its expression in terms of theta-functions can be found in
\cite{Fay, Farkas}, it depends only on the complex structure of
${\cal E}$.

\bigskip
{\bf Abelian differential of third kind.}

On the Riemann surface ${\cal E}$, there exists a unique abelian
differential of the third kind $dS_{q,r}(p)$, with two simple
poles at $p=q$ and at $p=r$, such that: \beq \Res_{p
\to q} dS_{q,r}(p) = 1 = -\Res_{p \to r} dS_{q,r}(p) \quad {\rm
and} \quad \forall i \,\,\,\oint_{{{\cal A}_i}} dS_{q,r}(p) = 0
\eeq We have: \beq dS_{q,r}(p)  = \int_{q'=r}^{q} B(p,q') \eeq
where the integration path does not intersect any ${\cal A}_i$ or
${\cal B}_i$.

$dS_{q,r}(p)$ is a differential on ${\cal E}$ in terms of $p$, but
it is a multivalued function of $q$ (and of $r$). After crossing a
cycle ${\cal B}_i$, it has no discontinuity, and after crossing a
cycle ${\cal A}_i$, it has a discontinuity: \beq {\rm
disc\,}(dS_{q,r}(p)) = dS_{q_+,r}(p)-dS_{q_-,r}(p) = \oint_{q'\in
{\cal B}_i} B(p,q') \eeq Note that the discontinuity is
independent of $q$.

\bigskip
{\bf Riemann bilinear identity.}

If $\omega$ is a differential form on ${\cal E}$, such that
$\oint_{q\in {\cal A}_i} \omega(q)=0$, we have:
\bea\label{RiemannbilinearId} \sum_i \Res_{q\to z_i} \om(q)
dS_{q,r}(p) &=& \sum_{i=1}^g\oint_{q\in{\cal A}_i} {\rm
disc}_{{\cal A}_i}(\om(q) dS_{q,r}(p)) \cr && -
\sum_{i=1}^g\oint_{q\in{\cal B}_i} {\rm disc}_{{\cal B}_i}(\om(q)
dS_{q,r}(p)) \cr &=& \sum_{i=1}^g\oint_{q\in{\cal A}_i}
\om(q)\,{\rm disc}_{{\cal A}_i}( dS_{q,r}(p)) \cr &=& \sum_{i=1}^g
{\rm disc}_{{\cal A}_i}( dS_{q,r}(p))\,\oint_{q\in{\cal A}_i}
\om(q) \cr &=&0 \eea where the LHS is the sum over all residues on
a fundamental domain, the poles $z_i$ are all the poles of $\om$
as well as the pole at $q=p$. This identity is obtained by moving
the integration contours on the surface, and taking carefully into
account discontinuities along the nontrivial cycles (see
\cite{Fay, Farkas}).

\appendix{Two points function in the planar limit}

We present here a new derivation leading term of the 2-point
function's leading term $W_{2}(p_1,p_2)$.

This case is of special interest because it represents some initial
condition for the diagrammatic rules. In fact, the two correlation
functions $W_{2}(p_1,p_2)$ and $U_{1}(p_1,y;p_2)$, are the basis
of the whole structure of the $W_{k}^{(h)}$'s. Moreover, it allows
us to show through a simple example the way we proceed further for
the general case.

We first rederive the well known result that the two point function
is nothing else but the Bergmann Kernel (see \cite{MarcoF} for instance).

\bigskip

Let $o\in{\cal E}$ be an arbitrary point on the Riemann surface. Since the Abelian differential of the 3rd kind defined in \eq{defdS}
$dS_{q,o}(p)$ behaves as ${dx(p)\over x(p)-x(q)}$ when $q\to p$, one can write the Cauchy formula under the form:
\beq
W_{2}(p,p_1) = -\Res_{q\to p} dS_{q,o}(p) W_{2}(q;p_1)
\eeq

One can see from \eq{eqWk} with $k=1$, and from \eq{vanishingAcycles}, that the integrand in the RHS has poles only for $q \to p$ and $q\to p_1$,
Since $W_{2}$ has vanishing ${\cal A}$-cycles due to \eq{vanishingAcycles},
we can use the Riemann bilinear identity \eq{RiemannbilinearId}, and get:
\bea
W_{2}(p,p_1) &=&  \Res_{q\to p_1} dS_{q,o}(p) W_{2}(q;p_1)
\eea

For $k=1$, \eq{eqWk} reads: \bea\label{eqW1}
E_y(x(p),y(p))\,W_{2}(p,p_1) &=& - P_{1}(x(p),y(p);p_1) dx(p) \cr
&& + d_{p_1} \left({U_{0}(p_1,y(p))\over x(p)-x(p_1)} \,
{dx(p)\over dx(p_1)}\right) \cr \eea and thus we have: \bea
W_{2}(p;p_1) &=&  \Res_{q\to p_1} dS_{q,o}(p) W_{2}(q;p_1) \cr &=&
- \Res_{q\to p_1} dS_{q,o}(p) {P_{1}(x(q),y(q);p_1) dx(q) \over
E_y(x(q),y(q))}  \cr && +  \Res_{q\to p_1} dS_{q,o}(p) { d_{p_1}
\left({U_{0}(p_1,y(q))\over x(q)-x(p_1)} \, {dx(q)\over
dx(p_1)}\right)\over E_y(x(q),y(q))}  \cr \eea Since
$P_{1}(x(q),y(q);p_1)$ is a polynomial in $x(q)$ and $y(q)$, it
has no pole at $q=p_1$. For the second term we use \eq{uoo}: \bea
W_{2}(p;p_1) &=&  \Res_{q\to p_1} dS_{q,o}(p) W_{2}(q;p_1) \cr &=&
d_{p_1}\,\Res_{q\to p_1} dS_{q,o}(p) { E(x(p_1),y(q))\,dx(q)\over
(x(q)-x(p_1))(y(q)-y(p_1)) E_y(x(q),y(q))}  \cr &=&
d_{p_1}\,dS_{p_1,o}(p) \cr &=&   B(p_1,p) \cr \eea We thus recover
the well known result: the two-points function is equal to the
Bergmann kernel on the Riemann surface corresponding to the
algebraic equation $E(x,y)=0$ (cf \cite{MarcoF, KazMar,
eynmultimat, Kri}).

\beq \label{prop1} \encadremath{ W_{2}(p;p_1)=B(p,p_1) }\eeq

\vs

Let us now compute $U_{1}(p,y;p_1)$.
For $k=1$, \eq{eqUk} reads:
\bea\label{eqU1}
{(y(r) - y(q) )U_{1}(q,y(r);p_1) \over dx(q)}
&=&  - {W_{2}(q;p_1) U_{0}(q,y(r))\over dx(q)^2}
 - P_{1}(x(q),y(r);p_1)   \cr
 && + d_{p_1}\left( {U_{0}(p_1,y(r))\over (x(q)-x(p_1))\,dx(p_1)}\right)
\eea
take it for $q=r=p^{i}$:
\bea
0&=&  - {W_{2}(p^{i};p_1) U_{0}(p^{i},y(p^{i}))\over dx(p^{i})^2}
 - P_{1}(x(p^{i}),y(p^{i});p_1)   \cr
 && + d_{p_1}\left( {U_{0}(p_1,y(p^{i}))\over (x(p^{i})-x(p_1))\,dx(p_1)}\right)
\eea
using that $x(p)=x(p^{i})$, we have:
\bea\label{eqU1intermediate}
0&=&  - {W_{2}(p^{i};p_1) U_{0}(p^{i},y(p^{i}))\over dx(p)^2}
 - P_{1}(x(p),y(p^{i});p_1)   \cr
 && + d_{p_1}\left( {U_{0}(p_1,y(p^{i}))\over (x(p)-x(p_1))\,dx(p_1)}\right)
\eea
Now, write \eq{eqU1} with $q=p$ and $r=p^{i}$:
\bea
{(y(p^{i}) - y(p) )U_{1}(p,y(p^{i});p_1) \over dx(p)}
&=&  - {W_{2}(p;p_1) U_{0}(p,y(p^{i}))\over dx(p)^2}
 - P_{1}(x(p),y(p^{i});p_1)   \cr
 && + d_{p_1}\left( {U_{0}(p_1,y(p^{i}))\over (x(p)-x(p_1))\,dx(p_1)}\right)
\eea
and insert \eq{eqU1intermediate}, you get:
\bea
(y(p^{i}) - y(p) )U_{1}(p,y(p^{i});p_1)
&=& {W_{2}(p^{i};p_1) U_{0}(p^{i},y(p^{i}))\over dx(p)} \cr
&& - {W_{2}(p;p_1) U_{0}(p,y(p^{i}))\over dx(p)} \cr
\eea

Using \eq{uoo}, i.e. $U_{0}(p,y) = {E(x(p),y)\over y-y(p)}dx(p)$, this implies:
\bea
(y(p^{i}) - y(p) )U_{1}(p,y(p^{i});p_1)
&=& {W_{2}(p^{i};p_1) E_y(x(p^{i)},y(p^{i}))}
\eea
Since $U_{1}(p,y;p_1)$ is a polynomial of degree $d_2-1$ in $y$, we can reconstruct it through the interpolation formula:
\bea
U_{1}(p,y;p_1) &=& {E(x(p),y)\over (y-y(p))}\,\sum_{i=1}^{d_2} {1\over y-y(p^{i})}\,{(y(p^{i}) - y(p) )U_{1}(p,y(p^{i});p_1)\over E_y(x(p^{i)},y(p^{i}))} \cr
\eea
i.e.
\bea
U_{1}(p,y;p_1) &=& {E(x(p),y)\over (y-y(p))}\,\sum_{i=1}^{d_2} {W_{2}(p^{i},p_1)\over y-y(p^{i})}
\eea

and in particular, at $y=y(p)$, we have:
\bea
R^0_{1}(p,p_1) dx(p) ={U_{1}(p,y(p);p_1)\over E_y(x(p),y(p))} &=& \sum_{i=1}^{d_2} {W_{2}(p^{i},p_1)\over y(p)-y(p^{i})}
\eea
and for $i\neq 0$, we have:

\beq \label{prop2}
R^i_{1}(p,p_1)
dx(p)={U_{1}(p,y(p^{i});p_1)\over E_y(x(p),y(p^{i}))}
={W_{2}(p^{i},p_1)\over (y(p^{i})-y(p))} \eeq

\appendix{Computation of Eq. (5.39) }
In this appendix one proves recursively \eq{eqUW} for any k and h.

Let us suppose that this formula is know for any $U_{l}^{(m)}$
with $m \leq h-1$ and for any $U_{l}^{(h)}$ with $l \leq k-1$. One
writes it: \beq\label{eqUWbis}
\begin{array}{l}
U_{l}^{(m)}(p,y(p^{i});\bfp_L)= \cr {E_y(x,y(p^{i})) \over
y(p^{i})-y(p)} \sum_{r=1}^{min(d_2,k+h)} \sum_{ L_1 \bigcup
\dots \bigcup L_r = L}  \sum_{m_{\alpha} = 0}^m
\sum_{l_\alpha=|L_\alpha|}^{l+m} \sum_{j_{\alpha,\beta} \neq
j_{\alpha',\beta'} \in [1,d_2]-\{i\}} {1 \over \Omega} \cr
{W_{l_1+1}^{(m_1)}(p^{i}, \bfp_{L_1} ,p^{j_{1,1}}, \dots
,p^{j_{1,l_1-|L_1|}}) \left(\prod_{\alpha=2}^{r}
W_{l_{\alpha}+1}^{(m_{\alpha})}(p^{j_{\alpha,0}},\bfp_{L_{\alpha}}
,p^{j_{\alpha,1}}, \dots
,p^{j_{\alpha,l_\alpha-|L_\alpha|}})\right) \over
dx(p)^{r-l-1+\sum L_\alpha} \prod_{\alpha,\beta}
y(p^{i})-y(p^{j_{\alpha,\beta}})}
\end{array}
\eeq

Let us introduce some shortened notations so that one can write
this proof in a few pages.

Considering the sum on the RHS of \eq{eqUWbis}, one can see that
there are two different kinds of terms:
\begin{itemize}
\item If $l_1=|L_1|$, one can factorise the term
$W_{|L_1|+1}^{(m_1)}(p^{i}, \bfp_{L_1})$. Let us note the sum of
these terms $W(p^{i},p_L) W(p_L, p^{j})$ where we have noted
$W$ instead of $W_{|L_1|+1}^{(m_1)}$ to indicate that these are
formal notations;

\item the other terms correspond to the sum over all $l_1 \neq
|L_1|$. Let us denote them by $W(p^{i},p_L,p^{j}) W(p_L, p^{j})$.

\end{itemize}

Using these notations, one can shortly write \eq{eqUWbis}: \beq
U_{l}^{(m)}(p,y(p^{i});\bfp_L)= W(p^{i},p_L) W(p_L, p^{j}) +
W(p^{i},p_L,p^{j}) W(p_L, p^{j}) \eeq

Thus the interpolation formula gives: \beq
U_{l}^{(m)}(p^{i},y(p^{i});p_L) = W(p_L, p^{j}) + W(p,p_L)
W(p_L, p^{j}) + W(p,p_L,p^{j}) W(p_L, p^{j}) \eeq

where the first term corresponds to the sum where all
$j_{\beta}$'s are different from $i$ and 0 and there is no
$W_{l_i}$ whose argument is $p$ or $p^{i}$.

On the other hand, one knows the relation \ref{solUkh}: \beq
\begin{array}{rcl}
 {U}_{k}^{(h)}(p,y(p^{i});\bfp_K) &=&\sum_{m=0}^{h} \sum_{j=0}^{k}  {{W}_{j+1}^{(m)}(p^{i},\bfp_J) {U}_{k-j}^{(h-m)}(p^{i},y(p^{i});\bfp_{K-J}) \over (y(p^{i})-y(p)) dx(p)}\cr
&& - \sum_{m=0}^{h} \sum_{j=0}^{k}  {{W}_{j+1}^{(m)}(p,\bfp_{J})
{U}_{k-j}^{(h-m)}(p,y(p^{i});\bfp_{K-J}) \over (y(p^{i})-y(p))
dx(p)}\cr && - {U_{k+1}^{(h-1)}(p,y(p^{i});p,\bfp_k) \over
(y(p^{i})-y(p)) dx} +
{U_{k+1}^{(h-1)}(p^{i},y(p^{i});p^{i},\bfp_k) \over
(y(p^{i})-y(p)) dx}\cr
\end{array}
\eeq

Remark that the terms in the RHS of this equation correspond to
the criterion of the hypothesis and one can then express them as
a product of $W$'s following the notations introduced earlier.
This reads: \beq
\begin{array}{l}
U_{k}^{(h)}(p,y(p^{i});p_K) =\cr
 W(p^{i},p_K) W(p_K, p^{j}) + W(p^{i},p_K) W(p,p_K) W(p_K, p^{j}) \cr
 + W(p^{i},p_K) W(p,p_K,p^{j}) W(p_K, p^{j}) - W(p,p_K) W(p^{i},p_K) W(p_K, p^{j}) \cr
 - W(p,p_K) W(p^{i},p_K,p^{j}) W(p_K, p^{j}) + W(p^{i},p_K, p^{j}) W(p_K, p^{j}) \cr
 + W(p^{i},p,p_K) W(p_K, p^{j}) + W(p,p_K) W(p^{i},p_K, p^{j}) W(p_K, p^{j}) \cr
 + W(p^{i},p,p_K,p^{j}) W(p_K, p^{j}) + W(p,p_K,p^{j}) W(p^{i},p_K, p^{j}) W(p_K, p^{j}) \cr
 - W(p,p^{i},p_K) W(p_K, p^{j}) - W(p^{i},p_K) W(p,p_K, p^{j}) W(p_K, p^{j}) \cr
 - W(p,p^{i},p_K,p^{j}) W(p_K, p^{j}) - W(p^{i},p_K,p^{j}) W(p,p_K, p^{j}) W(p_K, p^{j}) \cr
=  W(p^{i},p_K) W(p_K, p^{j}) + W(p^{i},p_K,p^{j}) W(p_K,
p^{j})\cr
\end{array}
\eeq

So one has proven the formula for $U_{k}^{(h)}$.

Because this formula is true for h=0, it is true for any k and h.

\eop

\appendix{Derivation of Eq. (5.40)}
One wants to show that: \beq \label{equality3}
\begin{array}{l}
\sum_{m=0}^{h} \sum_{j=0; mj\neq kh}^{k}  {W}_{j+1}^{(m)}(p,p_J)
{U}_{k-j}^{(h-m)}(p,y(p);p_{K-J}) + {U_{k+1}^{(h-1)}(p,y(p);p,p_k)
\over dx} = \cr \sum_{i=1}^{d_2} \sum_{m=0}^{h} \sum_{j=0; mj\neq
kh}^{k}  {W}_{j+1}^{(m)}(p^{i},p_J)
{U}_{k-j}^{(h-m)}(p^{i},y(p);p_{K-J}) \cr + \sum_{i=1}^{d_2}
{U_{k+1}^{(h-1)}(p^{i},y(p);p^{i},p_k) \over dx}\cr
\end{array}
\eeq

Let us compute the difference D between the two sides of the
equation by the introduction of \eq{eqUW} written with some few
different notations which are defined as follows:
\begin{itemize}
\item $l=r+h-\sum_\alpha h_\alpha$; \item $u_\beta =
\sum_{\epsilon = 1}^{\beta} (k_\epsilon - |K_\epsilon|) - \beta$.
\end{itemize}

One can then write: \beq
\begin{array}{l}
D=\sum_{m=0}^{h} \sum_{j=0; mj\neq kh}^{k}
{W}_{j+1}^{(m)}(p,p_J)E_y(x,y(p)) \sum_{i=1}^{d_2}{ 1 \over
y(p)-y(p^{i})} \cr \times \sum_{r=1}^{d_2} \sum_{o =1}^r
\sum_{h_{o} = 0}^{h-m} \sum_{k_{o}=0}^{k+h-j-m} \sum_{j_2 \neq
\dots \neq j_l \in [1,d_2]-\{i\}} \sum_{ K_1 \bigcup \dots \bigcup
K_r = K} {1 \over \Omega} \cr \times
{W_{k_1+1}^{(h_1)}(p^{i},p_{K_1},p^{j_{r+1}}, \dots
,p^{j_{r+u_1}}) \left(\prod_{\beta=2}^{r}
W_{k_{\beta}+1}^{(h_{\beta})}(p^{j_{\beta}},p_{K_{\beta}},p^{j_{r+u_{\beta-1}+1}},
\dots ,p^{j_{r+u_{\beta}}})\right) \over \prod_{\gamma=2}^l
y(p^{i})-y(p^{j_{\gamma}})}\cr - \sum_{i=1}^{d_2}
\sum_{m=0}^{h} \sum_{j=0; mj\neq kh}^{k}
{W}_{j+1}^{(m)}(p^{i},p_J) {E_y(x,y(p)) \over y(p)-y(p^{i})}
\cr \times \sum_{r=1}^{d_2} \sum_{o =1}^r \sum_{h_{o} = 0}^{h-m}
\sum_{k_{o}=0}^{k+h-j-m} \sum_{j_2 \neq \dots \neq j_l \in
[1,d_2]-\{i\}} \sum_{ K_1 \bigcup \dots \bigcup K_r = K} {1 \over
\Omega} \cr \times {W_{k_1+1}^{(h_1)}(p,p_{K_1},p^{j_{r+1}},
\dots ,p^{j_{r+u_1}}) \left(\prod_{\beta=2}^{r}
W_{k_{\beta}+1}^{(h_{\beta})}(p^{j_{\beta}},p_{K_{\beta}},p^{j_{r+u_{\beta-1}+1}},
\dots ,p^{j_{r+u_{\beta}}})\right) \over \prod_{\gamma=2}^l
y(p)-y(p^{j_{\gamma}})}\cr + E_y(x,y(p)) \sum_{i=1}^{d_2}{ 1
\over y(p)-y(p^{i})} \cr \times \sum_{r=1}^{d_2} \sum_{o =1}^r
\sum_{h_{o} = 0}^{h-1} \sum_{k_{o}=0}^{k+h} \sum_{j_2 \neq \dots
\neq j_l \in [1,d_2]-\{i\}} \sum_{ K_1 \bigcup \dots \bigcup K_r =
K} {1 \over \Omega} \cr \times \left[
{W_{k_1+1}^{(h_1)}(p^{i},p,p_{K_1},p^{j_{r+1}}, \dots
,p^{j_{r+u_1}}) \left(\prod_{\beta=2}^{r}
W_{k_{\beta}+1}^{(h_{\beta})}(p^{j_{\beta}},p_{K_{\beta}},p^{j_{r+u_{\beta-1}+1}},
\dots ,p^{j_{r+u_{\beta}}})\right) \over \prod_{\gamma=2}^l
y(p^{i})-y(p^{j_{\gamma})}} \right. \cr +
{W_{k_1+1}^{(h_1)}(p^{i},p_{K_1},p^{j_{r+1}}, \dots
,p^{j_{r+u_1}})
W_{k_{2}+1}^{(h_{2})}(p,p^{j_{2}},p_{K_{2}},p^{j_{r+u_{1}+1}},
\dots ,p^{j_{r+u_{2}}})  \over \prod_{\gamma=2}^l
y(p^{i})-y(p^{j_{\gamma}})}\cr \times \left.
\prod_{\beta=3}^{r}
W_{k_{\beta}+1}^{(h_{\beta})}(p^{j_{\beta}},p_{K_{\beta}},p^{j_{r+u_{\beta-1}+1}},
\dots ,p^{j_{r+u_{\beta}}})\right] \cr -\sum_{i=1}^{d_2}
{E_y(x,y(p)) \over y(p)-y(p^{i})} \cr \times \sum_{r=1}^{d_2}
\sum_{o =1}^r \sum_{h_{o} = 0}^{h-1} \sum_{k_{o}=0}^{k+h}
\sum_{j_2 \neq \dots \neq j_l \in [1,d_2]-\{i\}} \sum_{ K_1
\bigcup \dots \bigcup K_r = K} {1 \over \Omega} \cr \times \left[
{W_{k_1+1}^{(h_1)}(p^{i},p,p_{K_1},p^{j_{r+1}}, \dots
,p^{j_{r+u_1}}) \left(\prod_{\beta=2}^{r}
W_{k_{\beta}+1}^{(h_{\beta})}(p^{j_{\beta}},p_{K_{\beta}},p^{j_{r+u_{\beta-1}+1}},
\dots ,p^{j_{r+u_{\beta}}})\right) \over \prod_{\gamma=2}^l
y(p)-y(p^{j_{\gamma}})} \right.\cr +
{W_{k_1+1}^{(h_1)}(p,p_{K_1},p^{j_{r+1}}, \dots
,p^{j_{r+u_1}})
W_{k_{2}+1}^{(h_{2})}(p^{i},p^{j_{2}},p_{K_{2}},p^{j_{r+u_{1}+1}},
\dots ,p^{j_{r+u_{2}}})  \over \prod_{\gamma=2}^l
y(p^{i})-y(p^{j_{\gamma}})}\cr \times \left.
\prod_{\beta=3}^{r}
W_{k_{\beta}+1}^{(h_{\beta})}(p^{j_{\beta}},p_{K_{\beta}},p^{j_{r+u_{\beta-1}+1}},
\dots ,p^{j_{r+u_{\beta}}})\right]\cr
\end{array}
\eeq

The difference between the two first terms leaves only the terms
corresponding to $u_1\neq 0$ in the first one minus $u_1\neq 0$ in
the second one.

The difference between two last terms will allow us to compensate
the preceding ones. Indeed, the terms with $p^{i}$ and $p$
together in the same correlation function straightforwardly vanish
and one gets the exact opposite to the two first terms remaining.

Thus D=0 and the equality \ref{equality3} is proven.

\eop


\end{document}